\documentclass[aps,prb,10pt,twocolumn,amsmath,amssymb,showpacs,superscriptaddress]{revtex4-1}

\usepackage{graphicx}
\usepackage{dcolumn}
\usepackage{xfrac}
\usepackage{bm} 
\usepackage{amsmath}
\usepackage{subfigure}
\usepackage{gensymb}
\usepackage{hyperref}

\def\nn{\nonumber\\}
\def\beq{\begin{equation}}
\def\eeq{\end{equation}}
\def\bea{\begin{eqnarray}}
\def\eea{\end{eqnarray}}

\def\bomega{{\bm\Omega}}
\def\Komega{\K_n^{(\Omega)}}
\def\Kaomega{\K_{na}^{(\Omega)}}
\def\KBomega{\K_{n,\textrm{BZB}}^{(\Omega)}}

\def\Kchi{\K_n^{(\chi)}}
\def\Kred{{\cal K}}
\def\Kbzb{\Kred^\textrm{BZB}}

\def\Bnj{{\cal B}_{nj}}

\def\vf{\hat{\bf v}_F}
\def\ef{E_F}

\def\g2{\overline{\Gamma}}
\def\m2{\overline{\textrm{M}}}
\def\x2{\overline{\textrm{X}}}

\def\a{{\bf a}}
\def\b{{\bf b}}
\def\A{{\bf A}}
\def\k{{\bf k}}
\def\K{{\bf K}}
\def\G{{\bf G}}
\def\rr{{\bf r}}
\def\M{{\bf M}}
\def\ket#1{\vert#1\rangle}

\def\ip#1#2{\langle#1\vert#2\rangle}

\def\wt#1{\widetilde{#1}}

\def\nhat{\hat{\bf n}}

\def\bna{_{na}}

\def\Vt{\widetilde{V}}
\def\bnj{_{nj}}
\def\ch{\chi^{\phantom{I}}}

\DeclareMathOperator{\sgn}{sgn}

\newcommand{\eq}[1]{Eq.~(\ref{eq:#1})}
\newcommand{\eqs}[2]{Eqs.~(\ref{eq:#1}) and~(\ref{eq:#2})}
\newcommand{\equ}[1]{Equation~(\ref{eq:#1})}

\begin{document}

\title{Chiral degeneracies and Fermi-surface Chern numbers in bcc Fe}

\author{Daniel Gos\'albez-Mart\'{\i}nez}
\affiliation{Centro de F\'{\i}sica de Materiales, 
Universidad del Pa\'{\i}s Vasco, 20018 San Sebasti\'an, Spain}
\affiliation{
Donostia International Physics Center, 20018 San Sebasti\'an, Spain}
\author{Ivo Souza}
\affiliation{Centro de F\'{\i}sica de Materiales, 
Universidad del Pa\'{\i}s Vasco, 20018 San Sebasti\'an, Spain}
\affiliation{Ikerbasque Foundation, 48013 Bilbao, Spain}
\author{David Vanderbilt} 
\affiliation{ Department of Physics and Astronomy, Rutgers University,
  Piscataway, New Jersey 08854-8019, USA }

\date{\today}
\begin{abstract}
     
  The degeneracies in the spinor bandstructure of bcc~Fe are studied
  from first principles.  We find numerous isolated band touchings
  carrying chiral charges of magnitude one (Weyl points) or two
  (double-Weyl nodes), as well as nonchiral degeneracy loops (nodal
  rings).  Some degeneracies are located on symmetry lines or planes
  in the Brillouin zone and others at generic low-symmetry points,
  realizing all possible scenarios consistent with the magnetic point
  group.  We clarify the general theory relating the chiral band
  touchings to the Chern numbers of the Fermi sheets enclosing them,
  and use this approach to determine the Chern numbers on the Fermi
  surface of bcc~Fe.  Although most Fermi sheets enclose Weyl nodes,
  in almost all cases the net enclosed charge vanishes for symmetry
  reasons, resulting in a vanishing Chern number. The exceptions are
  two inversion-symmetric electron pockets along the symmetry line
  $\Delta$ parallel to the magnetization. Each of them surrounds a
  single Weyl point, leading to Chern numbers of $\pm 1$.  These small
  topological pockets contribute a sizable amount to the nonquantized
  part of the intrinsic anomalous Hall conductivity, proportional to
  their reciprocal-space separation. Variation of the Fermi level (or
  other system parameters) may lead to a touching event between Fermi
  sheets, accompanied by a transfer of Chern number between them.

\end{abstract}
\pacs{75.47.-m,75.50.Bb,71.18.+y,73.43.-f}

\maketitle


\section{Introduction}
\label{sec:intro}

The degeneracies in three-dimensional (3D) bandstructures that are not
lifted by the spin-orbit interaction are receiving increasing
attention in connection with topological states of matter.
Murakami~\cite{murakami-njp07} showed that in the absence of inversion
symmetry~($P$) a gapless crystalline phase can exist over a finite
region of parameter space between a topological insulator and a normal
insulator.  This {\it Weyl semimetal} phase is characterized by
topologically-protected isolated touching points --~the Weyl points
(WPs)~-- between valence and conduction bands, and its low-energy
excitations are described by a generalized Weyl equation of the form
$H(\k)=k_i\nu_{ij}\sigma_j$, where $\bf k$ is the wavevector relative
to the touching point and $\{ \sigma_i\}$ are the three Pauli matrices
plus the $2\times 2$ identity matrix. Similar phases were later
devised where time-reversal symmetry~($T$) is broken instead of
parity~$P$.\cite{wan-prb11,burkov-prl11}

Weyl semimetals are expected to have interesting transport properties.
In particular, the magnetic variety can have a nonzero intrinsic
anomalous Hall conductivity (AHC) proportional to the separation in
$k$~space between WPs of opposite chirality,\cite{yang-prb11} where
the chirality of a WP is $c=\sgn(\det \nu)=\pm 1$.  Like other
topological phases, Weyl semimetals display a bulk-boundary
correspondence: pairs of WPs of opposite chirality at the Fermi level
$\ef$ lead to metallic ``Fermi arc'' surface states connecting their
projections onto the surface Brillouin zone
(BZ).\cite{wan-prb11,turner-book13,haldane-arxiv14}

Real-world examples of Weyl semimetals have been recently
identified. Compounds in the TaAs family were predicted to be
$P$-breaking Weyl semimetals,\cite{huang-natcomm15,weng-prx15} and the
observation of Fermi-arc surface states using angle-resolved
photoemission spectroscopy was reported soon afterwards for both TaAs
and NbAs.\cite{xu-arxiv15a,lv-arxiv15,xu-arxiv15b} Another promising
candidate is BiTeI, which is known to undergo a normal-to-topological
transition under pressure, and must have a Weyl phase for some
interval of pressure\cite{liu-prb14b} even if it has not yet been
observed. There has also been recent progress in realizing a related
phase that preserves both $P$ and $T$ symmetries, the {\it Dirac
  semimetal}.\cite{liu-science14,neupane-natcom14} Here the bands are
everywhere Kramers-degenerate, so that a total of four bands meet at
the Fermi point.

More generally, isolated band touchings can occur at arbitrary
energies in 3D bandstructures with broken $PT$ symmetry. Under those
conditions accidental degeneracies that occur away from symmetry lines
and planes have codimension three, which implies that the parameters
$(k_x,k_y,k_z$) provide just enough degrees of freedom to bring a pair
of bands together at generic isolated points in the
BZ.\cite{murakami-njp07} With some exceptions (see below), isolated
degeneracies in a 3D parameter space are robust topological objects,
acting as monopole sources and sinks of Berry
curvature.\cite{berry-prsla84} This accounts for their remarkable
stability, as well as for many of the interesting phenomena associated
with them.

Weyl points, defined as linear crossings carrying a topological
(chiral) charge $\pm 1$, are not the only type of isolated degeneracy
in 3D bandstructures. Quadratic or cubic crossings carrying a charge
of $\pm 2$ or $\pm 3$ are also possible along certain symmetry axes;
they can be viewed as several WPs of the same chirality brought
together by point group symmetry.\cite{fang-prl12} We will use the
term {\it point node} (PN) to denote a generic isolated degeneracy,
and reserve the terms {\it Weyl point} and {\it double-Weyl node} for
the specific types that we will encounter in bcc Fe (triple-Weyl nodes
are disallowed in tetragonal ferromagnets, as they can only occur
along a six-fold axis).

In addition, it is possible to arrange, via external fine tuning, for
isolated band touchings of other types.  This can occur, for example,
in the context of a normal-to-topological transition in a $P$-broken,
$T$-invariant system; at the critical parameter values at which pairs
of Weyl nodes with opposite chiralities are first created or finally
annihilated, there is a quadratic band touching of zero overall chiral
charge.\cite{murakami-njp07,murakami-prb08} Since these are not
generic, however, they will not play any further role in our
considerations.

In some cases symmetry can glue bands together at a high-symmetry
point or along a high-symmetry line. For ferromagnetic crystals with
the spin-orbit interaction included in the Hamiltonian, a
group-theoretical
analysis\cite{falikov-pr68,cracknell-jpc69,cracknell-prb70} has shown
that such degeneracies can be present when the structure is hexagonal
(e.g., hcp Co), but not when it is tetragonal. Indeed we have not
encountered this type of degeneracy in our present study of the spinor
bandstructure of bcc Fe.

Finally, the spinor energy bands of a ferromagnet can also remain
degenerate along entire loops lying on BZ planes that are invariant
under reflection (i.e., mirror planes).  We will refer to this type of
degeneracy as a {\it line node} or {\it nodal ring}.

In metals, chiral degeneracies in the bandstructure may induce nonzero
Chern numbers on the Fermi-surface (FS) sheets. This possibility was
first considered by Haldane in connection with the FS formulation of
the intrinsic AHC.\cite{haldane-prl04,haldane-arxiv14}
Topologically-nontrivial Fermi sheets have also been discussed in the
context of topological superconductivity,\cite{qi-prb10,hosur-prb14}
and of the chiral magnetic effect and related effects in Weyl
semimetals.\cite{zhou-cpl13} Metals with topological Fermi sheets are
sometimes called {\it topological
  metals},\cite{haldane-arxiv14,weng-prx15} and the ideal Weyl
semimetal corresponds to the limiting case in which the topological
pockets collapse onto isolated Fermi points (and no additional Fermi
sheets are present).

In spite of all these formal developments, little is known about the
occurence of PNs, nodal rings, and especially topologically-nontrivial
Fermi sheets in everyday real materials.  Are they extremely rare, or
very common, in the electronic structure of typical $T$-broken
crystals such as ferromagnetic metals?  How does their presence affect
physical observables associated with the $k$-space Berry curvature,
notably the AHC?  A few studies of chiral degeneracies have been
conducted recently on model bandstructures of
photonic\cite{lu-natphot13} and ferromagnetic\cite{chen-prb13}
crystals, and on {\it ab~intio} bandstructures of $P$-broken
semiconductors and Weyl
semimetals.\cite{hirayama-prl15,huang-natcomm15,weng-prx15,huang-arxiv15b}
Otherwise, however, there has been remarkably little discussion in the
literature, and in particular, virtually no systematic searches for
topological Fermi sheets using first-principles methods.

In this paper, we clarify the formal relations between chiral PNs in
the bandstructure of metals with broken $PT$ symmetry, and the Chern
numbers of the individual Fermi sheets.  Furthermore, we examine in
detail the role of chiral PNs and topological Fermi sheets in the
intrinsic AHC of $T$-broken metals.  To that effect, we decompose the
AHC band-by-band and in terms of Fermi sheets, and show how the two
decompositions are related by $k$-space dipole moments of the occupied
chiral PNs, which vanish upon summing over all bands. This analysis
clarifies further (see also Ref.~\onlinecite{vanderbilt-prb14}) why
the presence of chiral degeneracies below $\ef$ is not an impediment
to a purely FS formulation of the nonquantized AHC, contrary to a
recent claim.\cite{chen-prb13} We also elucidate the relation between
two alternative FS formulations of the AHC, one in terms of Berry
curvatures and Berry phases,\cite{haldane-prl04} the other in terms of
Berry phases only.\cite{wang-prb07}

In order to see how these ideas play out in real materials, we decided
to give a complete census of all the degeneracies and their
topological charges, and then use that information to determine the
Fermi-sheet Chern numbers, in one or more simple ferromagnetic
metals. Taking bcc~Fe as our first test case, we found to our surprise
that chiral PNs are astonishingly ubiquitous in the bandstructure.
For example, we find 90 of them between bands six and seven alone! In
the present work we therefore chose to concentrate entirely on bcc Fe
as a paradigmatic system to illustrate the concepts and search
strategies. Interestingly, we find that bcc Fe is, at least within our
density-functional theory calculation using a generalized-gradient
approximation (see Sec.~\ref{sec:methods}), a $T$-broken topological
metal in the sense of Haldane.\cite{haldane-arxiv14} That is, a pair
of Fermi pockets have equal and opposite nonzero Chern numbers and
contribute significantly to the AHC.

The complex bandstructure of bcc Fe proves to be a flexible arena for
exploring the physics of topological metals beyond simple Weyl
semimetals. For example, we demonstrate that Chern numbers can be
transferred between Fermi sheets by varying the Fermi level or an
external parameter such as the magnetization direction. Topological
transitions of this kind have been discussed previously in the
literature for model
Hamiltonians,\cite{klinkhamer-ijmpa05,volovik-lectnotes07} but the
detailed mechanisms have not been investigated for real metals.  Even
though the touching events leading to the Chern-number exchange occur
locally between sheets on adjacent bands, in bcc~Fe we find that these
events are often concerted, e.g., such that the Chern number gets
transferred between two sheets on the {\it same} band via an
intermediary ``passive'' sheet in the band below.

We emphasize from the outset the crucial role played by the spin-orbit
interaction in the present study, where the Bloch states are treated
as spinors. Iron is a mostly collinear ferromagnet with fairly weak
spin-orbit coupling (SOC), but for our purposes ``weak SOC'' is
completely different from ``zero SOC,'' as it changes qualitatively
the nodal structure of the bands and the connectivity of the Fermi
surface.\cite{gold-jltp74,singh-prb75,lonzarich-book80} Moreover, SOC
induces a $k$-space Berry curvature on the Bloch states by
transmitting the breaking of time-reversal from the spins to the
orbital degrees of freedom. Without SOC, the up-spin (minority) and
down-spin (majority) bands are decoupled, and in each spin channel the
spatial wavefunctions are $T$-invariant (in addition to being
$P$-invariant).  Without SOC, the Berry curvature vanishes identically
in iron as a result of this effective $PT$ symmetry, and generic
like-spin degeneracies have codimension two, not three, i.e., they are
line nodes instead of isolated point nodes.  For opposite-spin
crossings one gets entire nodal surfaces on which
$E_{n\uparrow}(\k)=E_{m\downarrow}(\k)$. However, even a weak SOC gaps
these curves and surfaces almost everywhere, reducing the locus of
degeneracies to a collection of discrete points (and occasionally, on
mirror planes, a few loops). The weakness of the spin-orbit
interaction makes for a challenging computational problem, since true
band crossings coexist with, and must be distinguished from, minute
spin-orbit-induced avoided crossings.

The paper is organized as follows.  Some definitions, notation, and
basic relations are given in Sec.~\ref{sec:definitions}.  In
Sec.~\ref{sec:sheets} we work out the relations that allow the
determination of the Chern numbers of the Fermi sheets from a
knowledge of the population of chiral degeneracies in the energy
bands.  In Sec.~\ref{sec:ahc} we consider the role played by chiral
PNs in the theory of the intrinsic AHC, particularly in relation to
FS-based formulations. In Sec.~\ref{sec:methods} we describe the
electronic structure methods that were used in our calculations on
bcc~Fe. The numerical results are presented and discussed in
Sec.~\ref{sec:results}, and we conclude in Sec.~\ref{sec:summary} with
a summary.

\section{Definitions and basic relations}
\label{sec:definitions}

\subsection{Berryology}
\label{sec:berryology}

The $k$-space Berry connection of band~$n$
\beq 
\A_n(\k)=i\ip{u_{n\k}}{\bm{\nabla}_\k u_{n\k}} 
\eeq
is defined in terms of the cell-periodic Bloch states
$\ket{u_{n\k}}=e^{-i\k\cdot\rr}\ket{\psi_{n\k}}$. A geometric phase
(Berry phase) can be associated with a closed path ${\cal C}$ in
$k$-space by taking the circuit integral of the connection,
\beq
\label{eq:berry-phase}
\varphi_n({\cal C})=\oint_{\cal C} d\k\cdot\A_n(\k)\,.
\eeq
The Berry curvature is the curl of the connection,
\beq
\label{eq:curv}
\bomega_n(\k)={\bm\nabla}_\k\times {\bf A}_n(\k)\,,
\eeq
so that according to Stokes' theorem $\nhat\cdot\bomega_n(\k)$ has the
interpretation of a Berry phase per unit area, for a small planar loop
with unit surface-normal $\nhat$. The Berry connection is
gauge-dependent, meaning that its value at~$\k$ can be changed
continuously by modifying the phase choice for the Bloch eigenstates
around~$\k$. The Berry phase is gauge-invariant apart from a $2\pi$
indeterminacy, and the Berry curvature is fully gauge-invariant. Like
the energy bands $E_n(\k)$, the Berry curvature has the periodicity of
the reciprocal lattice, ${\bomega}_n(\k+\G)=\bomega_n(\k)$.

{\it Chern's theorem.} The Berry-curvature flux through a closed
oriented surface ${\cal S}$ in $k$-space is equal to $2\pi C_n({\cal
  S})$, where $C_n({\cal S})$ is an integer known as the Chern number:
\beq 
\label{eq:C-n-S}
C_n({\cal S})=
\frac{1}{2\pi}\oint_{\cal S} dS\,\nhat\cdot\bomega_n(\k)\,.  
\eeq
This statement is valid provided that band~$n$ remains nondegenerate
over the entire surface.

A two-dimensional BZ is effectively a closed surface by virtue of the
periodicity of $k$ space, so that Chern's theorem applies, and the
same is true for a 2D section of a 3D BZ cut along reciprocal-lattice
vectors. Consider a cubic lattice (simple cubic, bcc, or fcc): the
Chern number of band~$n$ on a BZ slice taken at fixed $k_z$ is
\beq
\label{eq:C-n-kz}
C_n(k_z)=\frac{1}{2\pi}\int_{\text{slice}} d^2k\,\Omega_{n,z}(\k)\,,
\eeq
where the integral is over a primitive cell on the $(k_x,k_y)$ plane
at fixed $k_z$.  Viewed as a function of $k_z$, the slice Chern number
$C_n(k_z)$ is a piecewise constant integer function. It can only
change at critical $k_z$ values where band $n$ comes in contact with a
contiguous band $n\pm 1$ at isolated points; when that happens,
$C_{n\pm 1}(k_z)$ changes by a compensating amount, and the process
can be viewed as an exchange of an integer amount of Chern number
between the two bands.\cite{kohmoto-prb92} The periodicity condition
\beq
\label{eq:per}
C_n(k_z+2\pi/a)=C_n(k_z)
\eeq
($a$ is the cubic lattice constant) implies that the smallest nonzero
number of integer steps over one period is two.

Another example of a closed surface is an isolated Fermi sheet.  Even
though some Fermi sheets may look open because of being cut at the BZ
boundary, they are in fact closed manifolds in the topological sense,
when equivalence under reciprocal-lattice translations is factored in
(as for example for the well-known shape of the Fermi surface in Cu).
Thus Chern's theorem applies and isolated Fermi sheets always have a
topological index.\cite{haldane-prl04,haldane-arxiv14} (The
possibility that two sheets touch, either as a result of fine tuning
or because of symmetry, should be kept in mind.)

One can also associate a Chern number with a composite group of bands
over a closed surface. We will consider the group formed by the
$n$~lowest bands, and define, for a BZ slice,
\beq
\label{eq:Ctil-n-kz}
\wt{C}_n(k_z)=\sum_{l=1}^n\,C_l(k_z)\,.
\eeq
The index $\wt{C}_n(k_z)$ inherits the properties already discussed
for $C_n(k_z)$, but it only reacts to touching events between the
uppermost band $n$ in the group and band $n+1$ above, and is instead
oblivious to band crossings within the group.

A related quantity of interest is the Berry flux through a bounded
oriented surface ${\cal S}$ in the BZ,
\beq
\label{eq:berry-flux}
\phi_n({\cal S})=\int_{\cal S}dS\,\hat{\bf n}\cdot\bomega_n(\k)\,.
\eeq
For example, ${\cal S}$ could be a patch on a BZ slice, threaded by a
nonquantized flux $\phi_n$, or an entire slice, now viewed as an open
rectangle rather than a closed 2-torus (quantized flux $\phi_n=2\pi
C_n$). If a smooth gauge is chosen for the Bloch states on the
boundary~${\cal C}$ of~${\cal S}$, then the Berry phase computed on
${\cal C}$ is equal, modulo $2\pi$, to the Berry flux through ${\cal
  S}$,\footnote{Although a $2\pi$ ambiguity is inherent to the
  definition of the Berry phase, its presence in \eq{flux2phase} may
  seem puzzling from the point of view of Stokes' theorem. Note
  however that Stokes' theorem cannot be applied globally to a BZ
  slice with nonzero Chern number using a periodic gauge, as it is not
  possible to extend a smooth gauge choice from the boundary into the
  interior: there will always be vortex-like singularities left at
  isolated points, which provide additional $\pm 2\pi$
  contributions.\cite{kohmoto-annphys85,hatsugai-prl93} Gauge
  discontinuities (pairs of vortices and anti-vortices) can in fact be
  generated even when the slice Chern number vanishes.}
\beq
\label{eq:flux2phase}
\varphi_n({\cal C})=\phi_n({\cal S})\,\text{mod}\, 2\pi\,.
\eeq

\subsection{Isolated band touchings (point nodes)}

We will denote by $W_{n\alpha}$ the $\alpha^{\rm th}$ PN between bands
$n$ and $n+1$, located at $(\k_{n\alpha},E_{n\alpha})$ and with
integer chiral charge $\chi_{n\alpha}$.  It follows from \eq{curv}
that ${\bm\nabla}\cdot\bomega_n\!=\!0$ everywhere except at the PNs,
which act as monopole sources of Berry flux.  Our sign convention for
$\chi$ is that the PN acts as a source of $2\pi\chi$ Berry flux in the
lower band, and as a sink of $2\pi\chi$ Berry flux in the upper
band. Thus,
\bea
\label{eq:div}
{\bm\nabla}\cdot\bomega_n(\k)&=&
2\pi\sum_{\alpha}\,\chi_{n\alpha}\delta^3(\k-\k_{n\alpha})\nn
&-&2\pi\sum_{\alpha}\,\chi_{n-1,\alpha}\delta^3(\k-\k_{n-1,\alpha})\,.
\eea

For any connected subvolume $\cal V$ of the BZ with boundary ${\cal
  S}$, the divergence theorem applied to band $n$ gives
\beq
\label{eq:div-thm}
\oint_{\cal S} dS\,\nhat\cdot\bomega_n=
\int_{\cal V} dV\,{\bm\nabla}\cdot\bomega_n\,,
\eeq
where the unit normal $\nhat$ points away from ${\cal V}$. If we use
this unit normal to orient the surface ${\cal S}$, then according to
\eq{C-n-S} the left-hand side equals $2\pi C_n({\cal S})$, and using
\eq{div} the right-hand side becomes the total Berry flux pumped into
band $n$ from PNs connecting to band $n+1$ above, minus the total
Berry flux sucked into band $n-1$ from PNs connecting to band $n-1$,
\beq
\label{eq:C-n-S-chi}
C_n({\cal S})=\sum_{W_{n\alpha}\in {\cal V}}\,\chi_{n\alpha}
      -\sum_{W_{n-1,\alpha}\in {\cal V}}\,\chi_{n-1,\alpha}\,.
\eeq

In the context of \eq{C-n-kz} we can apply the divergence theorem to
the BZ subvolume between two parallel slices. Because the fluxes
through the side faces cancel out in pairs, this gives the difference
between the Chern numbers on the top and bottom slices as the sum of
the chiral charges in between. For slices separated by $\Delta
k_z=2\pi/a$ the periodicity condition (\ref{eq:per}) implies that
\beq
\label{eq:sum-rule-a}
\sum_\alpha\,\chi_{n\alpha}-\sum_\alpha\,\chi_{n-1,\alpha}=0\,,
\eeq
where the sum is now over PNs in the full BZ.  Repeating the argument
for \eq{Ctil-n-kz} we obtain a stronger sum rule for the net charge of
all PNs connecting two adjacent bands,
\beq
\label{eq:sum-rule-b}
\sum_\alpha\,\chi_{n\alpha}=0\,.
\eeq
Alternatively, \eq{sum-rule-b} can be proved by induction starting
from \eq{sum-rule-a}, which for $n=1$ gives $\sum_\alpha
\chi_{1\alpha}=0$.

A single PN with $\chi\not=0$ cannot be eliminated from the
bandstructure by varying a control parameter, as this would violate
the above ``charge neutrality'' condition.  A chiral node can only
appear or disappear as part of a concerted chirality-conserving event,
as for example when two WPs of opposite chirality merge and
annihilate. Chiral degeneracies can be detected by measuring the
quantized Berry flux coming out of a small enclosing box. If a
box~${\cal S}$ encloses a single node of charge $\chi_{n\alpha}$ then,
according to \eqs{C-n-S}{C-n-S-chi},
\beq 
\label{eq:chi}
\chi_{n\alpha}=\frac{1}{2\pi}\oint_{\cal S}dS\,
\nhat\cdot\bomega_n(\k)\,.  
\eeq

\subsection{Lines of degeneracy (line nodes)}
\label{sec:line-nodes}

Line nodes can be thought of as ``flux tubes'' carrying a quantized
Berry flux of $\pi$. That is, the Berry phase of \eq{berry-phase}
around any small loop ${\cal C}$ encircling the line node is equal to
$\pi$.\cite{burkov-prb11,kim-prl15} Line nodes do not carry a net
chiral charge, and are therefore less robust than chiral PNs against
translationally-invariant perturbations. They are often ``protected''
by crystal symmetries, and lowering the symmetry can gap the line node
almost everywhere (possibily leaving behind a few PNs whose charges
sum up to zero).

\section{Fermi-surface Chern numbers}
\label{sec:sheets}

\subsection{Definitions}
\label{sec:defs}

We now turn our attention to the Chern numbers of Fermi sheets and
their relationship to the populations of enclosed chiral PNs.  We are
considering a metal with a spin-split Fermi surface as a result of
broken inversion or time reversal symmetry (or more precisely, broken
$PT$ symmetry).  In our nomenclature the ``Fermi surface'' $S_n$ of
band $n$ is the full set of points $E_n(\k)\!=\!\ef$, while the
``Fermi sheet'' $S_{na}$ is the $a$'th connected piece of the Fermi
surface.  (Here ``connectedness'' is defined without reference to the
BZ boundaries, so that a Fermi pocket centered at a zone corner is a
single Fermi sheet).  Note that $S\bna$ separates a region of
occupation $f\!=\!n-1$ from a region of occupation $f\!=\!n$. The
Chern number of $S\bna$ is, according to \eq{C-n-S},
\beq 
\label{eq:C-na}
C\bna=
\frac{1}{2\pi}\oint_{S\bna} dS\,\vf\cdot\bomega_n(\k)\,,
\eeq
where we have chosen $\vf$ as the unit normal to $S\bna$, i.e.,
parallel to the Fermi velocity and pointing toward the higher-energy
side, which is unoccupied in band $n$.

Throughout this section we will assume that all Fermi sheets are
isolated, as is required in principle in order to define their Chern
numbers from \eq{C-na}.  While this is the generic case for a
ferromagnet in the absence of any symmetry,\cite{haldane-arxiv14} bcc
Fe magnetized along [001] has a mirror plane at $k_z\!=\!0$, and as we
shall see, some Fermi sheets are glued together at isolated points on
that plane. The implications for the calculated Chern numbers will be
discussed in Sec.\ref{sec:chern-glue}.

\begin{figure}
\begin{center}
\includegraphics[width=1.0\columnwidth]{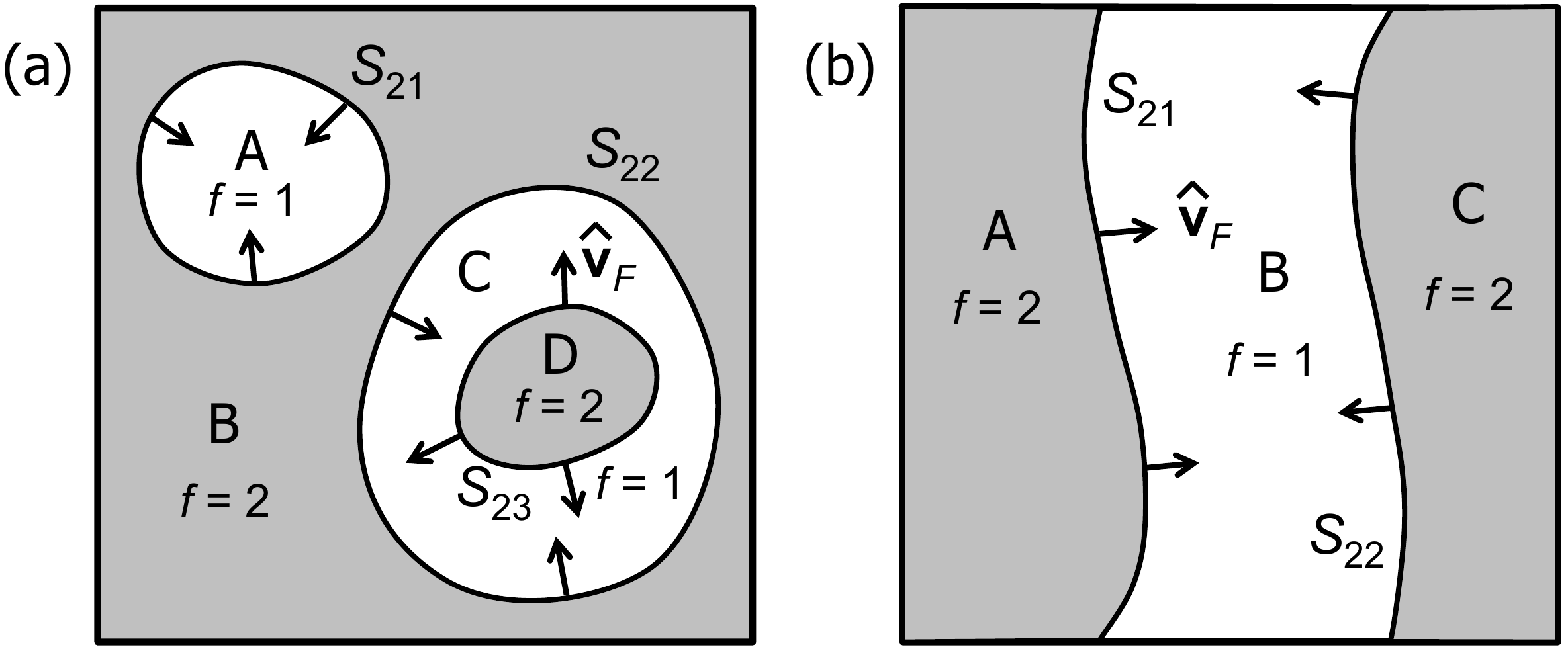}
\end{center}
\caption{Examples of Fermi sheet structures.}
\label{fig1}
\end{figure}

In order to proceed we need notions of ``interior'' and ``exterior.''
By definition, the \textit{exterior} is the side pointed to by $\vf$
(the unoccupied side), while the \textit{interior} is the occupied
side.  This can sometimes be counterintuitive, as for a hole pocket,
where the interior defined here is the geometric exterior and vice
versa.  If $S\bna$, when regarded all by itself, divides the BZ into
two distinct connected regions, we define these regions to be the
\textit{global} interior and exterior:
\[ I^*\bna = \hbox{global interior of $S\bna$} \,, \]
\[ E^*\bna = \hbox{global exterior of $S\bna$} \,. \]
To illustrate this, consider Fig.~\ref{fig1}(a), which shows a BZ with
three Fermi sheets separating band~1 from band~2; these are labeled
$S_{21}$, $S_{22}$, and $S_{23}$.  So, for example, $I^*_{21}$ is the
union of regions B, C and D, while $E^*_{23}$ is the union of A, B
and~C.

There may be some sheets, however, for which the concepts of global
interior and exterior are not applicable.  Consider, for example,
either one of the sheets shown in Fig.~\ref{fig1}(b).  Such Fermi
sheets are of the special kind having a ``Luttinger anomaly'' as
discussed around Eq.~(7) in Ref.~\onlinecite{haldane-arxiv14}, and
have the property that the integrated unit normal $\oint \vf\,dS$ is
nonzero.  These ``directed sheets'' need special treatment.  We take
the approach here of pairing them.  That is, we take the two adjacent
directed sheets $S_{21}$ and $S_{22}$ in Fig.~\ref{fig1}(b) and
henceforth consider them to comprise a single ``Fermi sheet'' whose
global interior is A and whose global exterior is B.  Henceforth, the
notation $S_{na}$ can refer either to a single non-directed sheet or
to a pair of oppositely directed sheets.

It is also useful to have a notion of \textit{immediate} interior and
exterior:
\[ I\bna = \hbox{immediate interior of $S\bna$} \,, \]
\[ E\bna = \hbox{immediate exterior of $S\bna$}\,. \]
To define the immediate interior, we decompose the subspace of the BZ
in which band $n$ is occupied into a set of disjoint subvolumes, each
of which is internally connected; call these $V\bnj$, where $j$ runs
over the number of disjoint subvolumes.  Similarly, let $\Vt\bnj$ be
the disjoint subvolumes making up the unoccupied part of the BZ for
band $n$.  Together, the $V\bnj$ and $\Vt\bnj$ cover the BZ once and
only once.  The immediate interior of $S\bna$ is then just the
subvolume $V\bnj$ that is immediately adjacent to $S\bna$.  More
precisely, it is the region $V\bnj$ for which $S\bna\in\delta V\bnj$,
where $\delta$ means ``the boundary of.'' (In general $\delta V\bnj$
may consist of several Fermi sheets, one of which is $S\bna$.)  A
similar definition applies to the ``immediate exterior,'' which is the
subvolume $\Vt\bnj$ adjacent to $S\bna$.  In Fig.~\ref{fig1}(a), for
example, the immediate exterior $E_{21}$ of sheet $S_{21}$ is region~A
(it is geometrically on the inside because it is a hole pocket).  The
immediate interior of $S_{21}$ is region B only!  The immediate
interior of $S_{22}$ is also region~B, while its immediate exterior is
only region~C, and so on.

Before continuing, we briefly note that there are two ways of thinking
about the volumes $V_{nj}$.  In one point of view, the nominal
boundaries of the BZ are ignored, so that, for example, an electron
pocket centered on a zone corner point is regarded as a single volume,
and all boundaries of the $V_{nj}$ are Fermi sheets.  This is the
viewpoint adopted in the present section.  Alternatively, one can
first establish a choice of parallelepiped or Wigner-Seitz BZ, and
decompose the interiors and exteriors into subregions within this BZ,
so that the $\delta V\bnj$ generally also include patches of the BZ
boundary.  We shall adopt the latter viewpoint when discussing one of
the FS formulations of the anomalous Hall conductivity in
Sec.~\ref{sec:haldane}, where application of the divergence theorem
requires the specification of definite BZ boundaries.

\subsection{Divergence theorem using global regions}
\label{sec:sheets-global}

Because $\delta I^*\bna$ has only one connected piece, namely $S\bna$,
we can apply the divergence theorem to find [see \eq{C-n-S-chi}]
\beq
C\bna = \sum_{W_{n\alpha}\in I^*\bna} \ch_{n\alpha}
       -\sum_{W_{n-1,\alpha}\in I^*\bna} \ch_{n-1,\alpha}\,.
\label{eq:gi}
\eeq
A similar argument applied to the global exterior implies that
\beq
C\bna = \; -\sum_{W_{n\alpha}\in E^*\bna} \ch_{n\alpha}
       +\sum_{W_{n-1,\alpha}\in E^*\bna} \ch_{n-1,\alpha}\,.
\label{eq:ge}
\eeq
The reversal of signs between these two equations is due to the fact
that for ${\cal V}=I^*\bna$ in \eq{div-thm} we have $\nhat=\vf$ on the
left-hand-side, while for ${\cal V}=I^*\bna$ we have $\nhat=-\vf$. In
view of the sum rule of \eq{sum-rule-b},
Eqs.~(\ref{eq:gi}-\ref{eq:ge}) are consistent with each other; we can
use either one to compute the Chern number from the point nodes,
depending on which is easier.

\subsection{Divergence theorem using immediate regions}
\label{sec:sheets-immediate}

We can apply the divergence theorem to an arbitrary connected
subvolume $V\bnj$ or $\Vt\bnj$, obtaining
\beq
\sum_{S\bna\in\delta V\bnj} C\bna 
=\sum_{W_{n\alpha}\in V\bnj} \ch_{n\alpha} \;\; 
-\sum_{W_{n-1,\alpha}\in V\bnj} \ch_{n-1,\alpha}
\label{eq:ii}
\eeq
or
\beq
\sum_{S\bna\in\delta \Vt\bnj} C\bna=
-\sum_{W_{n\alpha}\in \Vt\bnj} \ch\bna \;\; 
+\sum_{W_{n-1,\alpha}\in \Vt\bnj} \ch_{n-1,\alpha}\,.
\label{eq:ie}
\eeq
Unfortunately these equations do not immediately determine the FS
Chern numbers unless the boundary of the region in question is
composed of only a single sheet.

For example, applying \eq{ii} to subvolume B in Fig.~1(a) leaves the
sum of two Chern numbers ($C_{21}+C_{22}$) on the right-hand side, so
neither can be determined directly.  However, in cases like this an
iterative analysis will typically work; for example, applying \eq{ie}
to region A determines $C_{21}$, and together with the previous
result, this also determines $C_{22}$.

A similar problem arises in Fig.~1(b), but now it is more serious.
Recall that we agreed to pair directed sheets; having done so, we can
apply \eq{ii} and obtain the total Chern number summed over the two
sub-sheets.  But now there is no way to disentangle the Chern numbers
on the individual directed sub-sheets without an additional
calculation.  This does not necessarily have to be done on the actual
sub-sheet; in Fig.~1(b), for example, a calculation on one planar
surface lying between the two directed sheets would suffice.

\subsection{Sum rules on Chern numbers}
\label{sec:C-sum-rules}

Summing \eq{ii} or (\ref{eq:ie}) over all ~$n$ and all subvolumes, we
obtain the sum rule
\beq
\quad \sum\bna\, C\bna = 0
\label{eq:sumC}
\eeq
which, as Haldane points out, must be satisfied on gauge-invariance
grounds.\cite{haldane-prl04,haldane-arxiv14}

When $P$ symmetry is present (but $T$ is broken) the sum rule holds
for each band separately,
\beq
\label{eq:sumC-Pinv}
\sum_a\,C\bna = 0\,.
\eeq
This follows from combining \eqs{eig-P}{chi-P} with either \eq{ii} or
\eq{ie}: for every PN at $\k$, there is a partner at $-\k$ with the
same energy but opposite charge.  If, moreover, a Fermi sheet $S\bna$
encloses a parity-invariant momentum $\k_0=\G/2$, then $C\bna=0$.

\section{Anomalous Hall conductivity}
\label{sec:ahc}

The role played by degeneracies in the intrinsic AHC has been recently
debated\cite{chen-prb13,vanderbilt-prb14} in connection with
Fermi-surface formulations.\cite{haldane-prl04,wang-prb07} This debate
probably has its roots in the simplifying assumption, made more or
less explicitly at certain points in Refs.~\onlinecite{haldane-prl04}
and \onlinecite{wang-prb07}, that the band under consideration is
everywhere nondegenerate.  Instances where this isolated-band
assumption was made in Ref.~\onlinecite{haldane-prl04} include the
sentence below Eq.~(12), and also Eq.~(20); examples in
Ref.~\onlinecite{wang-prb07} are the sentence below Eq.~(7) and most
of Sec.~II.B.

In this section we show that the Fermi-surface formulations of
Refs.~\onlinecite{haldane-prl04} and \onlinecite{wang-prb07} remain
valid when the chiral degeneracies that are generally present in the
occupied band manifold are carefully accounted for. To be precise, the
nonquantized part of the intrinsic AHC is still given by the same bulk
Fermi-surface expressions derived in those works.  The presence of
isolated PNs carrying topological charges does {\it not} lead to
additional nonquantized, non-Fermi-surface contributions to the AHC,
as claimed in Ref.~\onlinecite{chen-prb13}.

\subsection{Fermi-sea formulation}
\label{sec:fermi-sea}

The AHC of a 3D crystal is conveniently expressed as
\beq
\label{eq:ahc-K}
\sigma_{ij}=-\frac{e^2}{2\pi h}\sum_{l=x,y,z}\,\epsilon_{ijl}K_l\,,
\eeq
where $e^2/h$ is the quantum of conductance and ${\bf K}$ is a
wavevector. When the Fermi level lies in an energy gap, $\K$ becomes
quantized to a reciprocal lattice
vector~$\G$.\cite{kohmoto-prb92,haldane-prl04}

Specializing to the intrisic (Karplus-Luttinger) contribution we can
work band by band and write
\beq
\label{eq:K-decomp-a}
{\bf K}=\sum_n\,{\bf K}_n\,. 
\eeq
The contribution from band~$n$ is
\beq
\label{eq:K-n}
{\bf K}_n
=\frac{1}{2\pi}\int_{I_n} d^3k\,\bomega_n(\k)\,,
\eeq
where the Berry curvature $\bomega_n(\k)$ is given by \eq{curv} and
$I_n=\cup_a I_{na}$ is the BZ region where band~$n$ is occupied.
Equations~(\ref{eq:ahc-K})--(\ref{eq:K-n}) form the standard Fermi-sea
expression for the intrinsic AHC.\cite{xiao-rmp10,nagaosa-rmp10}

While it is natural to view $\K_n$ as ``the intrinsic AHC contributed
by band $n$,'' it should be kept in mind that each $\K_n$ by itself is
not a physical observable; only the total $\K$ is.  For example, the
individual $\K_n$ are not invariant under arbitrary band-mixing gauge
transformations within the occupied band manifold.  In the next
section we discuss an alternative decomposition due to
Haldane,\cite{haldane-prl04} \eq{K-decomp-b} below, that is equally
valid and in some ways more informative than \eq{K-decomp-a}.

\subsection{Haldane's Fermi-surface formulation}
\label{sec:haldane}

\subsubsection{Dipole of the chiral-charge distribution in $k$ space}
\label{sec:bz-dipole}

Let us rewrite \eq{K-n} using an integration by parts of the form
\beq
\label{eq:corollary}
\int_{\cal V} dV ({\bm\nabla} g)\cdot\bomega_n=
\oint_{\delta{\cal V}} dS\,\nhat\cdot g\bomega_n
-\int_{\cal V} dV g{\bm\nabla}\cdot\bomega_n\,,
\eeq
which follows from replacing $\bomega_n(\k)$ with $g(\k)\bomega_n(\k)$
in \eq{div-thm}.  Setting ${\cal V}=I_n$, $g=k_i$, and using \eq{div}
we find
\beq
\label{eq:K-n-b}
\K_n=\Komega+\Kchi
\eeq
\beq
\label{eq:K-n-Omega}
\Komega=\frac{1}{2\pi}\oint_{\delta I_n} d^2k\,
\k\left[\nhat\cdot{\bomega}_n(\k)\right]
\eeq
\beq
\label{eq:K-n-chi}
\Kchi=\sum_{W_{n-1,\alpha}\in I_n}\, 
\k_{n-1,\alpha}\chi_{n-1,\alpha}-
\sum_{W_{n\alpha}\in I_n}\, \k_{n\alpha}\chi_{n\alpha}\,.
\eeq
These equations express $\K_n$ as the dipole moment of a distribution
of chiral charge inside the BZ.\footnote{\equ{K-n-b} is analogous to a
  formal real-space expression in Ref.~\onlinecite{martin-prb74} for
  the electric polarization of an insulating crystal (more precisely,
  it is an expression for the first-order change in polarization under
  a perturbation).}  $\Kchi$ is the contribution from the discrete
charges associated with the occupied PNs, and $\Komega$ is the
contribution from a continuous distribution of charge across the
surface $\delta I_n$ (with $\nhat\cdot\bomega_n(\k)/2\pi$ playing the
role of an areal density of ``bound chiral charge'').

We now adopt the second point of view described at the end of
Sec.~\ref{sec:defs}, in which the boundary $\delta I_n$ generally
consists of FS sheets $S_{na}$ together with the portions of the BZ
boundary (BZB) where band~$n$ is occupied; this is demanded by the
presence of the linear term $\k$ in \eq{K-n-Omega}.  We choose a
parallelepiped BZ, as opposed to a Wigner-Seitz one, to insure a
simple relation between opposing faces.  Thus, the surface integral in
\eq{K-n-Omega} runs over portions of the BZB, as well as over the
Fermi sheets per se.  Explicitly,
\beq
\label{eq:Ksum}
\Komega=\sum_a\Kaomega + \KBomega
\eeq
where
\beq
\label{eq:K-na-Omega}
\Kaomega=\frac{1}{2\pi}\oint_{S_{na}} d^2k\,
\k\left[\vf\cdot{\bomega}_n(\k)\right]
\eeq
is the contribution from the $a$'th Fermi sheet and 
\beq
\label{eq:K-n-BZB}
\KBomega=\frac{1}{2\pi}\oint_{{\rm BZB}_n} d^2k \,
\k\,[\nhat\cdot\bomega_n(\k)]
\eeq
is the contribution from the portions of the BZB where band $n$ is
occupied, denoted by BZB$_n$. In this last equation $\nhat$ points to
the outside of the BZ.

\equ{K-n-b} differs from Eq.~(20) in Haldane's
Ref.~\onlinecite{haldane-prl04} in that the latter does not include
the term $\Kchi$.  This extra term appears when the band has PN
degeneracies, and is needed to ensure that the cell dipole $\K_n$
remains invariant when the origin in $k$~space is shifted. Under a
shift of $-\delta\k$ \eq{K-n-b} changes by
\beq
\label{eq:K-n-shift}
\delta\K_n=\delta\k
\left[
  \sum_a\,C\bna
  +\sum_{W_{n-1,\alpha}\in I_n}\,\chi_{n-1,\alpha}
  -\sum_{W_{n\alpha}\in I_n}\,\chi_{n\alpha}
\right]\,,
\eeq
where $C\bna$ is the Fermi-sheet Chern number defined in \eq{C-na},
and $\delta\KBomega=0$ due to cancelling contributions from opposing
BZ faces.  The quantity in square brackets vanishes by virtue of
\eq{ii}.

In some cases the volume and surface chiral-charge distributions are
separately neutral. This happens for a completely filled band in a
generic crystal [\eq{sum-rule-a}], and for any band in a
centrosymmetric crystal [\eq{sumC-Pinv}].  Even then, the two separate
terms in \eq{K-n-b} are not fully invariant under another type of
transformation: a rigid shift of the BZ cell. Consider what happens
when a PN with $\chi_{n\alpha}=+1$ leaves the BZ on one side and
re-enters on the opposite side, at a point separated by $-\G$ from the
exit point: $\Kchi$ given by \eq{K-n-chi} jumps by $\G$, and it will
become clear in a moment that $\KBomega$ changes by $-\G$.

To summarize, the situation is as follows:

\begin{itemize}

\item The AHC contribution from band $n$ is most naturally defined as
  the integral of the Berry curvature over the occupied portion of the
  band, as in \eq{K-n}.

\item The contribution from band $n$ can be expressed as
  $\K_n=\Komega+\Kchi$, where the inclusion of the $\Kchi$ term is
  needed to preserve the invariance of $\K_n$ under an origin shift,
  or under a shift of the BZ cell.

\item As we shall see shortly, $\Komega$ is quantized for any fully
  occupied band.  On the other hand, $\Kchi$, and hence $\K_n$, has a
  nonquantized contribution even for a completely filled band if the
  band in question has chiral degeneracies with higher or lower bands.

\item Nevertheless, the $\Kchi$ contributions from \eq{K-n-chi} add up
  to zero when summing over all bands.\footnote{This is also true when
    the sum runs over an isolated subset of bands, i.e., a set of
    adjacent bands that can have PN degeneracies among themselves, but
    which remain separated from lower or higher bands by finite gaps
    everywhere in the BZ.} Hence
  \beq
  \label{eq:K-decomp-b}
  \K=\sum_n\, \Komega 
  \eeq 
  correctly gives the total AHC, in a way that the nonquantized part
  of $\K$ is now ascribed exclusively to the partially filled bands.

\end{itemize}

Let us reexamine the conclusions of Ref.~\onlinecite{chen-prb13} in
the light of the preceeding discussion.  Essentially the claim, stated
in the abstract, is that when chiral denegenacies are present in the
Fermi sea, the nonquantized part of the AHC is not entirely a (bulk)
Fermi-surface property. In the main text, the authors purport to show
that the additional nonquantized, non-FS contributions originate in
the chiral PNs of the completely filled bands.

If taken at face value, the above statements would seem to imply that
a purely FS formulation of the nonquantized AHC is not possible. But
this cannot be correct, since the nonquantized part of $\K$ can be
completely ascribed to the partially filled bands by means of
\eq{K-decomp-b}, even when chiral degeneracies are present. The
nonquantized part of $\K$ can moreover be expressed as a bulk FS
property, as shown below following Haldane's original
argument.\cite{haldane-prl04} In their analysis, the authors of
Ref.~\onlinecite{chen-prb13} appear to have overlooked the fact that
the nonquantized Fermi-sea contributions $\Kchi$ from the occupied PNs
sum up to zero over all bands.

\subsubsection{Fermi-surface expression for $\Komega$}
\label{sec:Komega}

In Ref.~\onlinecite{haldane-prl04}, Haldane further manipulated his
Eq.~(20) [our \eq{K-n-Omega}] to arrive at his Eq.~(21), in which the
BZB term was written in a more explicit form.  For completeness, we
shall repeat this derivation here using our notation.

We choose a parallelepiped BZ defined by a triplet of primitive
reciprocal lattice vectors $\b_j$, such that the reduced coordinate
$\kappa_j=(\a_j\cdot\k)/2\pi$ goes from $\kappa_{j0}$ to
$\kappa_{j0}+1$ inside the BZ, where $\kappa_{j0}$ fixes the corner of
the BZ cell.  Recalling that the integration in \eq{K-n-BZB} runs over
the portions of the six BZ faces where band $n$ is occupied, we can
decompose the BZB term into contributions from the three sets of
opposing faces according to
\beq
\label{eq:KBomegared}
\KBomega=\sum_{j=1}^3\,\Kbzb_{nj}\,\b_j \,,
\eeq
obtaining
\beq
\label{eq:Cn}
\Kbzb_{nj}=\frac{1}{2\pi}\oint_{\textrm{BZB}_n} d^2k\,\kappa_j\,
\nhat\cdot\bomega_n(\k)\,.
\eeq
When computing $\Kbzb_{nj}$, the integrations on the four side
surfaces of the BZB cancel in pairs, while the contributions from the
surfaces related by $\b_j$ fail to cancel because of the $\kappa_j$
factor. The result is
\beq 
\label{eq:C-occ}
\Kbzb_{nj}=\frac{1}{2\pi}\int_{\Bnj} d^2k\,
\hat{\bf a}_j\cdot\bomega_n(\k) \,,
\eeq
where $\Bnj$ is the portion of the BZB at $\kappa_j=\kappa_{j0}+1$
that is occupied in band $n$, and $\hat{\bf a}_j$ is the outward unit
normal. In the notation of \eq{berry-flux}, $\Kbzb_{nj}$ is $1/2\pi$
times the Berry flux
\beq
\label{eq:phi-nj}
\phi_{nj}=\int_{\Bnj} d^2k\,\hat{\a}_j\cdot\bomega_n(\k)\,
\eeq
exiting the BZ cell through the occupied portions of the BZ face
pointed to by $\b_j$.

If band $n$ is fully occupied, $\Kbzb_{nj}$ is just the Chern number
obtained by integrating the Berry curvature over the entire BZ face in
direction $j$.  If the band is also isolated, the integers
$\Kbzb_{nj}$ are independent of the choice of cell origin
$\bm{\kappa}_{0}$, and $\KBomega$ is a unique ``Chern vector'' $\sum_j
\Kbzb_{nj}\,\b_j$ describing the topology of the filled band.  If it
is fully occupied but not isolated, $\Kbzb_{nj}$ is still quantized to
be a triplet of integers, but these may change discontinuously if
$\bm{\kappa}_{0}$ is shifted in such a way that one of the BZ
boundaries passes over a chiral PN.  If band $n$ is only partially
occupied, then the integral in \eq{C-occ} is only over the occupied
portions of the BZ face at $\kappa_j=\kappa_{j0}+1$, and $\Kbzb_{nj}$
need not be an integer.

To summarize so far, \eq{K-decomp-b} has become
\beq
\K = \sum_{na} \Kaomega
   + \frac{1}{2\pi} \sum_{nj} \b_j \phi_{nj}
\label{eq:Ksummary}
\eeq
where $\Kaomega$ is the $\k$-weighted integral of the surface-normal
Berry curvature on sheet $S_{na}$, \eq{K-na-Omega}, and $\phi_{nj}$ is
the Berry flux passing throught the occupied portion $\Bnj$ of the BZ
face, \eq{phi-nj}.  After summing over all bands there is no ambiguity
modulo a quantum in either of the contributions above, and
\eq{Ksummary} correctly gives the total intrisic AHC modulo nothing.
\equ{Ksummary} is still not quite a Fermi-surface property, because
the $\phi_{nj}$ have to be obtained by integrating over portions of
the band lying at energies below $\ef$ on the BZB.

In order to arrive at Eq.~(21) of Haldane's
Ref.~\onlinecite{haldane-prl04}, we now abandon the goal of computing
the AHC exactly, and only ask for its nonquantized part.  Two
modifications can be made to \eq{Ksummary} that only affect the result
by quantized amounts, and lead to a FS expression for the nonquantized
part of $\K$.  First, the sums over~$n$ can be restricted to partially
occupied bands, recalling that completely filled bands only contribute
to the second term, and by a quantized amount. The second is to invoke
\eq{flux2phase} in order to replace the Berry fluxes $\phi_{nj}$ with
sums of Berry phases that are only defined modulo~$2\pi$,
\beq
\label{eq:varphi-nj}
\varphi_{naj}=\oint_{{\cal C}_{naj}}d\k\cdot{\bf A}_n(\k)\,.
\eeq
The oriented curve ${\cal C}_{naj}$ consists of one or more planar
circuits at the intersections of the sheet $S_{na}$ with the BZ face
at $\kappa_j=\kappa_{j0}+1$, so that $\cup_a{\cal C}_{naj}=\delta{\cal
  B}_{nj}$. If we view the BZ face as an open parallelogram with
edges, those circuits may include non-FS segments along the edges. If
instead we view it as a closed 2-torus, a nonvanishing ${\cal
  C}_{naj}$ consists exclusively of Fermi loops and we arrive at
Haldane's FS expression, Eq.~(21) of Ref.~\onlinecite{haldane-prl04},
in the form
\beq
\K :={\sum_{na}}' \, \Kaomega
   + \frac{1}{2\pi} {\sum_{naj}}'\, \b_j \varphi_{naj} \,,
\label{eq:haldane}
\eeq
where a prime on a sum indicates that $n$ only runs over bands
crossing $\ef$.  The symbol $:=$ indicates that the quantity on the
left-hand-side is equal to the right-hand side modulo a reciprocal
lattice vector~$\G$.  Since the Berry phases $\varphi_{naj}$ are
defined for loops lying on the Fermi surface, this is now a true
Fermi-surface property.

\subsection{Tomographic Fermi-surface formulation}
\label{sec:fermi-loops}

\equ{haldane} for the nonquantized part of the AHC involves Berry
curvatures on the Fermi sheets as well as Berry phases around Fermi
loops on the BZB.  An alternative FS formula was obtained in
Ref.~\onlinecite{wang-prb06} that only involves Berry phases.  In this
section we recover this Fermi-loop expression starting from
\eq{haldane}. The present derivation is therefore complementary to the
one given in Ref.~\onlinecite{wang-prb06}, which starts from the
Fermi-sea formulation of Sec.~\ref{sec:fermi-sea} and does not make an
explicit connection to Haldane's expression.

We work sheet by sheet, treating each sheet as one connected piece.
Sheets with nonzero Chern numbers are grouped together in such a way
that their combined Chern number vanishes. This is needed to obtain an
AHC contribution that is origin-independent in the sense of
\eq{K-n-shift}. Thus, in the following $S_{na}$ denotes either a
single sheet with $C_{na}=0$, or a group of sheets whose Chern numbers
add up to zero.

Writing the contribution from $S_{na}$ to \eq{haldane} as
\beq
\label{K-naj}
\K_{na}=\sum_{j=1}^3\,\Kred_{naj}\b_j\,,
\eeq
we find
\beq
\label{eq:K-naj}
2\pi\Kred_{naj}=
\int_{S_{na}}d^2k\,\kappa_j\,\vf\cdot\bomega_n(\k)
+\oint_{{\cal C}_{naj}}d\k\cdot{\bf A}_n(\k)\,.
\eeq
If the sheet is hole-like, the Fermi loops ${\cal C}_{naj}$ should be
traversed in the negative direction of circulation.

Recalling the interpretation of \eq{K-n-b} as a dipole moment in $k$
space, we can similarly view the first term in \eq{K-naj} as an
intracell dipole moment of a surface distribution of bound chiral
charge, and the second term as an intercell ``charge-transfer''
term. Separately, each term depends on the choice of cell vectors and
on the placement of the cell boundaries, but their sum is independent
of those arbitrary choices.

Since the total $\Kred_{naj}$ given by \eq{K-naj} is cell-invariant,
we are allowed to average the right-hand-side over several different
BZ cells. We choose for this purpose the range of cells obtained by
sliding a parallelepiped BZ along the full length of its edge
$\b_j$. The dipole term then averages to zero (because the net charge
per cell $C_{na}$ vanishes), and we are left with the average of the
intercell term.  As the cell slides over one period the forward-facing
boundary cuts through an entire BZ, producing a tomographic scan of
the Fermi sheet. The contribution from each slice (viewed as a
2-torus) is given by the Berry phases of the inscribed Fermi loops.
Averaging over a discrete set of slices spanning the entire BZ we find
\beq
\label{eq:K-sheet}
\Kred_{naj}=\frac{1}{n_{\rm slice}}\sum_{i=1}^{n_{\rm slice}}\,
\frac{\varphi_{naj}(i)}{2\pi}\,,
\eeq
and summing over all Fermi sheets we recover the FS expression in
Ref.~\onlinecite{wang-prb07} for the nonquantized AHC,
\beq
\label{eq:Kred-j}
\Kred_j:={\sum_{na}}' \, \Kred_{naj}\,.
\eeq

Equations~(\ref{eq:K-sheet}-\ref{eq:Kred-j}) must be supplemented with
a prescription for choosing the branch cuts of the Berry phases. The
allowed choices are strongly constrained by the BZ-averaging
procedure: the first term in \eq{K-naj} changes continuously with the
position of the BZ cell (because the ``bound chiral charge'' is spread
across the sheet $S_{na}$), and this should be exactly balanced by the
change in the second term. Thus $\varphi_{naj}$ must change gradually
(by much less than $2\pi$) from one slice to the next.  Note that
enforcing smoothness only fixes the branch cut on a slice relative to
that on the previous slice. The freedom to choose the branch cut on
the first slice leaves the integer part of $\Kred_{naj}$ undetermined,
as expected of a FS formulation.

After summing \eq{K-sheet} over the indices~$n$ and~$a$, the overall
quantum of indeterminacy can be resolved by equating
$\sum_{na}\varphi_{naj}$ with the total Berry flux $\sum_n\phi_{nj}$
on the first slice (a Fermi-sea quantity).\cite{wang-prb07} It is
however not always possible to similarly resolve the quantum for the
AHC contribution from an {\it individual} Fermi sheet $S_{na}$. If the
sheet encloses chiral PNs, the integer part of $\Kred_{naj}$ obtained
by anchoring the Berry phase $\varphi_{naj}$ on the first slice to the
Berry flux $\phi_{naj}$ will depend on where that slice stands
relative to those PNs.

For a composite sheet comprising several disconnected pieces,
$\varphi_{naj}$ changes by $2\pi C_\alpha$ while traversing a single
piece $S_\alpha$ with Chern number $C_\alpha$.\cite{vanderbilt-prb14}
If we fix the branch choice by arbitrarily setting $\varphi_{naj}=0$
below the first piece, $\varphi_{naj}$ reaches $2\pi C_1$ at the top
of that piece, remains constant until hitting the next piece, and
finally drops back to zero at the top of the last piece.  We will
encounter this type of behavior when studying the AHC of bcc Fe in
Sec.~\ref{sec:ahc-results}.

\section{Computational details}
\label{sec:methods}
\subsection{First-principles calculations and Wannier-function
  construction}
\label{sec:abinitio}

Our electronic structure calculations including SOC are carried out
for ferromagnetic Fe in the bcc structure. We fix the lattice constant
at the experimental value $a=5.42$~bohr, and orient the magnetization
along the easy axis [001]. (That is, minority and majority spins point
up and down respectively.) In reality, $\alpha$-Fe distorts very
slightly from body-centered cubic to body-centered tetragonal on
cooling through the magnetic transition, but we ignore this lattice
strain effect and set $c/a=1$ in our calculations.

We use the plane-wave pseudopotential method as implemented in the
{\tt Pwscf} code from the {\tt Quantum-Espresso}
package,\cite{giannozzi-jpcm09} in a noncollinear spin framework.
Relativistic norm-conserving pseudopotentials are generated from
parameters similar to those in Ref.~\onlinecite{wang-prb06}, and an
energy cutoff of 120~Ry is used for the plane-wave expansion of the
wavefunctions. Exchange and correlation effects are treated within the
PBE generalized-gradient approximation.\cite{perdew-prl96} The
self-consistent total energy calculations are done with a $16\times
16\times 16$ Monkhorst-Pack mesh for the BZ integration, while for the
non-self-consistent calculations a $10\times 10\times 10$ mesh is
used.  A Fermi smearing of 0.02~Ry is used during the self-consistent
cycle, and the 28 lowest bands are calculated in the
non-self-consistent step.

Starting from the Bloch functions computed in the non-self-consistent
step, maximally-localized Wannier
functions\cite{marzari-prb97,souza-prb01} are constructed using the
{\tt Wannier90} code.\cite{mostofi-cpc08} The choice of trial orbitals
and energy windows is the same as in Refs.~\onlinecite{wang-prb06}
and~\onlinecite{wang-prb07}, resulting in eighteen partially occupied
spinor Wannier functions.  We also carry out some SOC-free
calculations where nine Wannier functions are generated separately for
each spin channel.

\subsection{Berry phases, curvatures, and Chern numbers}
\label{sec:plaquettes}

For the evaluation of Berry curvatures and Berry phases in $k$ space
we use the efficient Wannier-interpolation algorithms of
Refs.~\onlinecite{wang-prb06} and~\onlinecite{wang-prb07}.  In this
section we focus on the aspects that are specific to the present work,
and refer the reader to those papers for other details.

We begin by describing how to compute the slice Chern number
$C_n(k_z)$ of \eq{C-n-kz} [the same approach is used for the slice
Chern number $\widetilde{C}_n(k_z)$ of \eq{Ctil-n-kz}, and also for
the box Chern number of \eq{chi}]. The most direct approach would be
to evaluate the Berry curvature on a $k$-point mesh covering the
2D~BZ. The disadvantage is that for any finite mesh the result is not
exactly quantized, and we have encountered situations where the
deviations were significant even for very dense meshes (this can
happen close to critical $k_z$ values where the true Chern number
changes discontinuously).

The interpretation of the Berry curvature as a Berry phase per unit
area suggests an alternative strategy: tile the 2D~BZ with square
plaquettes, and compute the Berry phases $\varphi_n^\square(k_z)$
around their edges.\cite{fukui-jpsj05} The Chern number is
\beq
\label{eq:chern-plaquettes}
C_n(k_z)=\frac{1}{2\pi}\sum_\square^{\rm
  2D\,BZ}\,\varphi_n^\square(k_z)\,, 
\eeq
where the plaquette Berry phase $\varphi_n^\square(k_z)$ is evaluated
with the discretized Berry-phase formula\cite{king-smith-prb93} on a
counterclockwise path consisting of its four corners.  In our
implementation we use the Wannier-based version of the discretized
Berry-phase formula, Eq.~(25) in Ref.~\onlinecite{wang-prb07}. (In the
case of $\widetilde{C}_n(k_z)$ the multiband
generalization\cite{king-smith-prb93} of the second term in that
equation should be used instead.)  Note that because each link along
the discretized path is traversed twice in opposite directions when
evaluating \eq{chern-plaquettes}, the net contribution from the first
term in Eq.~(25) of Ref.~\onlinecite{wang-prb07} vanishes identically.

The above prescription is guaranteed to give an integer result for
$C_n(k_z)$ for arbitrary lattice spacings.\cite{fukui-jpsj05} The
correct Chern number is obtained by choosing each Berry phase
$\varphi_n^\square(k_z)$ in the principal branch between $-\pi$ and
$\pi$, provided that the magnitude of the Berry-curvature flux through
every plaquette is safely below $\pi$. In practice we start by tiling
the 2D BZ with a uniform array of equally-sized plaquettes
(e.g. $200\times 200$), and whenever $|\varphi_n^\square(k_z)|>\pi/3$
for some plaquette we divide it into subplaquettes so as to meet the
above requirement.\footnote{This algorithm may fail if, for example, a
  double-Weyl node lies very close to a plaquette. The Berry flux
  through the plaquette can then be of the order of $2\pi$, which
  would be mistaken for a flux close to zero. This ambiguity can be
  avoided by evaluating the flux directly from the gauge-invariant
  Berry curvature.}  The slice Chern number is then obtained by
summing the Berry phases over all (sub)plaquettes covering the
2D~BZ. Care must be taken to ensure that every link is traversed
twice, once in each direction.  If, for example, a plaquette has been
divided into four subplaquettes but a neighboring plaquette has not,
then when computing the Berry phase for the larger plaquette the
middle point of the shared edge should be included in the discretized
path along with the four corner points.

We use the same algorithm to determine the Berry flux through the
occupied portions of a BZ slice, \eq{phi-nj}.  For a partially filled
band the flux is not quantized, and the first term in Eq.~(25) of
Ref.~\onlinecite{wang-prb07} must be included; because the links along
the edge between occupied and empty regions are traversed only once,
that term gives a net contribution.  As far as the nonquantized part
of the Berry flux is concerned (i.e., the Berry phase around the Fermi
loops), this approach is equivalent to integrating the Berry
connection over a ragged path at the edge approximating the Fermi
loops. For improved numerical accuracy we triangulate the edge with
new $k$ points chosen to approximate the location of the Fermi loops;
this changes the shape of the plaquettes along the edge from squares
to triangles or irregular quadrilaterals.

\section{Numerical results for bcc iron
}
\label{sec:results}

\subsection{Band degeneracies}
\label{sec:degen}

\subsubsection{Overview of the spinor bandstructure}
\label{sec:bands}

\begin{figure*}
\begin{center}
\includegraphics[width=1.35\columnwidth]{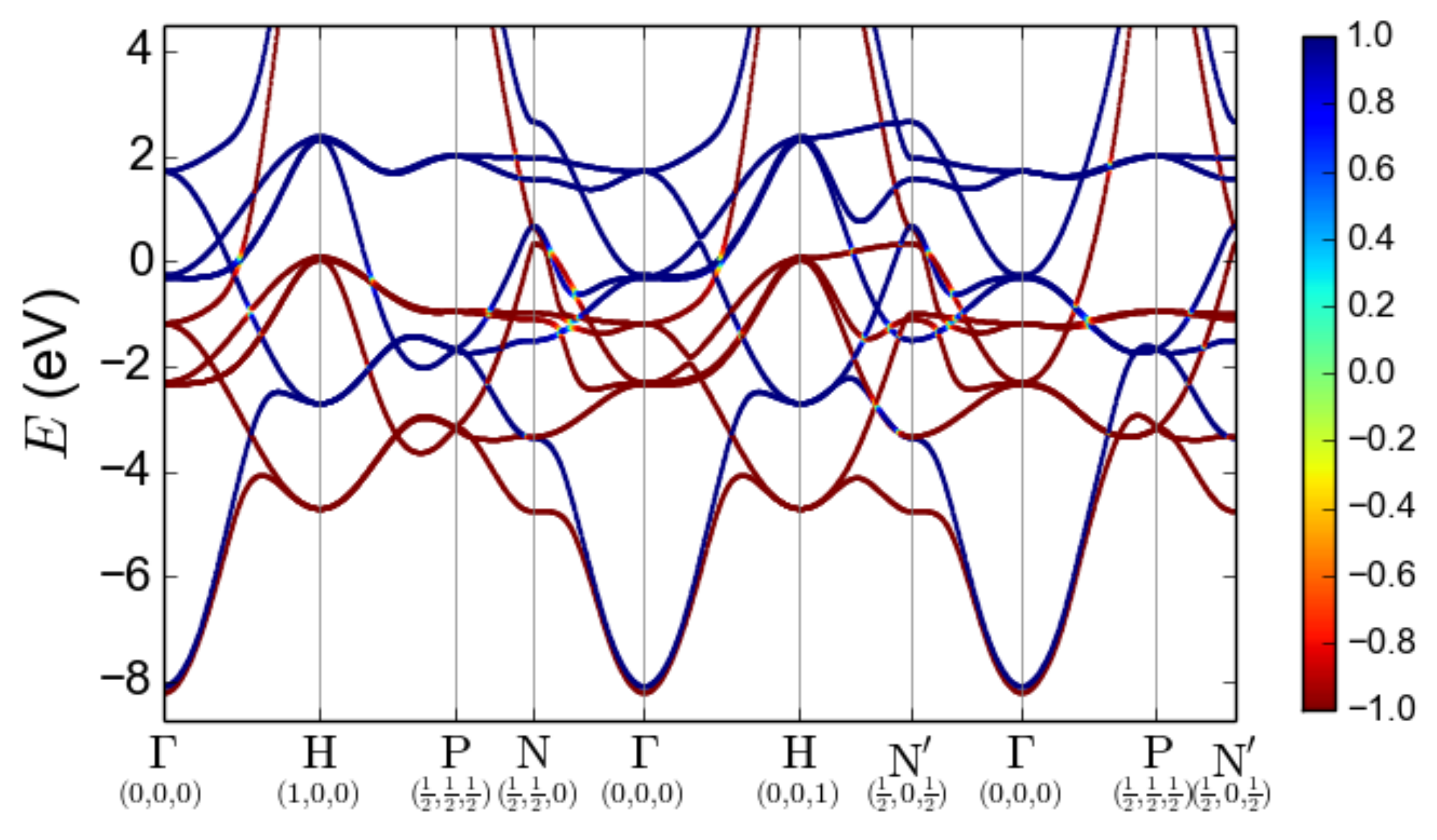}
\end{center}
\caption{Bandstructure of bcc Fe with spin-orbit coupling
  included. Energies are measured from the Fermi level, and the bands
  are color-coded by the expectation value of the spin component $S_z$
  in units of $\hbar/2$, ranging from red (majority down-spin
  character) to blue (minority up-spin) as indicated by the colorbar.}
\label{fig2}
\end{figure*}

Our electronic structure calculations including SOC are carried out
for ferromagnetic Fe in the bcc structure, as discussed in
Sec.~\ref{sec:abinitio}.  The energy bands, shown in Fig.~\ref{fig2},
are in good agreement with previous bandstructure calculations for
bcc~Fe where SOC was
included.\cite{singh-prb75,yamagami-jpsj98,yao-prl04} The smaller
exchange splitting at the band bottom (at around $-8$~eV) compared to
Ref.~\onlinecite{yamagami-jpsj98} is a consequence of using the PBE
generalized-gradient approximation rather than the local-density
approximation to treat exchange and correlation effects.

\begin{figure}
\begin{center}
\includegraphics[width=0.7\columnwidth]{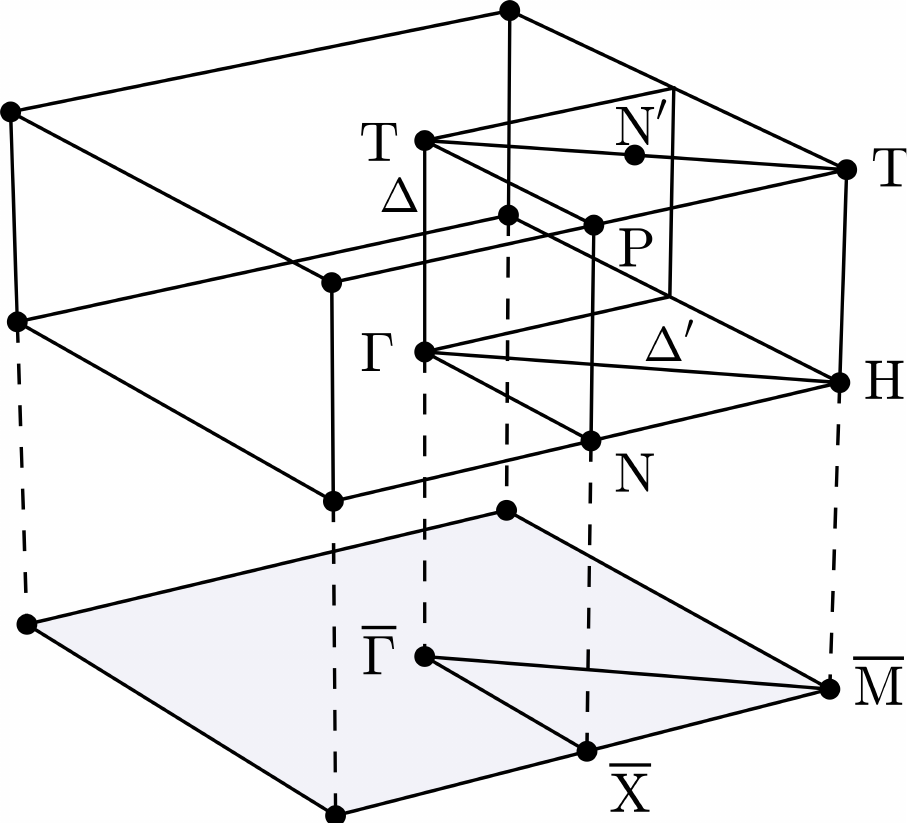}
\end{center}
\caption{Symmetry points and lines in reciprocal space for bcc~Fe with
  spin-orbit coupling included. The top volume is the half Brillouin
  zone between $k_z=0$ and $k_z=\pi/a$ (becomes a full BZ when
  expanded using inversion through $\Gamma$). The shaded area at the
  bottom is a two-dimensional projection. Therein, $\g2$ is a
  projection of $\Gamma$-${\rm T}$-${\rm H}$-${\rm T}$-$\Gamma$
  ($\Delta$), $\x2$ is a projection of ${\rm N}$-${\rm P}$-${\rm
    N}$-${\rm P}$-${\rm N}$, and $\m2$ is a projection of ${\rm
    H}$-${\rm T}$-$\Gamma$-${\rm T}$-${\rm H}$. Note that ${\rm T}$ is
  not actually a symmetry point, but ${\rm N'}$ is.}
\label{fig3}
\end{figure}

Even though an undistorted cubic structure is used, the combined
action of spin polarization and SOC reduces the symmetry of the
electronic states from cubic to tetragonal, and the conventional
labeling scheme for the high-symmetry points and lines in the BZ must
be modified accordingly. For example, the six points~${\rm N}$ inside
the BZ get split into two groups. We reserve the label~${\rm N}$ for
the two points lying on the $k_z=0$ plane, and label the four points
lying on the $k_z=\pi/a$ plane as~${\rm N}'$; within each separate
group, symmetry points are related to one another by four-fold
rotations. The $\Gamma{\rm H}$ high-symmetry line along the (001)
direction also becomes inequivalent to the other two, since it is no
longer related to them by any point-group symmetry; we label them
$\Delta$ and $\Delta'$ respectively. The symmetry points and lines are
indicated in the half-BZ shown in the top part of Fig.~\ref{fig3}.
The full tetragonal BZ is spanned by the vectors
$(2\pi/a)(1\overline{1}0)$, $(2\pi/a)(110)$, and $(2\pi/a)(001)$. Note
that only the first two are reciprocal-lattice vectors.

The lowering of symmetry due to SOC is reflected in the bandstructure
in Fig.~\ref{fig2}, but since SOC is weak the effect is small on the
scale of the figure. As we shall see in Sec.~\ref{sec:fs-fe-overview},
somewhat larger splittings occur away from the symmetry
lines.\cite{yamagami-jpsj98}

\subsubsection{Types of degeneracies}
\label{sec:degen-types}

Band degeneracies can be classified as {\it essential} or {\it
  accidental}.\cite{heine-book} An essential degeneracy is a band
crossing at a high-symmetry point or along a high-symmetry line in the
BZ where an irreducible representation (``irrep'') of dimension larger
than one exists. It has been shown that essential degeneracies do not
occur in the bandstructure of tetragonal ferromagnets when SOC is
included.\cite{falikov-pr68,cracknell-jpc69,cracknell-prb70} Thus, all
band degeneracies in bcc Fe are accidental.  While their exact
locations in $k$~space are not fixed by symmetry, most (but not all)
degeneracies in bcc Fe lie along symmetry lines or planes in the BZ.
Below we give an overview of the types of accidental degeneracies that
are present, classified by symmetry (they are in fact the generic
types of degeneracies in any tetragonal ferromagnet).

The magnetic point group of bcc Fe is $4/mm'm'$.\cite{cracknell-prb70}
The four-fold axis points along the magnetization direction [001], and
$m$ denotes mirror reflection about the (001) plane orthogonal to the
four-fold axis. The symmetry elements~$m'$ are reflections about the
inequivalent vertical planes (010) and (110), combined with time
reversal.

Turning to the symmetries in reciprocal space, inspection of
Fig.~\ref{fig3} reveals four-fold symmetry about the axis $\Delta$
that project onto $\g2$, and also about the one that projects onto
$\m2$.\footnote{Note that the projected BZ at the bottom of
  Fig.~\ref{fig3} has twice the area of the (001) surface BZ of the
  bcc lattice, and that the symmetry labels $\m2$ and $\x2$ denote
  different points in each case: $\x2$ in the projected BZ corresponds
  to $\m2$ in the surface BZ, and $\m2$ in the projected BZ is a
  periodic image of $\g2$ in the surface BZ. Thus, the four-fold axes
  that project onto $\g2$ and $\m2$ in Fig.~\ref{fig3} are really the
  same axis, apart from a vertical offset of $2\pi/a$.}  At an
arbitrary point on one of the four-fold axes no other symmetries are
present when SOC is included, and each band belongs to one of four
one-dimensional irreps of the little group $C_4$; bands belonging to
different irreps can cross, so that for critical values of $k_z$ (not
fixed by symmetry) PNs are generated.

The four irreps along the four-fold axes are labeled by their symmetry
eigenvalues $e^{i\pi(m+1/2)/2}$, and the labels of the crossing bands
carry information about the chiral charge at the degeneracy
point.\cite{fang-prl12} Two types of degeneracies can occur.  (i) If
the two bands have adjacent labels on the complex unit circle, they
disperse linearly in all directions away from the degeneracy point;
the chirality of the WP is positive ($\chi_{n\alpha}=+1$) if the label
of the lower band~$n$ changes by a factor of $i$ with increasing
$k_z$, and negative ($\chi_{n\alpha}=-1$) it it changes by a factor of
$-i$. (ii) If the labels are not adjacent, the crossing is a
double-Weyl node with $\chi_{n\alpha}=\pm 2$, where the dispersion
away from the node is linear along the symmetry axis and quadratic on
the orthogonal plane; in this case, the sign of the chiral charge
cannot be inferred from the symmetry labels alone.

The axis that projects onto $\x2$ is a two-fold axis, where there are
two one-dimensional irreps (little group $C_2$). Along this axis,
crossings between bands belonging to different irreps are always
linear ($\chi_{n\alpha}=\pm 1$), but the chirality of the WP cannot be
deduced from the symmetry labels.\cite{fang-prl12}

A different type of degeneracy occurs on the simple mirror plane at
$k_z=0$, i.e., the $\Gamma{\rm N H}$ plane in Fig.~\ref{fig3}.
(Equivalent planes are separated by integer multiples of $2\pi/a$;
because the structure is body-centered, the plane ${\rm TPN'}$ at
$k_z=\pi/a$ is not a mirror plane). There can be no chiral PN
degeneracies on a simple mirror plane, because the chiral charge is
odd under reflection [\eq{chi-mz}]. Instead, the generic degeneracies
are nonchiral nodal rings.  The spinor energy eigenstates carry mirror
symmetry labels $\pm i$; bands with different labels can cross, and
since the condition $E_{n\k}=E_{n+1,\k}$ is one constraint for the two
degrees of freedom $k_x$ and $k_y$, the crossing takes place along
lines.

Next we consider the vertical symmetry planes $\Gamma{\rm N'H}$,
$\Gamma{\rm NP}$, and ${\rm HNP}$ that project onto the lines
$\g2\,\m2$, $\g2\,\x2$, and $\x2\,\m2$ respectively. In real space
these are $m'$ planes (mirror composed with $T$), but because of the
time-reversal component, a generic point on the reciprocal-space plane
is not invariant under $m'$.  Instead, the symmetry operations that
leave the wavevector invariant (modulo $\G$) on those planes are of
the form $C_2T$, where $C_2$ is a two-fold rotation about an axis
normal to the plane.  Because the $C_2T$ operator is antiunitary it
does not admit multiple irreps, and so does not lead to nodal rings as
in the case of simple mirror planes. However, it does place additional
restrictions on the form of the Hamiltonian on the plane.  In
particular, it forces the Hamiltonian matrix to be real when expressed
in terms of symmetry-adapted Bloch basis orbitals $\ket{\chi_{n\k}}$
whose phases are chosen such that
$C_2T\ket{\chi_{n\k}}=\ket{\chi_{n\k}}$. Since degeneracies occur with
codimension two for real Hamiltonians, WPs generically occur on the
symmetry plane.  \footnote{The effective Hamiltonian for a pair of
  crossing bands can only contain $\sigma_1$ and $\sigma_3$ Pauli
  matrices; zeroing their coefficients generates a
  degeneracy. Alternative arguments for the codimension of two for
  states in a $C_2T$ symmetry plane can be found in
  Ref.~\onlinecite{fang-prb15}.}

Finally, we will encounter one more type of degeneracy, namely those
occurring at generic points in the interior of the BZ. Here a division
into distinct irreps again plays no role, since no symmetry is
present.  One still gets isolated touchings in general, however,
because the codimension is three for the occurrence of degeneracies in
Hermitian Hamiltonians.  That is, by adjusting the wavevectors
$(k_x,k_y,k_z)$ one can generically zero the prefactors of the Pauli
matrices $(\sigma_1,\sigma_2,\sigma_3)$ needed to express the
effective Hamiltonian in the vicinity of a putative crossing of two
bands.  In the absence of fine tuning, these are always simple Weyl
nodes with a chiral charge of $\pm1$.

We emphasize that all of the above considerations apply when SOC is
present.  When it is neglected, additional degeneracies can occur, as
will be discussed briefly in Sec.~\ref{sec:fs-fe-overview}.

\subsubsection{Survey of degeneracies}
\label{sec:degen-survey}

We have used a combination of numerical tools to locate and
characterize the degeneracies in the spinor bandstructure of bcc
Fe. In this section we present some representative results from our
survey.

\begin{figure}
\begin{center}
\includegraphics[width=0.9\columnwidth]{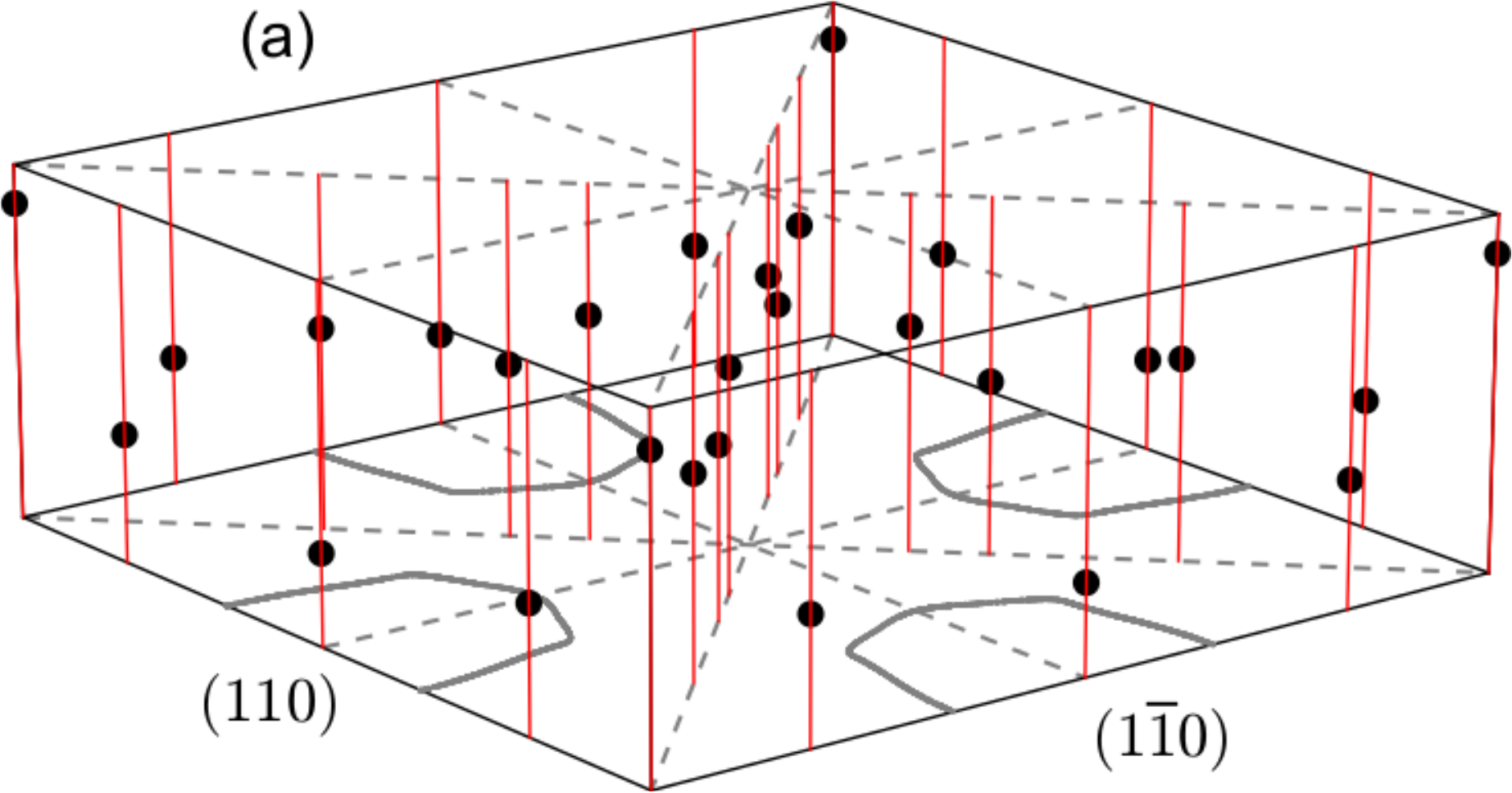}
\includegraphics[width=0.9\columnwidth]{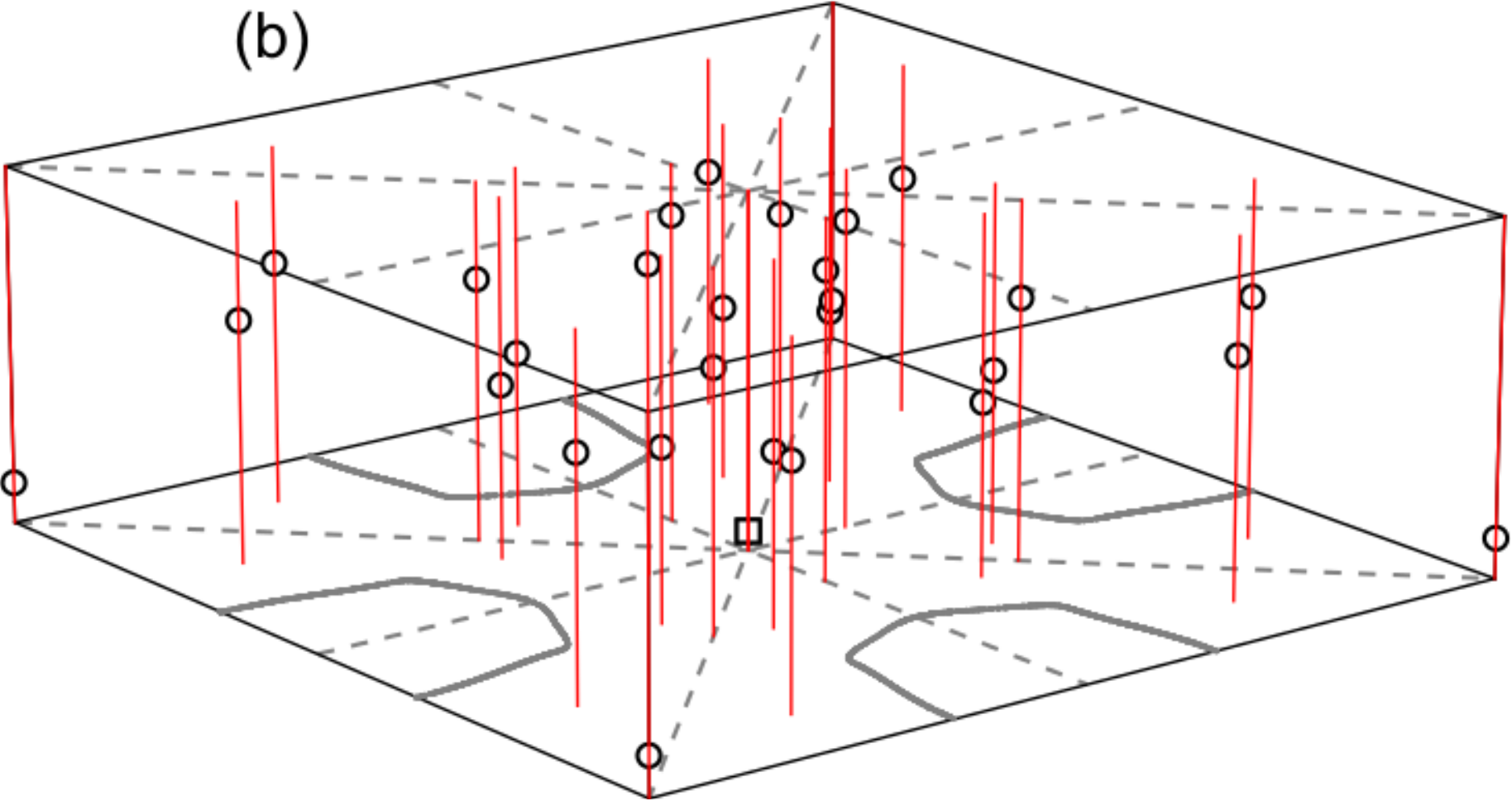}
\end{center}
\caption{(Color online.) Degeneracies between bands six and seven in
  the half Brillouin zone of Fig.~\ref{fig3}.  (a) Black dots denote
  Weyl points of positive chirality (monopole sources of Berry
  curvature on band six). (b) Open circles denote Weyl points of
  negative chirality, and the single open square represents a negative
  double-Weyl node. Each chiral degeneracy is threaded by a vertical
  line to help locate it with respect to the projected high-symmetry
  lines (dashed grey lines).  The solid grey lines on the $k_z=0$
  plane represent nonchiral nodal rings.}
\label{fig4}
\end{figure}

In order to locate all the degeneracies between a pair of bands, we
carry out a steepest-descent minimization of the gap function
$(E_{n+1,\k}-E_{n\k})^2$ starting from a uniform grid covering the BZ,
and look for gap-closing points.\footnote{Note that the gap function
  $(E_{n+1,\k}-E_{n\k})^2$ is differentiable everywhere in the BZ,
  including at degeneracy points where the bands disperse linearly in
  one or more directions. In our Wannier-based implementation the
  needed band gradients are readily available.\cite{yates-prb07}} (In
practice we flag as a potential degeneracy any point where the gap is
below some small threshold, of the order of $10^{-5}$~eV.) The chiral
charge of each degeneracy is determined by enclosing it in a small box
and evaluating the outward Berry flux [\eq{chi}]; in this way we are
able to discriminate between a true chiral band crossing with nonzero
quantized flux, and a tiny avoided crossing for which the flux
vanishes. Figure~\ref{fig4} shows a map of the degeneracies between
bands six and seven in the half-BZ of Fig.~\ref{fig3}.

\begin{figure}
\centering\includegraphics[width=0.7\columnwidth]{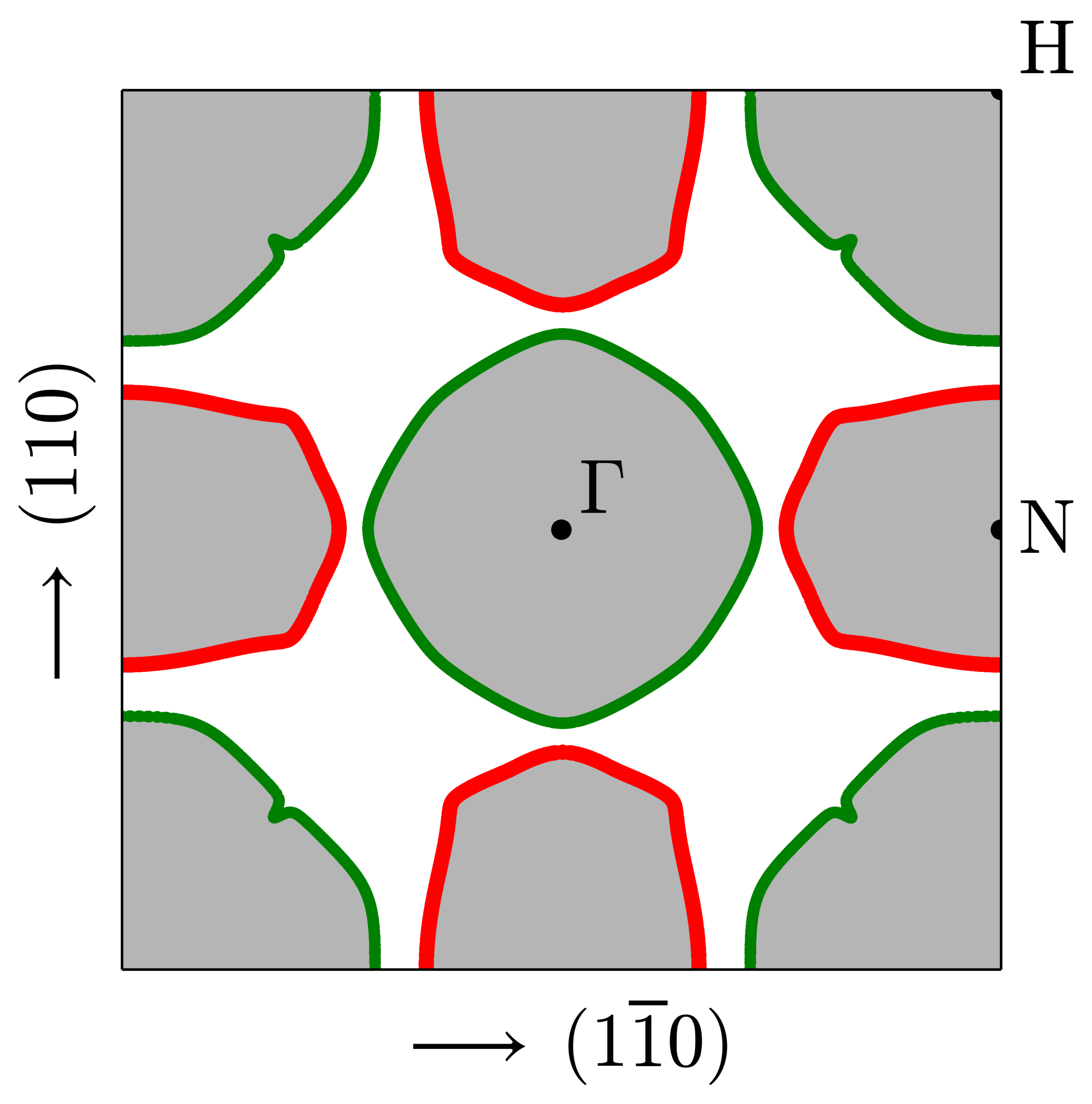}
\caption{(Color online.) Mirror symmetry labels ($+i$ in grey and $-i$
  in white) of band six on the mirror plane at $k_z=0$.  The
  degeneracy lines of band six with bands five and seven are drawn in
  thin (green) and thick (red) lines respectively.}
\label{fig5}
\end{figure}

In addition to finding isolated degeneracies, the above procedure
results in a rather dense set of nonchiral degeneracies between bands
six and seven lying on the mirror plane at $k_z=0$. These are
organized along closed loops, consistent with the line-node scenario
of Sec.~\ref{sec:line-nodes}. In order to confirm that they are nodal
rings protected by mirror symmetry, we have performed two separate
numerical tests. First we checked that these are true crossings
between states belonging to different irreps, by plotting in
Fig.~\ref{fig5} the mirror symmetry label of band six across the
$k_z=0$ plane. The boundaries between regions of different symmetry
(the grey and white regions) coincide with the locus of degeneracies
of band six with either band seven or band five, indicated with lines.
For example, when going between a grey and a white region across a
thick (red) line, bands six and seven cross with one another and
exchange symmetry labels.

The second test was to determine the Berry flux carried by each
ring. To this end we evaluated numerically the Berry phases of small
circular loops interlinked with the rings, obtaining the value~$\pi$
expected for nodal rings.  Note that the sign of the Berry phase of a
mirror-symmetric vertical loop is flipped by the mirror operation, but
because the Berry phase is only defined modulo $2\pi$, the nontrivial
value $\pi$ is still allowed.

Recently, the Fermi surfaces of several $T$-invariant crystals (some
$P$-broken,\cite{huang-natcomm15,weng-prx15} others
$P$-invariant\cite{weng-prb15,zeng-arxiv15}) were found to consist of
nodal rings on mirror planes, but only when SOC is absent.  Those
Fermi rings are crossings between {\it pairs} of spin-degenerate
bands, and SOC hybridizes the states of opposite spin, gapping the
rings everywhere except at a few isolated points despite the unbroken
mirror symmetry.\cite{weng-prx15,zeng-arxiv15} In bcc~Fe the spin
degeneracy is lifted by the exchange interaction, removing the
hybridization channels and stabilizing the nodal rings against SOC
(except when SOC destroys the mirror symmetry itself, e.g., on the
vertical symmetry planes).

Let us now analyze the isolated degeneracies away from the $k_z=0$
plane in Fig.~\ref{fig4}. There are 45 in total in the half-BZ, and
except for a double-Weyl node along the four-fold axis $\Delta$ all of
them are simple WPs. Their properties are listed in
Table~\ref{table1}. Each row represents one or more symmetry-related
PNs lying on the same BZ slice at fixed~$k_z$, and their multiplicity
(one, two, four, or eight) is determined by the projection onto the 2D
BZ at the bottom of Fig.~\ref{fig3}. Multiplicities of two and four
are generated by $C_4$ symmetry, and multiplicities of eight are
generated by $C_4$ symmetry together with $C_2T$ symmetry about the
vertical symmetry planes.  A mirror-equivalent set of nodes, but with
reversed chiral charges, appears in the $k_z<0$ half of the BZ.

The degeneracies between bands six and seven realize almost all
possibilities that can generically exist in a tetragonal ferromagnet:
nodes of chiral charge $\pm1$ that project onto $\g2$ or $\m2$
(multiplicity one), $\x2$ (multiplicity two), $\g2\,\m2$, $\g2\,\x2$,
or $\x2\,\m2$ (multiplicity four), or generic points (muliplicity
eight); and nodes of chiral charge $\pm2$ projecting onto $\g2$ or
$\m2$.  Only the very last possibility is missing between bands six
and seven in Table~\ref{table1}, but an opposite-spin crossing of this
type occurs, for example, between bands four and five at $k_z\simeq
0.376\times 2\pi/a$.

\def\upup{$\uparrow\uparrow$}
\def\updn{$\uparrow\downarrow$}
\def\dndn{$\downarrow\downarrow$}
\def\dent#1{\multicolumn{1}{c}{$#1$}}
\begin{table}
\begin{center}
\begin{ruledtabular}
\begin{tabular}{dddcdrcc}
\dent{k_x} & \dent{k_y} & \dent{k_z} & Proj. & \dent{E} & Spin & \dent{\chi} & 
Mult.\\
\hline
  0.0   &  0.0   &  0.027 &  $\g2$      & -1.16 & \dndn & $-2$ & 1 \\
  0.0   &  1.0   &  0.056 &  $\m2$      &  0.06 & \dndn & $-1$ & 1 \\
  0.5   &  0.5   &  0.130 &  $\x2$      & -0.96 & \dndn & $+1$ & 2 \\
  0.180 &  0.820 &  0.180 &  $\x2\,\m2$ & -0.34 & $\sim$\dndn & $+1$ & 4 \\
  0.074 &  0.322 &  0.242 &  gen.       & -1.01 & \updn & $-1$ & 8 \\
  0.0   &  0.327 &  0.243 & $\g2\,\m2$  & -0.99 & \updn & $+1$ & 4 \\
  0.0   &  0.583 &  0.284 & $\g2\,\m2$  & -0.87 & \dndn & $+1$ & 4 \\
  0.0   &  0.217 &  0.316 & $\g2\,\m2$  & -0.99 & \upup & $+1$ & 4 \\
  0.135 &  0.662 &  0.338 & gen.        & -1.07 & \dndn & $-1$ & 8 \\
  0.0   &  0.365 &  0.365 & $\g2\,\m2$  & -1.02 & \upup & $-1$ & 4 \\
  0.118 &  0.118 &  0.429 & $\g2\,\x2$  & -0.84 & \updn & $-1$ & 4 \\
  0.0   &  1.0   &  0.446 & $\m2$       & -0.94 & \updn & $+1$ & 1
\end{tabular}
\end{ruledtabular}
\caption{\label{table1}Census of chiral degeneracies between bands six
  and seven in the half BZ of Figs.~\ref{fig3}-\ref{fig4}. The
  coordinates $(k_x,k_y,k_z$) are in units of $2\pi/a$; energy $E$ is
  in eV relative to the Fermi level.  ``Proj.'' indicates the
  projection onto the 2D BZ at fixed $k_z$ (bottom of Fig.~\ref{fig3};
  ``gen.'' is a generic point). Last three columns indicate spin
  crossing type ($\downarrow$~and~$\uparrow$ are majority and minority
  spins respectively), chiral charge ($\chi>0$ if the band touching is
  a source of Berry curvature on band six), and multiplicity within
  the 2D BZ.}
\end{center}
\end{table}

\begin{figure}
\centering\includegraphics[width=0.90\columnwidth]{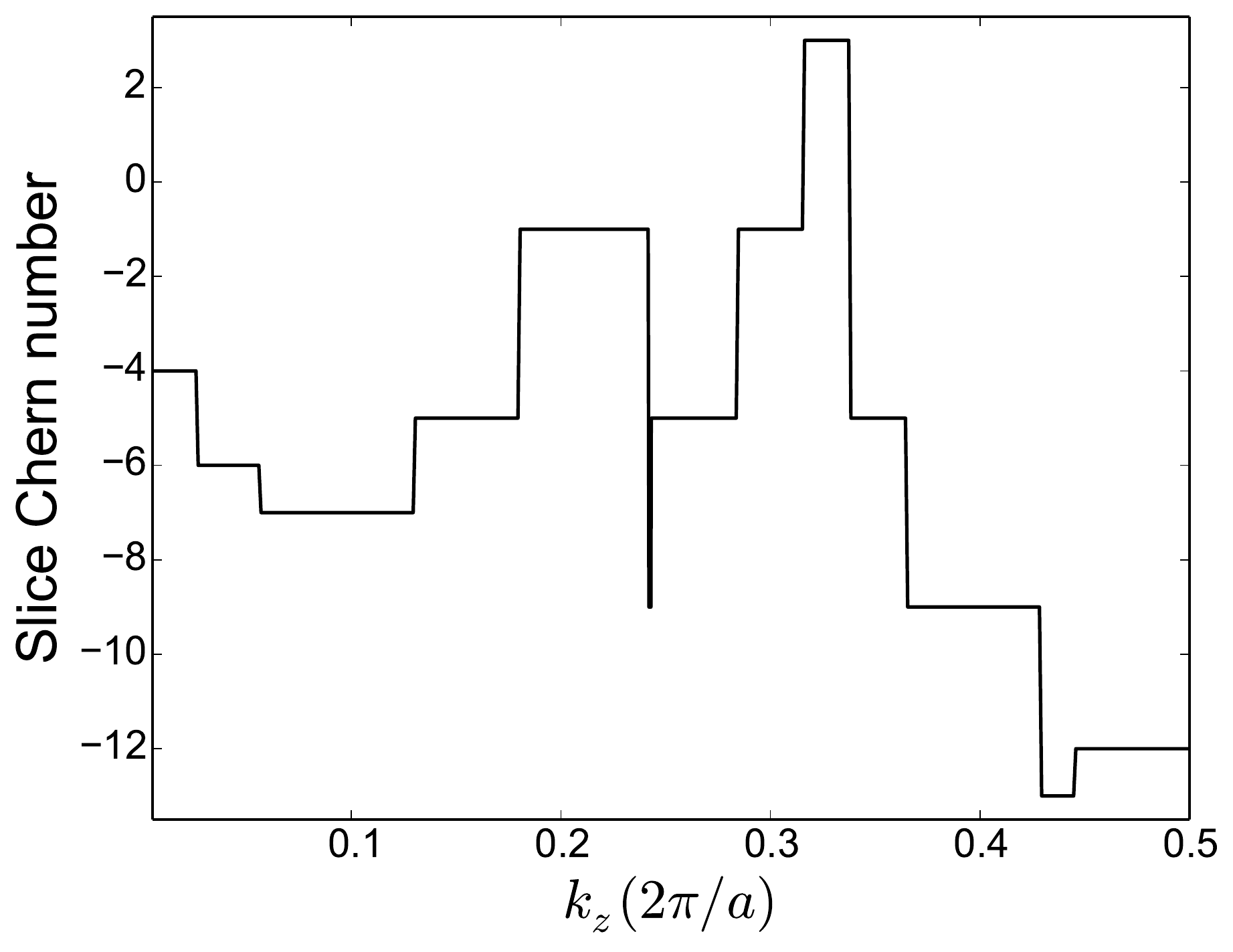}
\caption{Slice Chern number $\wt{C}$ [\eq{Ctil-n-kz}] associated with
  the six lowest bands of bcc~Fe. Since $\wt{C}$ is symmetric about
  $k_z=0$ and $k_z=\pi/a$ [\eqs{C-about-0}{C-about-pi}], only half of
  the range of $k_z$ is shown.}
\label{fig6}
\end{figure}

The correctness of the PN assignments in Table~\ref{table1} is
confirmed in Fig.~\ref{fig6}, where we plot between $k_z=0$ and
$k_z=\pi/a$ the joint slice Chern number $\wt{C}$ [\eq{Ctil-n-kz}] of
the six lowest bands. Each step discontinuity signals the presence on
the corresponding BZ slice of one or more chiral PNs connecting bands
six and seven.  The positions of the steps match the $k_z$ coordinates
listed in Table~\ref{table1}, and their sizes satisfy
\beq
\label{eq:Delta-C}
\Delta\wt{C}=(\text{Multiplicity})\times(\text{Chiral charge})\,
\eeq
for the chiral charges and multiplicities in Table~\ref{table1}.

\begin{figure}
\centering\includegraphics[width=0.90\columnwidth]{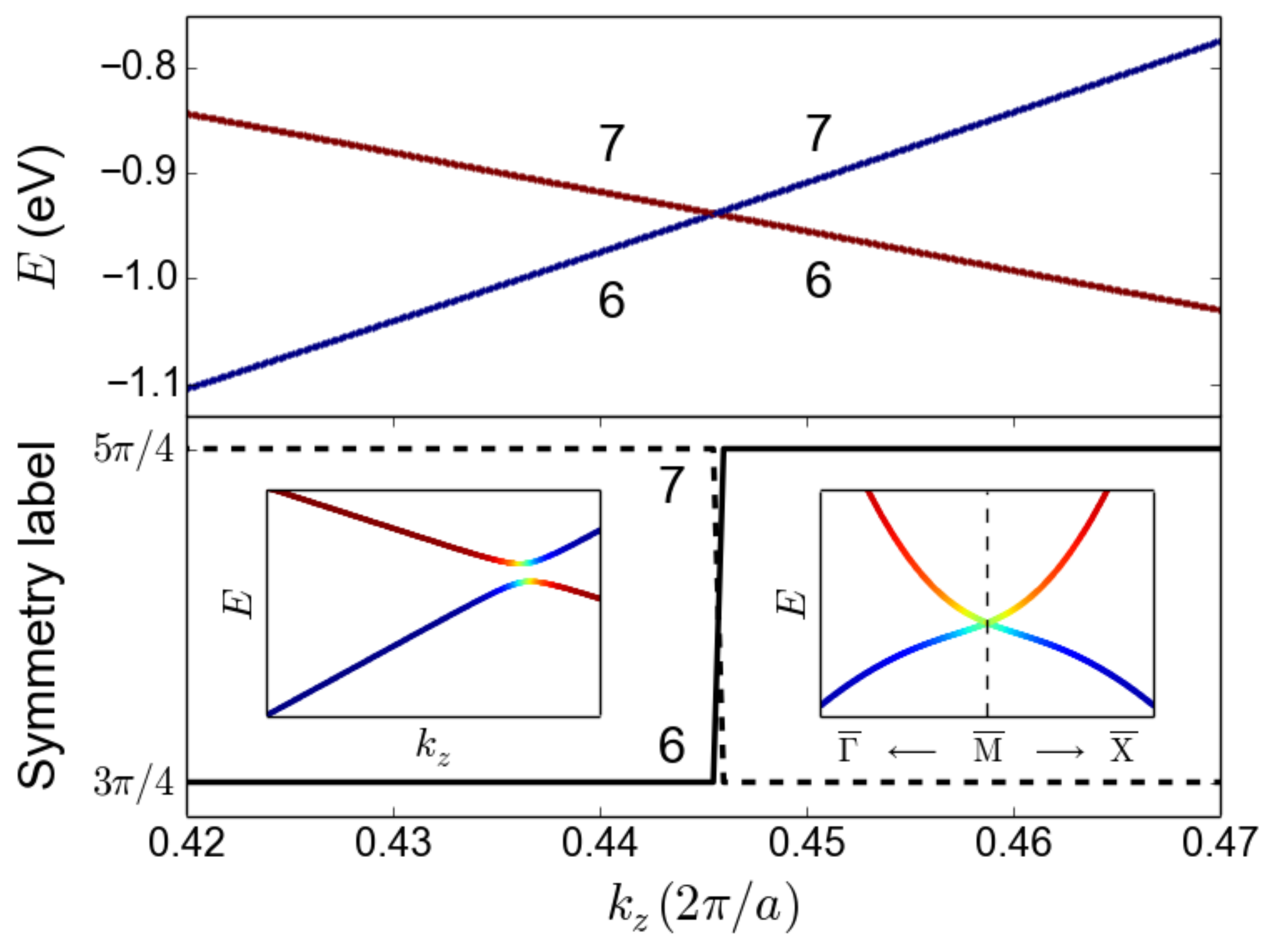}
\caption{(Color online.)  A Weyl point between bands six and seven
  along the four-fold axis that projects onto $\m2$ in
  Fig.~\ref{fig3}. The top panel and the right inset show the band
  dispersions as one passes through the touching point along the
  symmetry axis and on the orthogonal plane respectively. The left
  inset shows the dispersion along a vertical line that is shifted
  from the symmetry axis by $\delta k_x=0.05\times 2\pi/a$.  The
  bands are color-coded by the spin as in Fig.~\ref{fig2}. The main
  bottom panel shows the evolution along the symmetry axis of the
  phase of the $C_4$ eigenvalues (symmetry labels) of the crossing
  bands.}
\label{fig7}
\end{figure}

\begin{figure}
\centering\includegraphics[width=0.9\columnwidth]{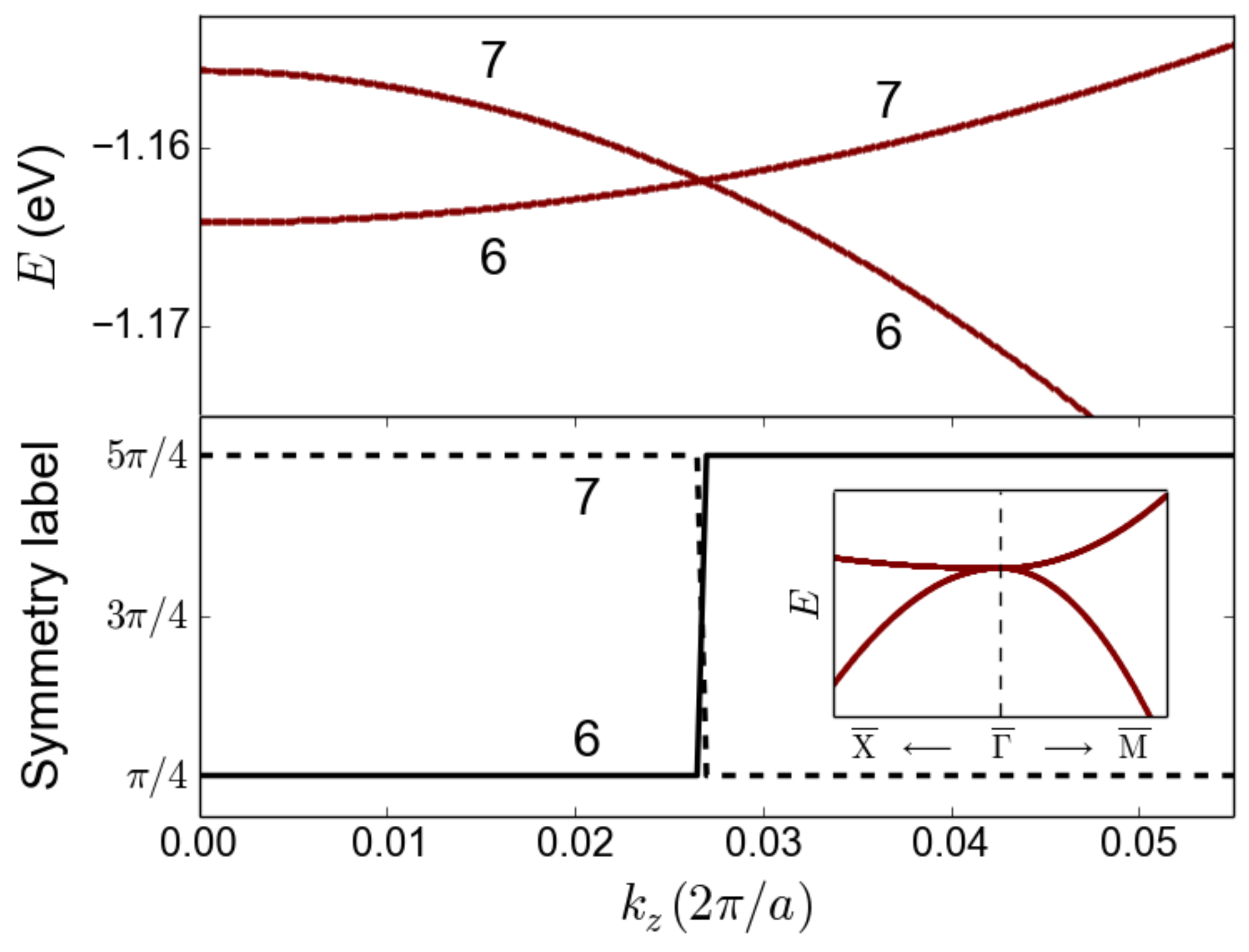}
\caption{(Color online.) A double-Weyl node between bands six and
  seven along the four-fold axis $\Delta$ in Fig.~\ref{fig3}.  The top
  and bottom panels and the right inset have the same meaning as in
  Fig.~\ref{fig7}.  Near the crossing the two bands have majority-spin
  character, as indicated by the red color.}
\label{fig8}
\end{figure}

Let us contrast the behavior of the energy bands and symmetry labels
in the vicinity of a simple WP and of a double-Weyl node, when both
lie along four-fold axes. Figure~\ref{fig7} pertains to the WP at
$k_z\simeq 0.446$ projecting onto $\m2$. The upper and lower panels
show respectively the linearly-dispersing bands and the evolution of
the symmetry labels along the symmetry axis. As expected for a WP of
positive chirality, the label of the lower band changes by a factor of
$i$ at the crossing.\cite{fang-prl12} Exactly at the crossing the two
bands retain their separate spin characters, but moving slightly away
from~$\m2$ (left inset) they hybridize and no longer touch.  The right
inset shows that the bands also disperse linearly away from the
touching point in the orthogonal directions (the leading behavior is
linear even though strong nonlinearities are present), but now with
strong spin mixing near the crossing.  The behavior around the
double-Weyl node on the $\Delta$ axis is shown in Fig.~\ref{fig8}, and
it is qualitatively different. Here the bands still disperse linearly
away from the node along the axis, but the dispersion is quadratic on
the orthogonal plane at the critical $k_z$. Moreover, the symmetry
labels of the crossing bands are noncontiguous on the complex unit
circle.\cite{fang-prl12}

This concludes our survey of the degeneracies between bands six and
seven; other bands display the same basic types of degeneracies.  For
future reference we list in Table~\ref{table2} the chiral degeneracies
between bands nine and ten.

\begin{table}
\begin{center}
\begin{ruledtabular}
\begin{tabular}{dddcdrcc}
\dent{k_x} & \dent{k_y} & \dent{k_z} & Proj. & \dent{E} & Spin & \dent{\chi} & Mult.\\
\hline
  0.0   &  1.0   &  0.047 &  $\m2$      &  2.35 & \upup & $+1$ & 1 \\
  0.0   &  0.0   &  0.065 &  $\g2$      & -0.25 & \upup & $+1$ & 1 \\
  0.188 &  0.812 &  0.188 &  $\x2\,\m2$ &  1.85 & \upup & $-1$ & 4 \\
  0.322 &  0.322 &  0.331 & $\g2\,\x2$  &  1.81 & \updn & $+1$ & 4 \\
  0.0   &  0.0   &  0.428 &  $\g2$      & -0.06 & \upup & $-1$ & 1 \\
  0.233 &  0.233 &  0.446 & $\g2\,\x2$  &  1.39 & \upup & $-1$ & 4 \\
  0.0   &  0.0   &  0.480 &  $\g2$      &  0.07 & $\sim$\upup & $+2$ & 1
\end{tabular}
\end{ruledtabular}
\caption{\label{table2}Census of chiral degeneracies between bands
  nine and ten in the half BZ of Fig.~\ref{fig3}. The table is
  organized in the same way as Table~\ref{table1}.}
\end{center}
\end{table}

\subsection{Fermi surface}
\label{sec:fs-fe}

\subsubsection{Overview and spin-orbit effects}
\label{sec:fs-fe-overview}

\begin{figure}
\begin{center}
\includegraphics[width=0.9\columnwidth]{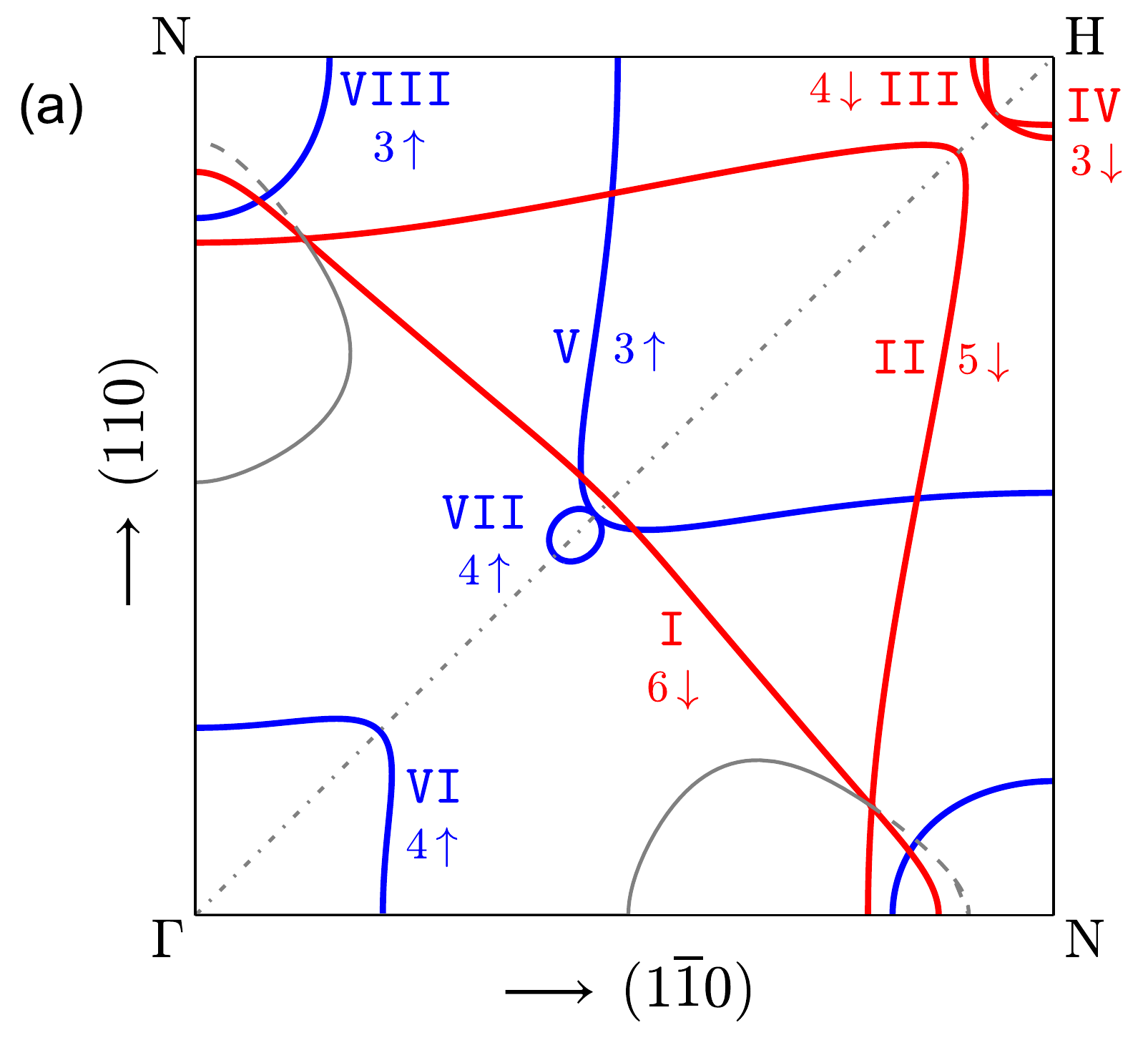}
\vskip 0.3cm
\includegraphics[width=0.9\columnwidth]{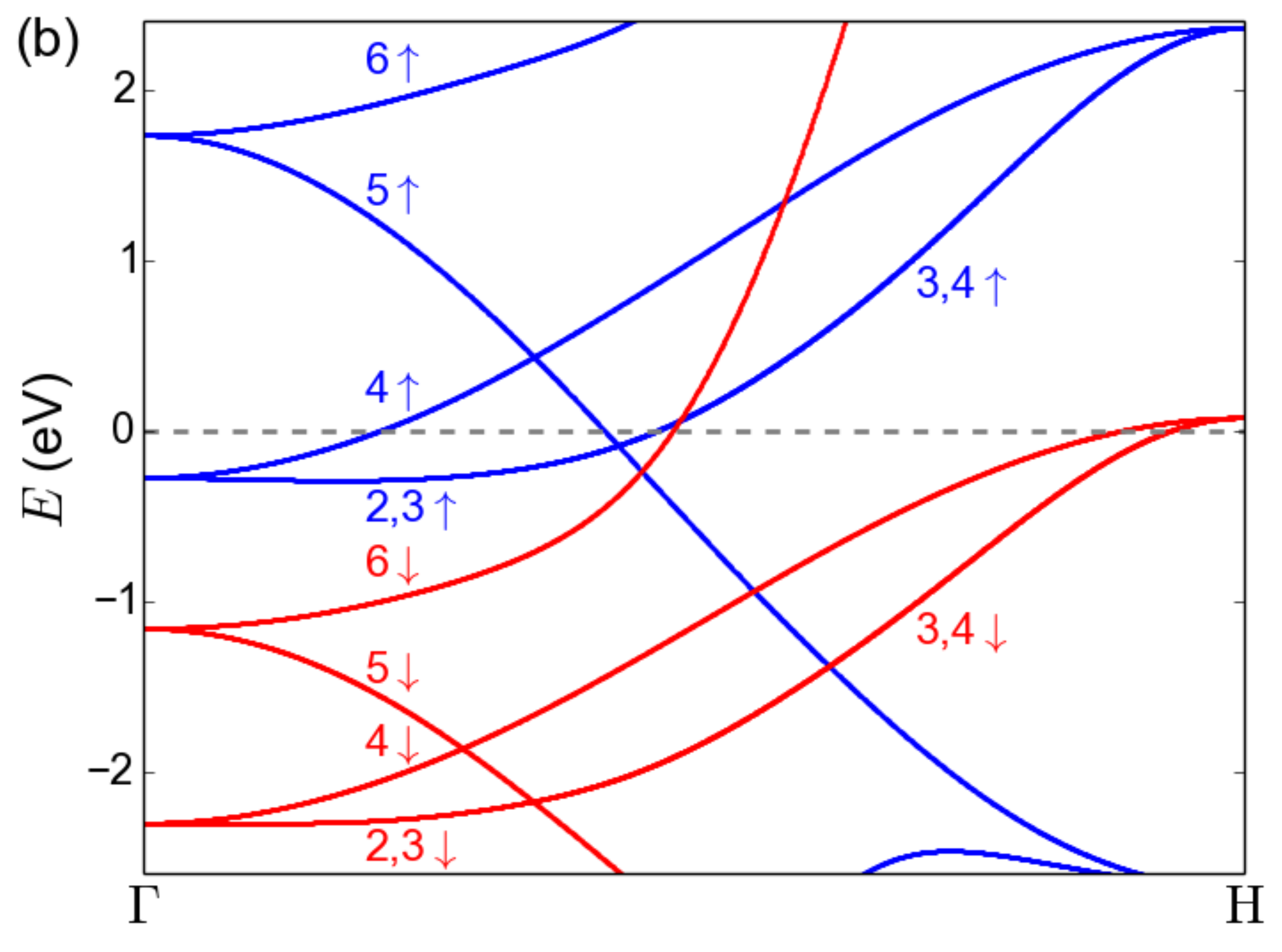}
\end{center}
\caption{(Color online.) (a) Fermi contours of bcc Fe on the
  $\Gamma{\rm N H}$ plane at $k_z=0$, calculated without spin-orbit
  coupling.  The majority-spin contours~I--IV are in red, the
  minority-spin contours~V--VIII in blue. Curved grey lines are
  accidental degeneracy lines (nodal rings) that cross the Fermi level
  at points of contact between like-spin Fermi sheets, where they
  change from solid ($E_{\rm degen}<\ef$) to dashed ($E_{\rm
    degen}>\ef$). The dashed-dotted line from~$\Gamma$ to~${\rm H}$ is
  a line of essential degeneracy.  (b) Energy bands close to the Fermi
  level along the $\Gamma{\rm H}$ line. In both panels, the
  spin-dependent band indices are indicated.}
\label{fig9}
\end{figure}

We begin by analyzing the FS of bcc~Fe with SOC switched off. Callaway
and co-workers\cite{singh-prb75,callaway-prb77} introduced a labeling
scheme for the Fermi sheets that has been widely adopted in the
literature.  The sheets are organized into eight groups.  In a
calculation without SOC the majority-spin Fermi sheets belong to
groups I--IV, and the minority-spin sheets to groups V--VIII.
Figure~\ref{fig9}(a) shows the labeled Fermi contours on the mirror
plane at $k_z=0$.  We now consider the possible touchings and
intersections between Fermi sheets, starting with the SOC-free case.

Crossings between Fermi sheets of opposite spin occur along entire
loops in the 3D BZ where the two constrains
$E_{n\uparrow}(\k)=E_{m\downarrow}(\k)=\ef$ are satisfied. Some of
those loops intersect the portion of the $k_z=0$ plane shown in
Fig.~\ref{fig9}(a), at the six points where sheets I or II (red) cross
V or VIII (blue).

Touchings between like-spin Fermi sheets can occur along lines of
degeneracy, at the isolated points where the degeneracy energy $E_{\rm
  degen}$ equals~$\ef$. Four such contact points can be seen in
Fig.~\ref{fig9}(a).  Two are linear touchings between majority-spin
sheets I-II, and they are located along nodal rings (the curved grey
lines) connecting bands five and six on the mirror plane.
\footnote{The explanation given in Sec.~\ref{sec:degen-types} for the
  formation of nodal rings on mirror planes carries over to the
  SOC-free bands. The mirror-symmetry eigenvalues are now $\pm 1$, and
  nodal rings can only form between like-spin bands.}  Each nodal ring
changes from a solid line ($E_{\rm degen}<\ef$) to a dashed line
($E_{\rm degen}>\ef$) at the Fermi touching points, and the band
indices of the two Fermi sheets that touch tell which bands are
degenerate along a given ring. The other two are quadratic touchings,
one between sheets III-IV for majority spins and the other between
sheets V-VII for minority spins.  Both of them lie on the
dashed-dotted grey line $\Delta'$ from~$\Gamma$ to~${\rm H}$, where
there is an essential degeneracy between pairs of bands in each spin
channel (without SOC the little group along~$\Delta'$ is $C_{4v}$,
which has a two-dimensional irrep).  The bands close to the Fermi
level are plotted along $\Delta'$ in Fig.~\ref{fig9}(b).  Note that
the essential degeneracy is between bands two and three near~$\Gamma$,
and between bands three and four near~${\rm H}$.

There is one more opportunity for like-spin Fermi touchings, namely at
generic points in the BZ along nodal rings located away from any
symmetry lines or planes: the removal of SOC restores an effective $T$
symmetry in each spin channel, which combined with parity produces an
effective $PT$ symmetry protecting such
rings.\cite{weng-prb15,kim-prl15} Several low-symmetry nodal rings are
present in the bandstructure near the Fermi level, but none of them
cross~$\ef$. For example, they occur below $\ef$ between majority-spin
bands three and four and also five and six, and between minority-spin
bands two and three; and above~$\ef$ between minority-spin bands four
and five.

\begin{figure}
\subfigure[\,Band 5]{\includegraphics[width=3.75cm]{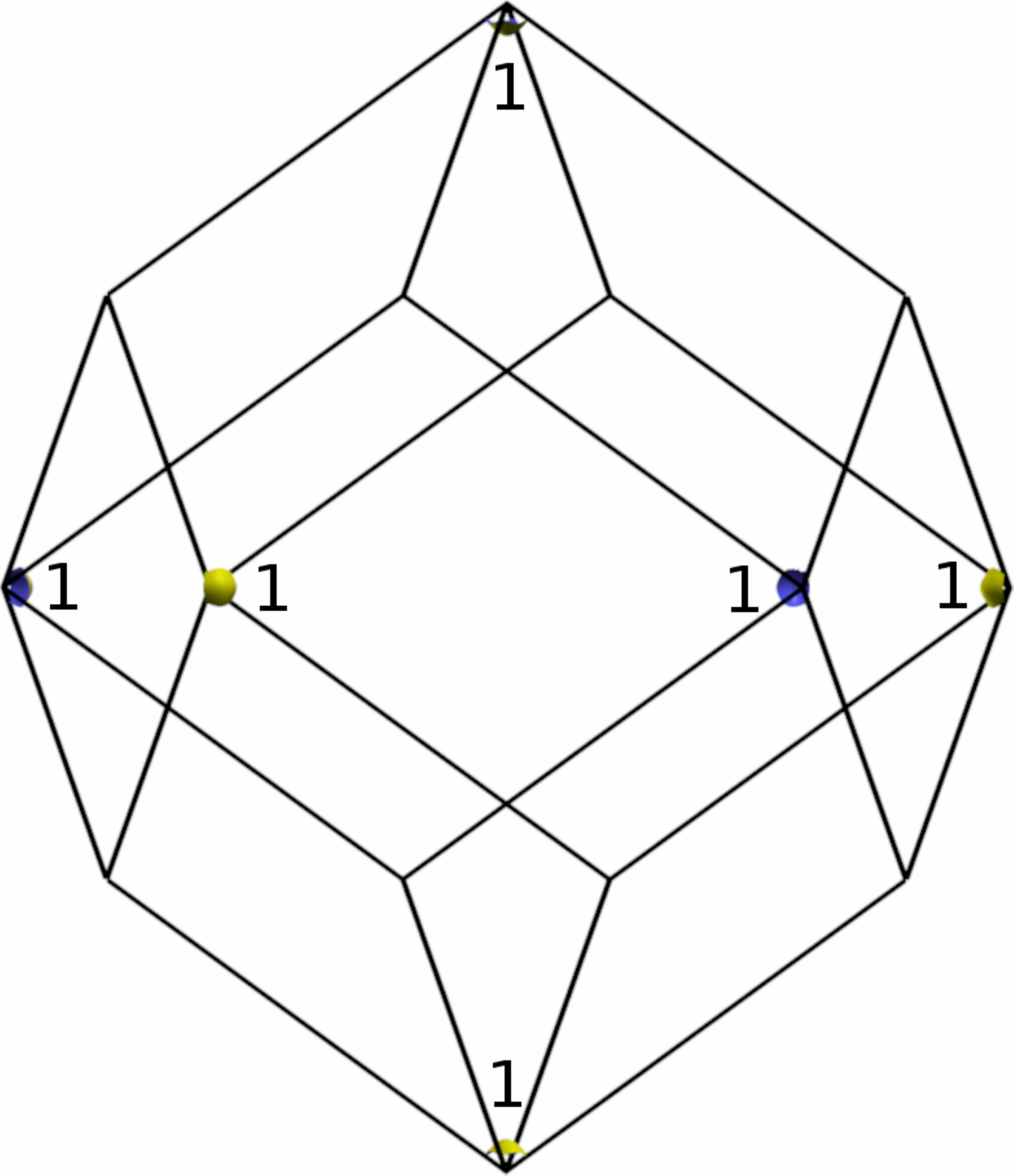}}
\hfill
\subfigure[\,Band 6]{\includegraphics[width=3.75cm]{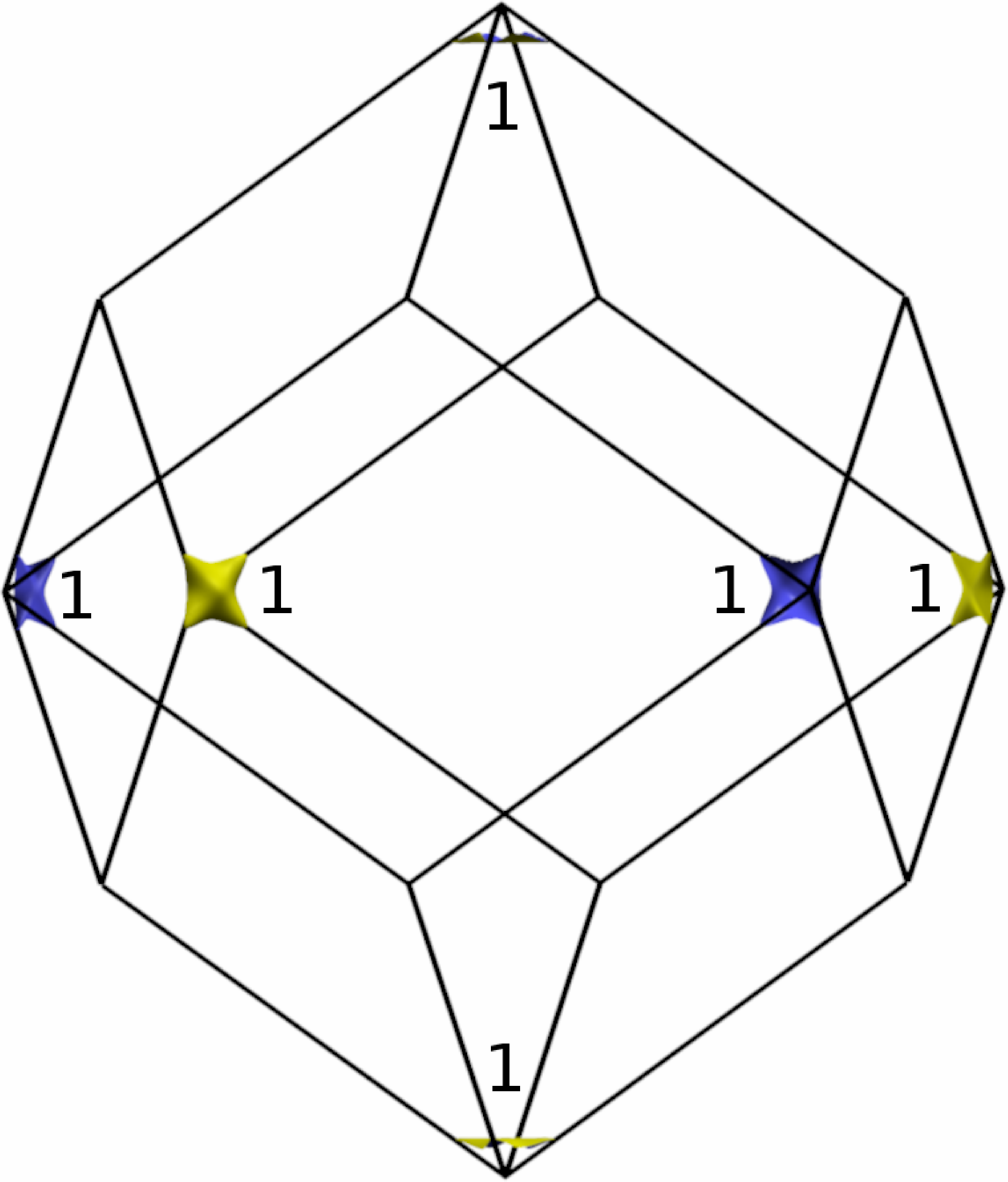}}
\vfill
\subfigure[\,Band 7]{\includegraphics[width=4cm]{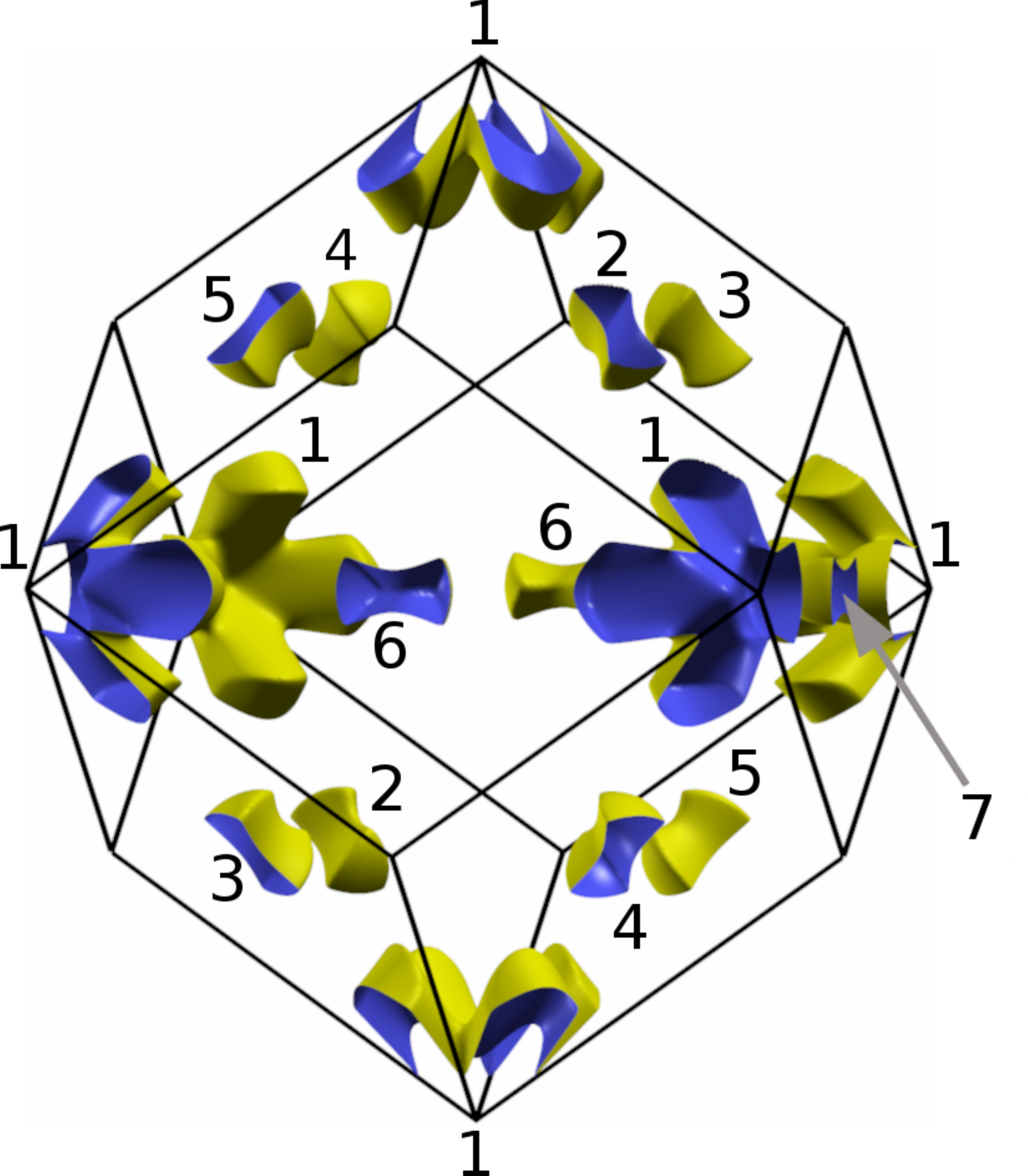}}
\hfill
\subfigure[\,Band 8]{\includegraphics[width=3.75cm]{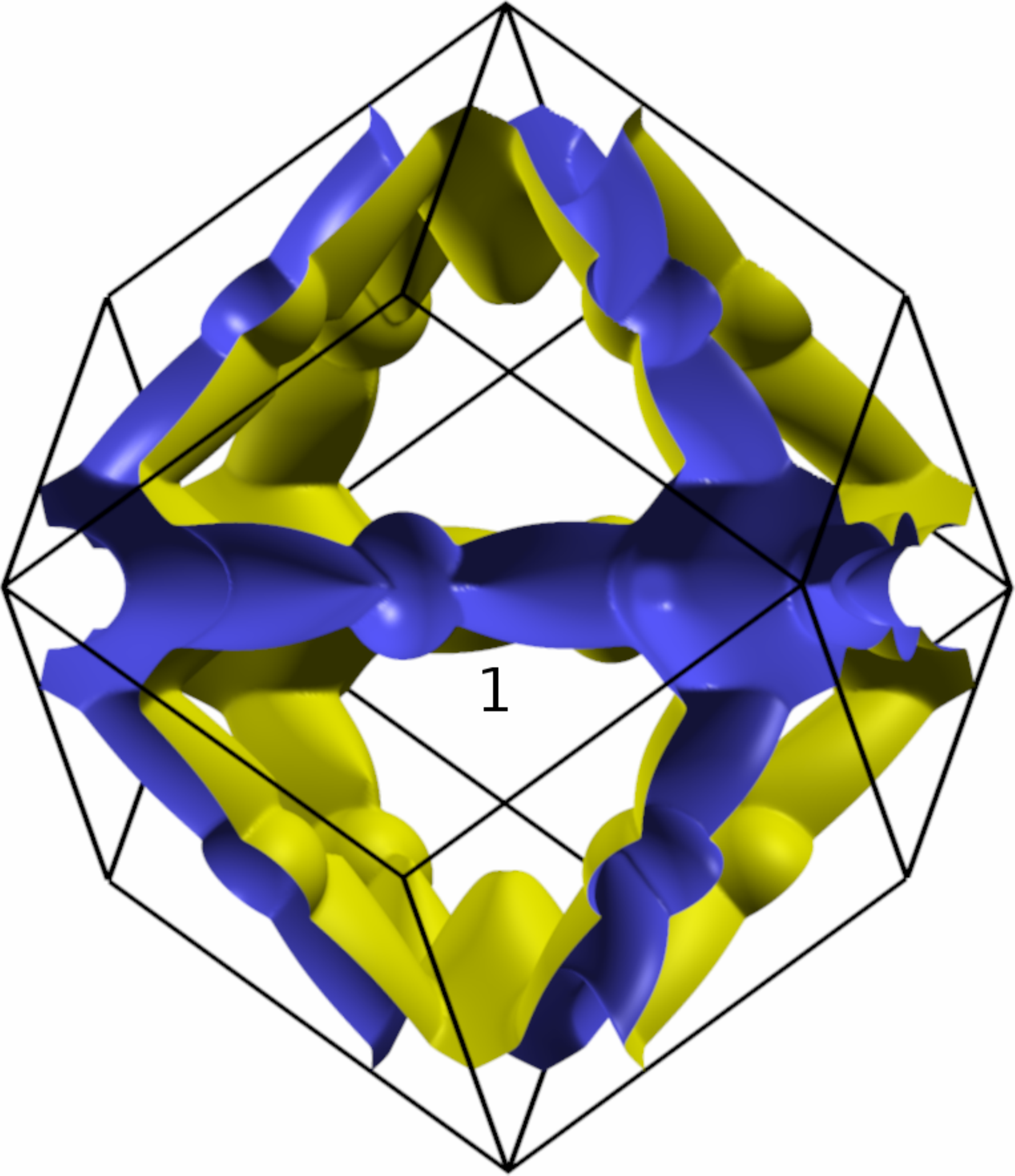}}
\vfill
\subfigure[\,Band 9]{\includegraphics[width=3.75cm]{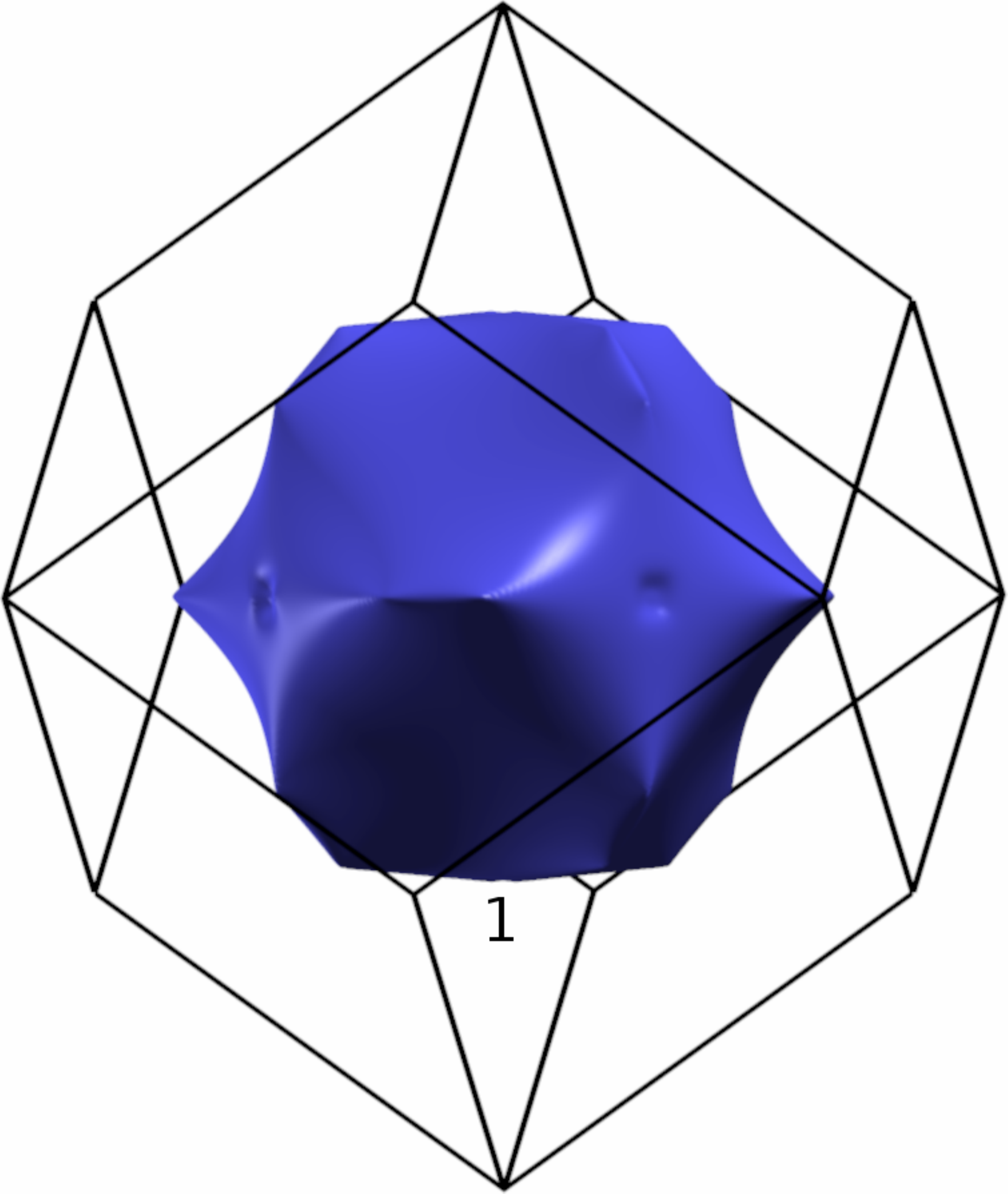}}
\hfill
\subfigure[\,Band 10]{\includegraphics[width=3.75cm]{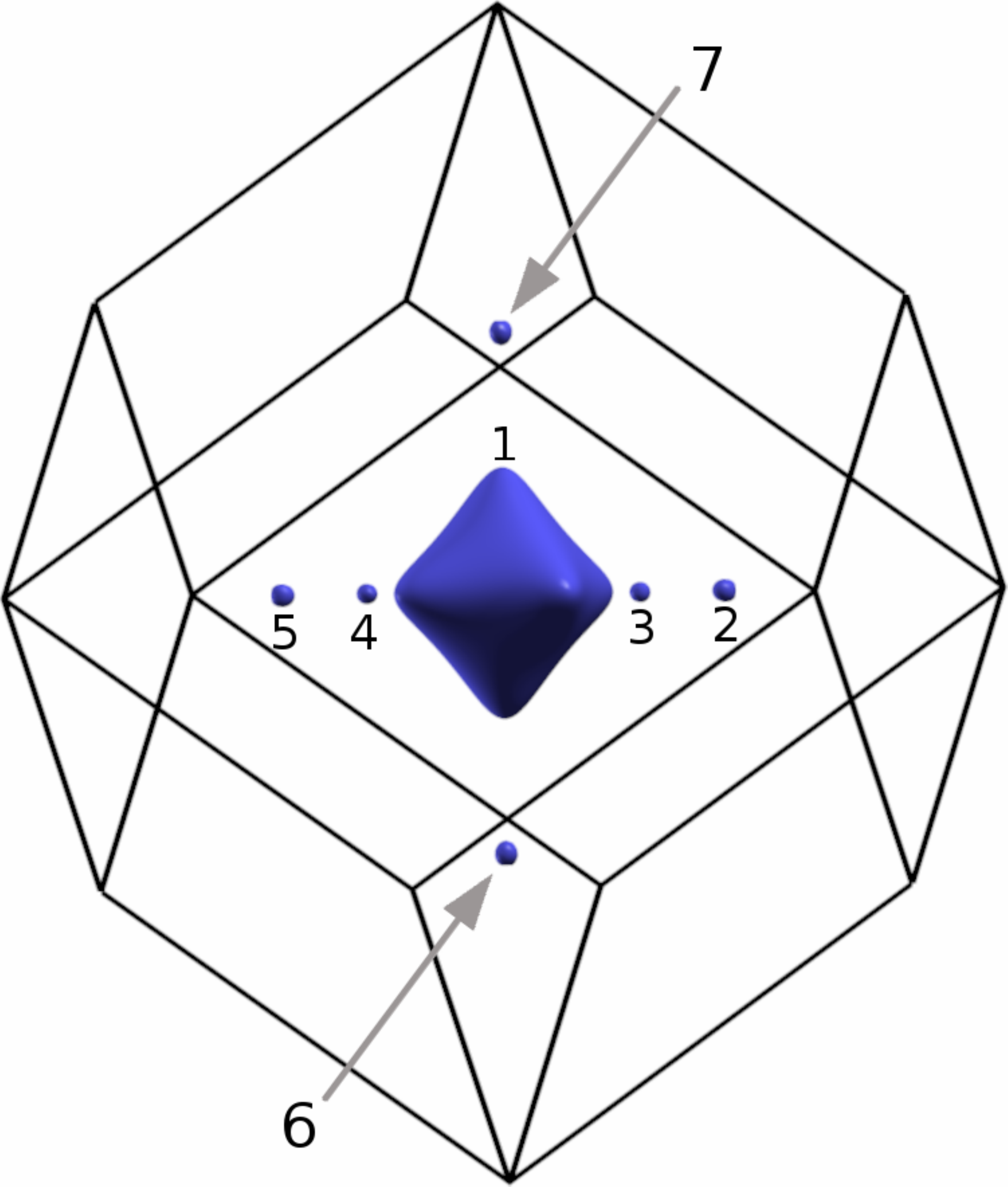}}
\caption{(Color online.)  Fermi surface of bcc Fe with spin-orbit
  coupling included.  The indices $a$ of the individual Fermi sheets
  $S_{na}$ on each band $n$ are indicated.}
\label{fig10}
\end{figure}

Now we turn to the FS with SOC included. Figure~\ref{fig10} shows the
calculated FS in the 3D BZ. Each Fermi sheet is labeled $S_{na}$ as in
Sec.~\ref{sec:sheets}, and sheets with the same band index~$n$ are
displayed together. The unoccupied sides are colored in blue, and the
occupied sides in yellow/gold. Thus the pockets in bands five to seven
are hole-like, and so is the connected tubular structure in band
eight, while the pockets in bands nine and ten are electron-like.

The spin-orbit interaction changes both the symmetry and the
connectivity of the FS, and its organization into groups of
symmetry-related sheets must be modified accordingly.  Without SOC,
group~VIII comprises the six pockets surrounding each of the points
${\rm N}$ in the BZ (the centers of the faces of the Wigner-Seitz cell
in Fig.~\ref{fig10}). With SOC, this group splits in two: group
VIII(a) formed by the four pockets surrounding the points~${\rm N'}$
in Fig.~\ref{fig3} [sheets two to five in Fig.~\ref{fig10}(c)], and
group VIII(b) containing the two pockets surrounding the points~${\rm
  N}$ [sheets six and seven in Fig.~\ref{fig10}(c)].

The other group that gets split by SOC is group VII comprising the six
``satellite'' electron pockets in band ten.  The four pockets along
the lines $\Delta'$ in Fig.~\ref{fig3} [pockets two to five in
Fig.~\ref{fig10}(f)] remain related by $C_4$ symmetry, so we group
them together as VII(a). Pockets six and seven along~$\Delta$ are no
longer related by symmetry to the other four; they are however related
to each other by both parity and mirror reflection, and we label them
VII(b).

\begin{figure*}
\begin{center}
\includegraphics[width=0.85\columnwidth]{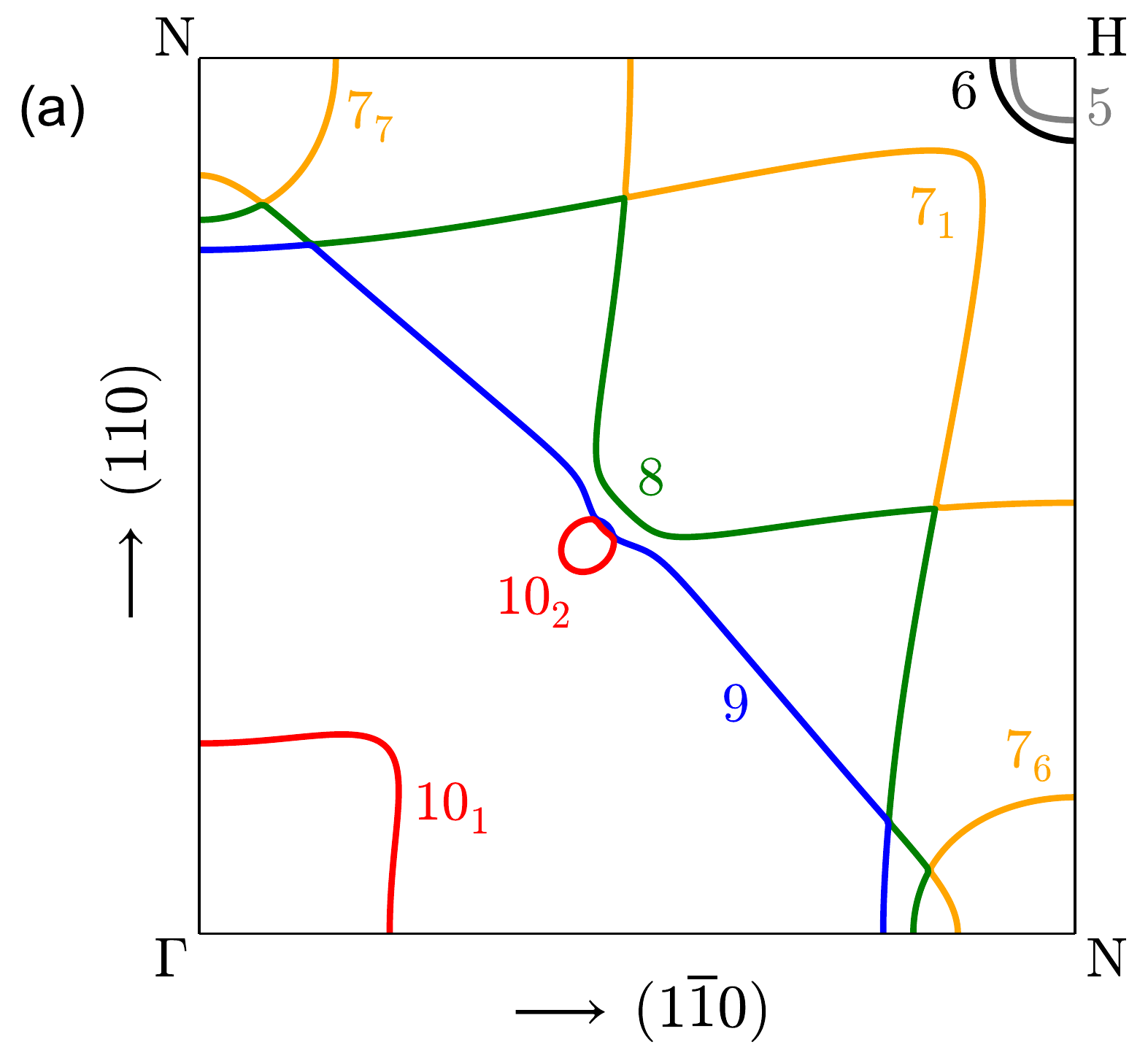}
\hspace{1.0cm}
\includegraphics[width=0.85\columnwidth]{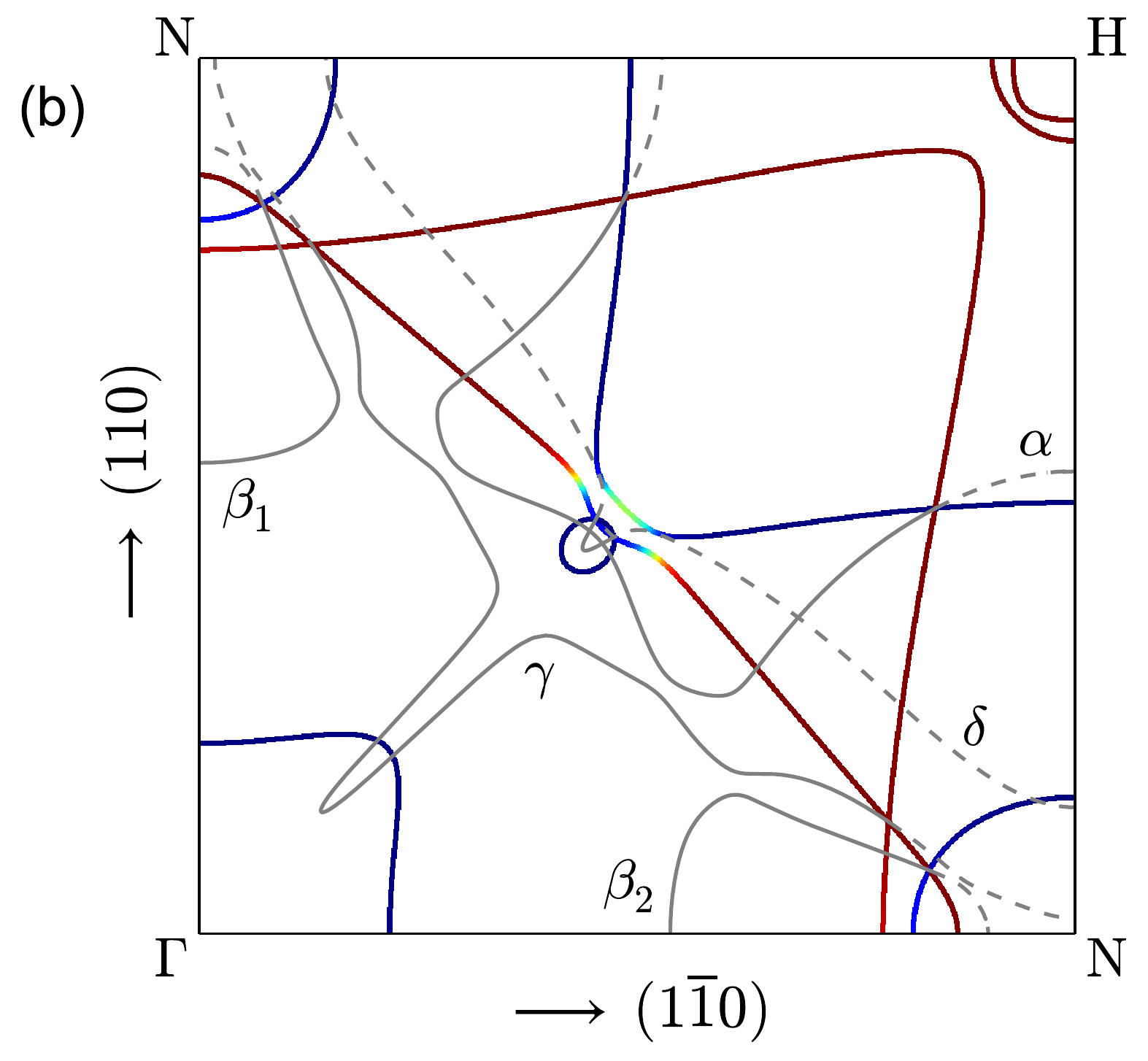}
\end{center}
\caption{(Color online.) Fermi contours of bcc Fe on the $\Gamma{\rm N
    H}$ plane at $k_z=0$, with spin-orbit coupling included. (a)~The
  contours are color-coded by the band index~$n$ and labeled~$n_a$,
  where~$a$ is the sheet index (it is omitted for bands with a single
  Fermi sheet).  (b)~The Fermi contours are color-coded by the spin in
  the same way as the energy bands in Fig.~\ref{fig2}, and the grey
  lines labeled~$\alpha$, $\beta_1$, $\beta_2$, $\gamma$, and $\delta$
  are degeneracy lines (nodal rings) that cross the Fermi level at
  points of contact between Fermi sheets, where they change from solid
  ($E_{\rm degen}<\ef$) to dashed ($E_{\rm degen}>\ef$).}
\label{fig11}
\end{figure*}

\begin{figure}
\begin{center}
\includegraphics[width=0.8\columnwidth]{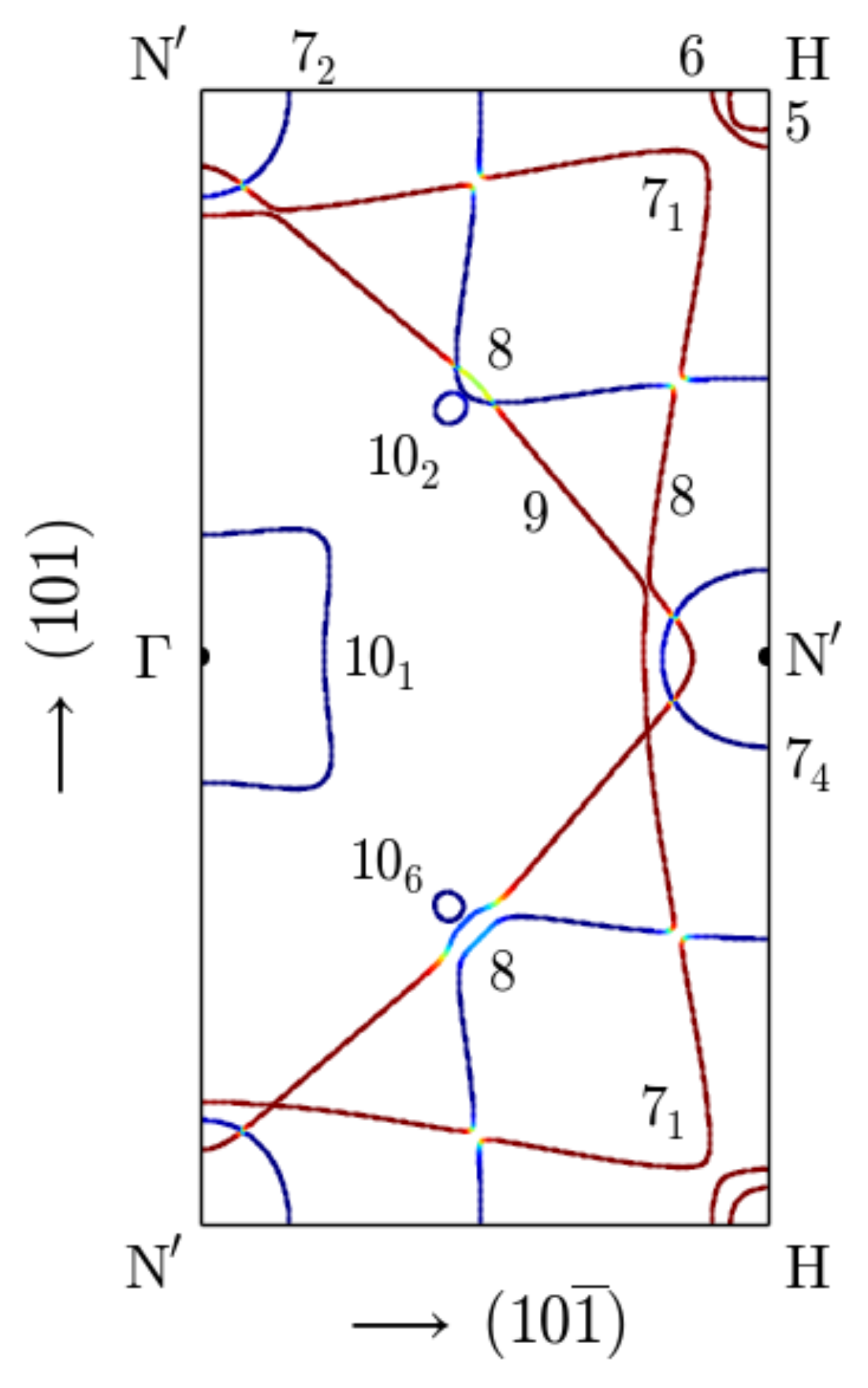}
\end{center}
\caption{(Color online.) Same as Fig.~\ref{fig11}(b), but for the
  $\Gamma{\rm N'H}$ plane at $k_y=0$. The Fermi contours are labeled
  in the same way as in Fig.~\ref{fig11}(a). Because the symmetry on
  this plane is reduced by the spin-orbit interaction, the displayed
  region has twice the area compared to Fig.~\ref{fig11} and there are
  no symmetry-protected nodal rings.}
\label{fig12}
\end{figure}

Figure~\ref{fig11}(a) displays the Fermi contours on the $k_z=0$
plane.  The contours are labeled as in Fig.~\ref{fig1}, but using the
more compact notation $n_a$ instead of $S_{na}$.  (The correspondence
with the labels for groups of symmetry-related Fermi sheets is given
in Table~\ref{table3} below.)  Comparison with Fig.~\ref{fig9}(a)
shows that some, but not all, of the gluing points between Fermi
sheets have been removed by SOC, with sizable gaps opening up in some
cases.  In particular, the quadratic touching between sheets III-IV
along $\Delta'$ has been lifted, and the one between sheets V-VII
turned into a pair of linear crossings between the pair $(9,10_2)$ on
either side of $\Delta'$.

Generic Fermi-sheet touchings require a locus of band degeneracies of
dimension $d\geq 1$ since the additional constraint $E_{\rm
  degen}=\ef$ reduces the dimensionality by one, and with SOC the only
eligible degeneracies are nodal rings on the $k_z=0$ plane.  The nodal
rings connecting bands six and seven in Figs.~\ref{fig4} and
\ref{fig5} lie entirely below $\ef$ and thus play no role here.
However, several nodal rings lying on the $k_z=0$ plane do cross
$\ef$, as indicated by the solid/dashed grey curves in
Fig.~\ref{fig11}(b).  The ones labeled $\alpha$, $\beta_1$, and
$\beta_2$ are degeneracy loops between bands seven and eight; $\alpha$
glues sheet 8 to sheet 7$_1$ at eight points, while~$\beta_1$
and~$\beta_2$ provides gluings to sheets~7$_6$ and~7$_7$ at four
points each.  Ring~$\gamma$ provides eight points of contact between
sheets~8 and~9, and ring~$\delta$ connects sheet~9 twice to each of
10$_2$, 10$_3$, 10$_4$, and 10$_5$.  We thus see that it is quite
common for Fermi sheets to be glued to one another on the $k_z=0$
plane in bcc~Fe. The gluing-together of Fermi sheets is protected by
mirror symmetry, and hence it does not require fine tuning.  By tuning
external parameters, it is also possible to arrange for sheets to
touch away from the $k_z=0$ plane. Two example will be discussed in
Secs.~\ref{sec:chern-glue} and~\ref{sec:results-fs-decomp}.

The situation is very different on the $k_y=0$ plane of
Fig.~\ref{fig12}, which is not a mirror plane with SOC present.  No
actual Fermi touchings occur there, although some of the avoided
crossings are very small and difficult to discern on the scale of the
figure. In particular, what look like two touchings between
sheets~$(8,9)$ in the lower half of the figure are in fact tiny
avoided crossings, as will become clear in Sec.~\ref{sec:ahc-results}.

\subsubsection{Fermi-sheet Chern numbers}
\label{sec:fs-fe-chern}

Without SOC all the band degeneracies in bcc Fe are nonchiral, the
Berry curvature vanishes identically, and as a result the Fermi
surface is topologically trivial. As we have seen, the inclusion of
SOC generates chiral band touchings that act as
sources and sinks of Berry curvature.  We will now determine the Chern
numbers induced on the Fermi sheets from the census of chiral PNs, as
described in Sec.~\ref{sec:sheets}.  For isolated Fermi sheets the
Chern number is unique and is correctly obtained from this census.
For those that are glued to neighboring ones by nodal rings lying on
the $k_z=0$ plane, on the other hand, it provides only part of the
story. The true sheet Chern number is ill-defined until the symmetry
protecting the nodal rings is broken, and when it is, the sheet Chern
number may have additional contributions from new PNs that appear at
special points along the vaporized nodal
ring.  This will be discussed in Sec.~\ref{sec:chern-glue}.

We use the formulation of Sec.~\ref{sec:sheets-immediate}, based on
the PN population in ``immediate regions.''  (Since no Luttinger
anomalies are present, the special treatment that is required in such
cases is not needed here.)  Inspection of Fig.~\ref{fig10} shows that,
for each Fermi sheet, it is possible to identify an occupied or empty
connected BZ region having that Fermi sheet as the sole boundary.  For
the hole-like sheets of band seven that means working with
exterior-like regions, while for the electron-like sheets of band ten
it means working with interior-like regions; bands five, six, eight
and nine have a single sheet each, so that either approach can be used
(in practice we choose the one leading to the smallest number of
enclosed PNs).  In this way we are able to determine the Chern number
of each Fermi sheet from a single evaluation of either \eq{ii} or
\eq{ie}, without the need for a recursive procedure.

\begin{table}
\begin{center}
\begin{ruledtabular}
\begin{tabular}{ccccccc}
Band & Sheet & Group & Type  & Enclosed     & Enclosed & Chern  \\ [-0.025in]
$n$  & $a$   &  label     &      & symm. points & PNs & number \\
\hline
5    & 1    & IV    & Hole    & H          & 0,0        & 0$\phantom{*}$\\
6    & 1    & III    & Hole    & H         & 0,2        & 0$\phantom{*}$\\
7    & 1    &  V   & Hole    & H           & 2,0        & 0*\\
7    & 2,3,4,5 & VIII(a) & Hole & \,\,N$'$ & 0,4        & 0$\phantom{*}$\\
7    & 6,7    &  VIII(b)   & Hole    & N   & 0,0        & 0*\\
8    & 1    &  II   & Hole      & H,N,N$'$ & 16,36      & 0*\\
9    & 1    &  I   & Elec.    & $\Gamma$   & 42,4       & 0*\\
10   & 1   &  VI    & Elec.    & $\Gamma$  & 2,0        & 0$\phantom{*}$\\
10   & 2,3,4,5 & VII(a) & Elec. & \,\,Along $\Delta'$ & 0,0 & 0*\\
10   & 6,7    & VII(b)   & Elec.& Along $\Delta$      & 1,0 & $-$1,+1$\phantom{*}$
\end{tabular}
\end{ruledtabular}
\caption{\label{table3}Chern numbers of the Fermi sheets of bcc Fe,
  with spin-orbit coupling included. Symmetry-related Fermi sheets are
  listed on the same row and are assigned a group label. The Chern
  numbers are determined from the populations of chiral points nodes
  (PNs) within ``immediate regions,'' which are exterior-like
  (interior-like) for hole-like (electron-like) Fermi sheets. For
  non-isolated Fermi sheets (labeled by an asterisk), this definition
  neglects the possible contribution of $\pi$ fluxes at touching
  points between Fermi sheets (see Sec.~\ref{sec:chern-glue}). The
  high-symmetry points enclosed by each Fermi sheet are indicated, and
  if none are enclosed the symmetry line where the pocket lies is
  indicated instead. The numbers $i,j$ of enclosed PNs with
  bands~$n-1$ and $n+1$ are also indicated.}
\end{center}
\end{table}

Our implementation of \eq{ii} for electron pockets is as follows.  In
order to determine the internally connected occupied subvolumes
$V\bnj$, we cover the BZ with a uniform $300\times 300\times 300$ grid
and set up an undirected graph with nodes at those grid points where
band~$n$ is occupied, and whose edges are the links to the
nearest-neighbor points that are also occupied.
The problem becomes a standard one in graph theory, namely, to
identify the connected components of an undirected graph.  Once the
subvolumes $V\bnj$ have been identified in this way (the algorithm
``wraps around'' the BZ, so that the BZ boundaries are excluded from
the definition of $\delta V\bnj$), we go through the list of occupied
PNs $W_{n\alpha}$ and $W_{n-1,\alpha}$ and assign each of them to the
subvolume $V\bnj$ containing the closest grid point.  This allows us
to carry out the summations on the right-hand-side of \eq{ii} to
determine the Chern number $C\bna$ of each electron pocket
$S_{na}=\delta V\bnj$. The implementation of \eq{ie} for the hole-like
Fermi sheets is completely analogous.

The results are summarized in Table~\ref{table3}, and several features
can be readily understood from symmetry considerations. The vanishing
of the Chern numbers summed over all the sheets of any given band is a
consequence of inversion symmetry [\eq{sumC-Pinv}], and this explains
the values $C_{na}=0$ for bands six, eight, and nine with one Fermi
sheet each (the small hole pocket in band five does not enclose any
PNs, and so its Chern number vanishes trivially). According to
Sec.~\ref{sec:C-sum-rules}, in bands with several Fermi sheets
inversion symmetry still imposes $C_{na}=0$ on those which enclose
parity-invariant momenta (the points $\Gamma$, ${\rm H}$, ${\rm N}$
and ${\rm N'}$ in Fig.~\ref{fig3}). This accounts for all sheets in
band seven, as well as the central pocket~$10_1$ in
band ten. Note that the vanishing Chern numbers come about in
different ways for Fermi sheets belonging to groups VIII(a) and
VIII(b) in band seven: the latter do not enclose any PNs, while
the former enclose four PNs each with band eight (two
inversion-symmetric pairs).

This leaves the six satellite pockets in band ten. The pockets
$(10_2,10_3,10_4,10_5)$ located along $\Delta'$ do not enclose any
PNs. The pockets $(10_6,10_7)$, on the other hand, enclose a single WP
each. These are touchings with band nine, located along $\Delta$ at
$k_z = \mp 0.428\times 2\pi/a$ (see Table~\ref{table2}).
As their chiralities are reversed by mirror symmetry, the enclosing
pockets have opposite Chern numbers~$\mp 1$.

\begin{figure}
\centering\includegraphics[width=1.0\columnwidth]{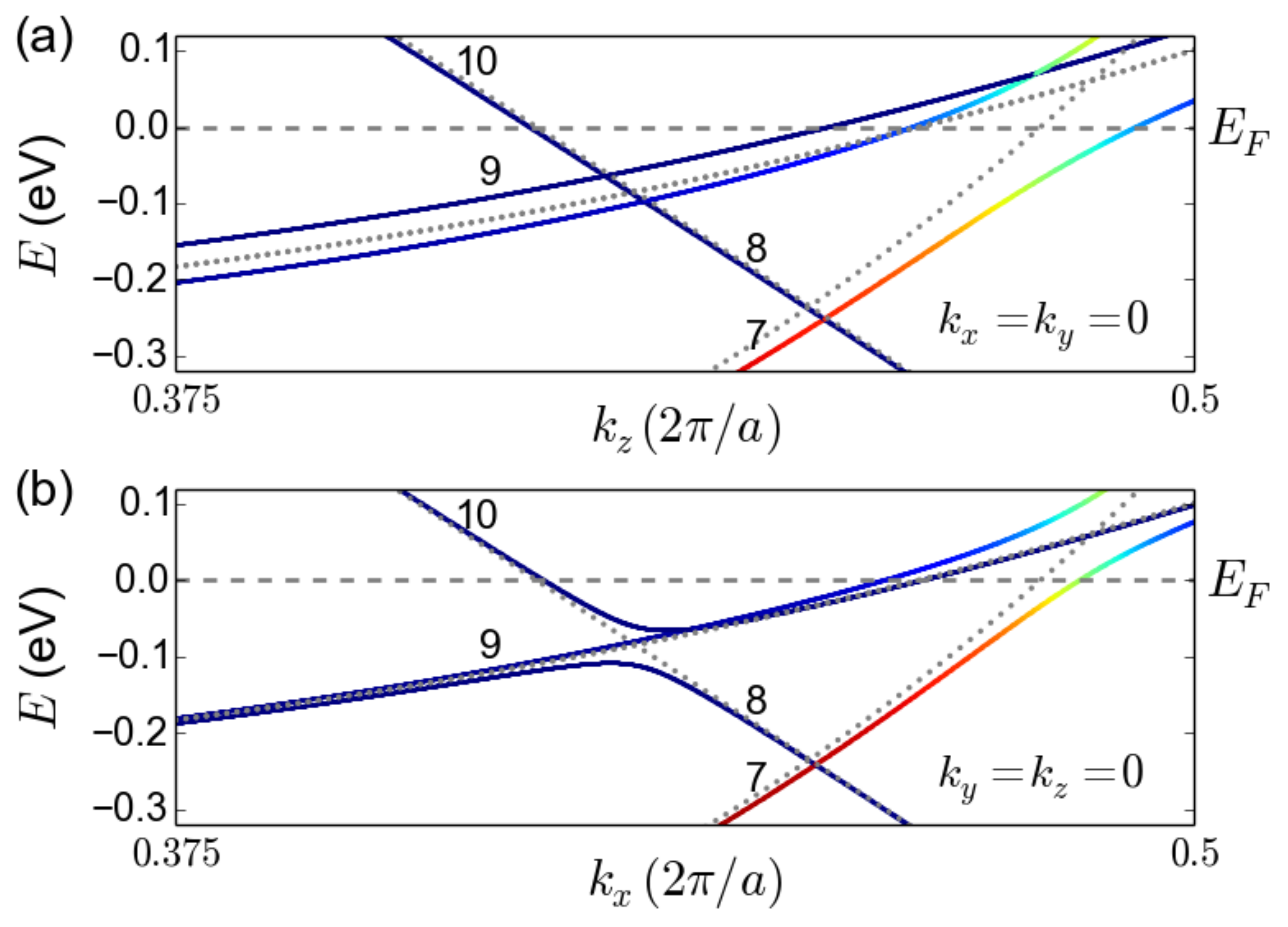}
\caption{(Color online.)  Detail of the spinor bandstructure of
  Fig.~\ref{fig2} along the line $\Delta$ close to the electron pocket
  $10_7$ [panel~(a)], and along the line $\Delta'$ close to the pocket
  $10_2$ [panel~(b)]. The spin-orbit-free energy bands of
  Fig.~\ref{fig9} are shown as dotted grey lines.}
\label{fig13}
\end{figure}

Figure~\ref{fig13} shows the energy bands near a pocket of type VII(b)
[panel (a)], and near one of type VII(a) [panel (b)].  Upon lowering
the Fermi level the electron pocket~$10_7$ in panel~(a) will shrink,
eventually turning into a hole pocket in band nine with the same Chern
number $+1$.  The situation with pocket $10_2$ in panel~(b) is less
clear, because of the degeneracy between bands nine and ten near the
bottom of band ten. That degeneracy can be lifted by tilting the
magnetization in certain directions (those for which no WPs are left
inside the now-isolated pocket, as will be discussed in
Sec.~\ref{sec:chern-glue}). Under those conditions pocket $10_2$
simply disappears as $\ef$ dips below the bottom of band ten; because
its Chern number vanishes, this does not violate the sum rule of
\eq{sumC}.

In order to understand how the two satellite pockets along $\Delta$
become different from the four along $\Delta'$ when SOC is included,
and why nonzero Chern numbers are induced on the former but not the
latter, it is instructive to follow the evolution of the relevant
bands in Figs.~\ref{fig13}(a-b) as SOC is turned on.  In the SOC-free
limit bands eight to ten [corresponding to minority-spin bands two to
four in Fig.~\ref{fig9}(b)] behave identically along $\Delta$ and
$\Delta'$, with a doubly-degenerate band (eight and nine on the left,
nine and ten on the right) crossing a singly-degenerate one (ten on
the left, eight on the right).  Along $\Delta$ in Fig.~\ref{fig13}(a),
SOC lifts the two-fold degeneracy between bands eight and nine, and
the crossing of band ten creates two nearby WPs, with the one between
bands nine and ten inducing a Chern number of $+1$ on the enclosing
pocket $10_7$.  Along $\Delta'$ in Fig.~\ref{fig13}(b), however, the
nature of the hybridization with band ten is very different. While no
mirror plane containing $\Delta$ survives when SOC is turned on, the
$M_z$ mirror plane containing $\Delta'$ does survive, and such a plane
can only harbour nodal rings, not chiral PNs.  The touching between
bands nine and ten in Fig.~\ref{fig13}(b) belongs to just such a nodal
ring -- in fact, the same one that appears as ring~$\delta$ in
Fig.~\ref{fig11}(b).  Since no chiral PNs are enclosed by pocket
$10_2$, its Chern number vanishes.

On their own, the pockets $(10_6,10_7)$ are an almost ideal
realization of the simplest $T$-broken Weyl semimetal. It is therefore
natural to ask whether their presence gives rise to ``Fermi-arc''
surface states connecting their projections onto the surface BZ, as in
actual Weyl semimetals. Although we have not explicitly calculated the
surface bands of bcc~Fe, that seems unlikely: almost everywhere on the
surface BZ there are projected bulk states at the Fermi level coming
from all the other (trivial) Fermi sheets, and their presence destroys
the stability of the putative Fermi arcs.

In summary, although chiral degeneracies abound in the spinor
bandstructure of bcc~Fe, most Fermi sheets are constrained by symmetry
to have zero Chern numbers. The only ones that are free from such
constraints are the isolated electron pockets $(10_6,10_7)$. In our
calculation their Chern numbers are $\mp 1$, turning bcc~Fe into a
topologically nontrivial metal.

\subsubsection{Chern numbers of non-isolated Fermi
  sheets}
\label{sec:chern-glue}

In Table~\ref{table3} we have assigned Chern numbers to all the sheets
making up the Fermi surface bcc~Fe, including those in groups I, II,
V, VII(a), and VIII(b) that are glued together
along nodal rings lying in the mirror plane [Fig.~\ref{fig11}(b)].
For these sheets, the meaning of the assigned Chern number requires a
more careful explanation, since Chern numbers are in principle only
well-defined for isolated Fermi sheets, and they can change when
sheets touch as a function of a control parameter.

\begin{figure}
\centering\includegraphics[width=0.7\columnwidth]{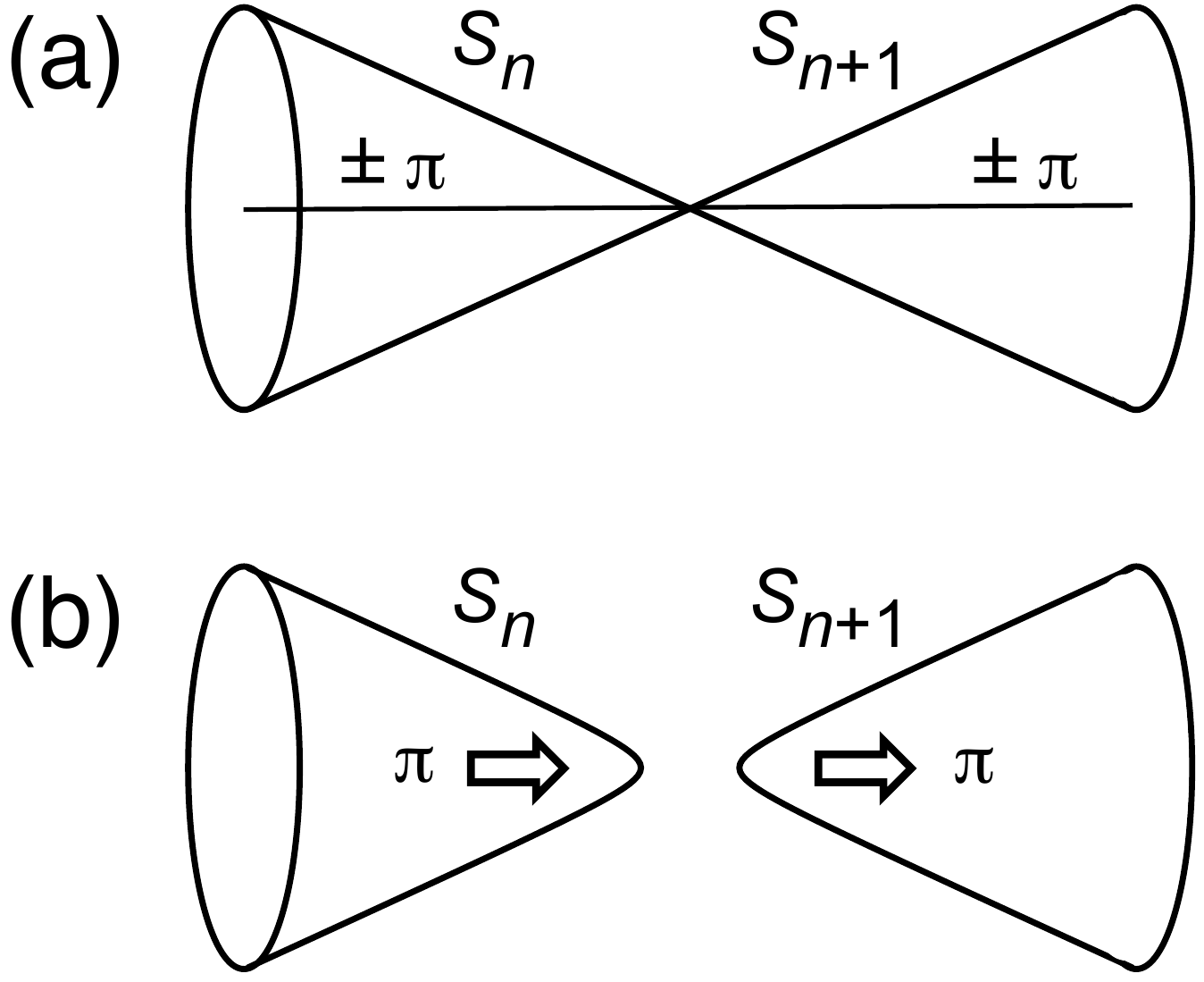}
\caption{(a)~Gluing together of two Fermi sheets along a
    symmetry-protected degeneracy line carrying a Berry flux of $\pi$
    modulo $2\pi$, indicated as $\pm \pi$. (b)~The symmetry has been
    broken weakly, gapping the degeneracy line and separating the
    Fermi sheets. A Berry flux of definite sign now exits one sheet
    and enters the other.}
\label{fig14}
\end{figure}

Figure~\ref{fig14}(a) shows a simplified sketch of a generic situation
in which Fermi sheets $S_n$ and $S_{n+1}$ are glued together at an
isolated ``gluing point'' where a nodal ring passes through $\ef$.  In
our case, the nodal ring lies in the $k_z=0$ plane and is protected by
$M_z$ symmetry. The local two-band $k\cdot p$ Hamiltonian in the
vicinity of the touching has a form like
\begin{equation}
H(k_x',k_y',k_z)=\ef+\alpha k_x' + \beta k_y' \sigma_3 + \gamma k_z \sigma_1
\label{eq:kdotp}
\end{equation}
where $k_x'$ and $k_y'$ are measured relative to the gluing point and
are parallel and normal to the nodal line respectively, and $\sigma_j$
are Pauli matrices.  The $E=0$ solutions are conical as shown in the
figure, with the nodal line passing through the conical intersection
point.  Each nodal line carries a Berry flux of $\pi$ modulo $2\pi$,
which is indicated as $\pm\pi$ in the figure; without breaking the
symmetry we cannot say which value applies.  The Fermi-sheet Chern
numbers reported in Table~\ref{table3} correspond to the integral of
the Berry curvature over the sheet in question, \eq{C-na}, but
neglecting these possible $\pi$-flux contributions.  This is the same
as the sum of the chiral charges of enclosed PNs that are unrelated to
the nodal ring, as will become clear shortly.

Now imagine that the symmetry that was protecting the nodal ring is
broken weakly; in our case this can be done by tilting the
magnetization~$\M$ away from the tetragonal axis.  Then the two bands
become gapped everywhere along the path except possibly at PNs along
it, corresponding to the addition of a term $\mu(k)\sigma_2$ to
\eq{kdotp}, where the argument $k$ is a reminder that $\mu$ can vary
along the path. Then a definite flux of $+\pi$ or $-\pi$ (depending on
the sign of $\mu$) flows along the previous path of the ring,
concentrated in a small vicinity (of size proportional to $\mu$)
around the path.  As shown in Fig.~\ref{fig14}(b), the previously
glued Fermi sheets now separate and become hyperboloids, and a
concentration of $\pi$ flux now exits one sheet and enters the
other. These $\pi$-flux contributions must be included when computing
the true Chern number of the sheet in the presence of the symmetry
breaking.

Note, however, that if the nodal ring enters a given Fermi sheet at
one point, it must exit at another, and the total contribution of the
$\pi$ fluxes will be of opposite sign and will cancel unless $\mu(k)$
crosses through zero and changes sign at some special point along the
ring, generating a new chiral PN at this location.  As long as this PN
is included in the census of enclosed PNs, the total Chern number of
the sheet will still be correctly given by the sum of enclosed chiral
charges.

In order to see how this works out in the present context, we have
investigated the consequences of breaking the $M_z$ mirror symmetry,
which is responsible for the nodal rings shown in Fig.~\ref{fig11}(b).
When we tilt $\M$ away from [001] the nodal rings evaporate, leaving
behind a few extra PNs, and the previously glued-together Fermi sheets
become detached. We can then safely determine their individual Chern
numbers from the populations of enclosed chiral degeneracies
(including the newly formed PNs).

Parameterizing the magnetization direction by polar and azimuthal
angles $\theta$ and $\phi$, we compute the properties at a series of
values of $\phi$ at a fixed polar angle of $\theta=20\degree$.  Since
parity remains a crystal symmetry even for arbitrary $(\theta,\phi)$,
the Chern numbers must still vanish individually for the Fermi sheets
surrounding parity-invariant momenta, i.e., sheets 7$_1$, $(7_6,7_7)$,
8, and~9. The Chern numbers of those sheets thus vanish unambiguously
in the limit $\theta\rightarrow0$.

According to Table~\ref{table3}, this leaves the four pockets
$(10_2,10_3,10_4,10_5)$ as the only ones that may acquire a non-zero
Chern number as a result of the breaking of $M_z$ symmetry.  We find
that the Chern number of any one pocket fluctuates between values of
$-1$, 0, and 1 as $\phi$ is varied; the Chern numbers of
inversion-related pairs of sheets are always opposite, as required by
parity; and that the sum of the four is always zero. As discussed
above, the nonzero individual values result from remnant WPs from the
vaporized nodal rings that are left lying inside the sheet in
question.  Those Weyl nodes move along the ring paths as a function
of~$\phi$, and at critical angles two nodes of opposite chirality join
sheet~9 with a pair of inversion-related pockets in band ten.  This
concerted touching event leads to a net transfer of Chern number
between the two pockets in the pair, mediated by sheet~9 whose Chern
number remains at zero. One such touching event between sheets~9
and~$10_2$, and the subsequent annihilation of a pair of remnant WPs
inside~$10_2$, are depicted in the Supplemental Material.\footnote{See
  Supplemental Material at
  \url{http://cfm.ehu.es/ivo/suppl-gosalbez15.pdf} for a depiction of
  touching events between Fermi sheets accompanied by a transfer of
  Chern number.}

\subsection{Anomalous Hall conductivity}
\label{sec:ahc-results}

The intrinsic AHC of bcc Fe has been calculated from first-principles
by several authors.  In Refs.~\onlinecite{yao-prl04} and
\onlinecite{wang-prb06} two different implementations of the Fermi-sea
approach of Sec.~\ref{sec:fermi-sea} were used, while in
Ref.~\onlinecite{wang-prb07} a Fermi-surface calculation was carried
out along the lines of Sec.~\ref{sec:fermi-loops}. In all these works
the focus was on the total intrinsic AHC (i.e., summed over all bands
in the Fermi-sea calculations, and over all Fermi sheets in the
Fermi-surface calculation). The resulting AHC values were found to be
in excellent agreement with one another, and in reasonable agreement
with measurements at room temperature.

In this section we provide a breakdown of the intrinsic AHC first into
band contributions and then into Fermi-sheet contributions.  Even
though only the total intrinsic AHC is a well-defined physical
observable, such decompositions provide insights into the role played
by band degeneracies.  It can be readily understood that nodal rings
do not contribute to the AHC: they form when $\M\parallel [001]$, in
which case the AHC vector $\K$ of \eq{K-decomp-a} is constrained by
symmetry to point along $k_z$, while instead the Berry flux carried by
the nodal rings is directed along the tangential direction on the
$(k_x,k_y)$ plane.  As for the chiral Weyl nodes, they do contribute
to the AHC by acting as sources and sinks of Berry curvature, as will
become apparent in the ensuing discussion.

\subsubsection{Band-by-band decomposition}
\label{sec:bbb}

We work with the AHC vector $\K=\sum_n\K_n$, where the contribution
from a single band is given by \eq{K-n}.  Following
Ref.~\onlinecite{wang-prb07} we calculate $\K_n$ in reduced
coordinates as an average over BZ slices of the Berry flux through the
occupied portions of each slice,
\beq
\label{eq:K-nj}
\Kred_{nj}=\frac{1}{2\pi}\a_j\cdot\K_n=\frac{1}{n_{\rm slice}}
\sum_{i=1}^{\rm n_{\rm slice}}\,\frac{\phi_{nj}(i)}{2\pi}\,,
\eeq
where the Berry flux $\phi_{nj}(i)$ [\eq{phi-nj}] is evaluated by
adding up the Berry phases around small plaquettes covering the
occupied portions of the BZ slice (see Sec.~\ref{sec:plaquettes}).
Since we take the magnetization to point along $\a_3=(0,0,a)$, the
only nonzero components of the AHC tensor are
$\sigma_{xy}=-\sigma_{yx}$, and $\Kred_{nj}$ only needs to be
evaluated for $j=3$. 

The calculated fluxes $\phi_{n3}$ are piecewise continuous functions
of $k_z$, jumping by integer multiples of $2\pi$ when passing through
chiral PNs.  In our calculations we initially divide the BZ into 800
evenly-spaced slices, and whenever $|\phi_{n3}(i+1)-\phi_{n3}(i)|>\pi$
we interpose four additional slices. This allows us to locate more
precisely the step discontinuities, and to distinguish them from rapid
but continuous variations in the flux.

\begin{table}
\begin{center}
\begin{ruledtabular}
\begin{tabular}{cdddd}
Band & \dent{\hspace{0.3cm}\Kred_{n3}^{(\Omega)}} & 
\dent{\hspace{0.4cm}\Kred_{n3}^{(\chi)}} 
 & \dent{\hspace{0.3cm}\Kred_{n3}} & 
\multicolumn{1}{c}{\hspace{-0.7cm}AHC}\\
$n$  & & & & \multicolumn{1}{c}{\hspace{-0.7cm}(S/cm)}  \\
\hline
1    &  2       &   0.51     &  2.51    &   -3394 \\
2    & -6       &   3.03     & -2.97    &    4018 \\
3    &  2       &   1.96     &  3.96    &   -5345 \\
4    &  6       &  -8.85     & -2.85    &    3840 \\
5    & -8.01    &   6.22     & -1.79    &    2413 \\
6    & -7.80    &   3.27     & -4.53    &    6111 \\
7    & 14.12    &  -6.44     &  7.68    &  -10368 \\
8    & -3.17    &  -0.31     & -3.48    &    4702 \\
9    & -0.53    &   1.33     &  0.80    &   -1076 \\
10   &  0.83    &  -0.72     &  0.11    &    -146 \\
Total& -0.56    &   0        & -0.56   &     755
\end{tabular}
\end{ruledtabular}
\caption{\label{table4}Band-by-band decomposition of the anomalous
  Hall conductivity (AHC) of bcc~Fe. In the three middle columns the
  dimensionless AHC is further decomposed according to \eq{K-nj-red}.}
\end{center}
\end{table}

Table~\ref{table4} shows the breakdown of the AHC into band
contributions. In the middle columns the dimensionless band
contribution given by \eq{K-nj} is further decomposed according to
\eq{K-n-b},
\beq 
\label{eq:K-nj-red}
\Kred_{n3}=\Kred_{n3}^{(\Omega)}+\Kred_{n3}^{(\chi)}\,,
\eeq
and in the last column the total contribution from band~$n$ is
converted to S/cm using $\sigma_{n,xy}=-(e^2/ha)\Kred_{n3}$. In
practice we calculate $\Kred_{n3}$ and $\Kred_{n3}^{(\chi)}$ from
\eqs{K-nj}{K-n-chi} respectively, and then obtain
$\Kred_{n3}^{(\Omega)}$ as the difference. Recall from
Sec.~\ref{sec:haldane} that depending on the placement of the cell
boundaries relative to the PNs, integer amounts may get transferred
between the two terms in \eq{K-nj-red}; the values in
Table~\ref{table4} are for a BZ cell located between $k_z=-\pi/a$ and
$k_z=\pi/a$.

All non-empty bands ($n\leq 10$) contribute to the AHC in
Table~\ref{table4}, as expected for a Fermi-sea formulation.  For
bands that are fully occupied ($n\leq 4$) the $\Omega$ term in
\eq{K-nj-red} becomes the slice Chern number of \eq{C-n-kz} evaluated
on the cell boundary at $k_z=\pm\pi/a$; the total $\Kred_{n3}$ of each
of those four bands is not quantizated, however, because of the
additional term in \eq{K-nj-red} contributed by the chiral
degeneracies.\footnote{This can be understood by noting that for a
  full band the summand in \eq{K-nj} becomes an integer slice Chern
  number, whose average is noninteger whenever PN-induced step
  discontinuities are present.}  Since those $\chi$ terms sum up to
zero over all bands, we can choose to focus exclusively on the
$\Omega$ terms, in which case the nonquantized part of the AHC is
apportioned entirely to the bands crossing~$\ef$ $(5\leq n\leq
10)$. This viewpoint will be adopted in the next section.

\subsubsection{Fermi-surface decomposition}
\label{sec:results-fs-decomp}

Following Sec.~\ref{sec:fermi-loops}, we write the AHC contribution
from each Fermi-sheet as an average over BZ slices of Fermi-loop Berry
phases [\eq{K-sheet}].
Instead of evaluating the Berry phases $\varphi_{naj}(i)$ directly on
the FS as done in Ref.~\onlinecite{wang-prb07}, we first compute Berry
fluxes and then remove the discontinuities, as explained below. In the
case of bands nine and ten with electron-like pockets, we determine
the flux $\phi_{nj}(i)$ through the occupied portions of each BZ slice
in the same way as in Sec.~\ref{sec:bbb}, by adding up the Berry
phases around small occupied plaquettes. \footnote{Contrary to
  Sec.~\ref{sec:bbb}, here we are only interested in the nonquantized
  part of the flux. For that purpose it is sufficient to apply the
  plaquette subdivision strategy of Sec.~\ref{sec:plaquettes} to the
  plaquettes along the edge between occupied and empty regions, and
  not to those that are completely surrounded by other occupied
  plaquettes.} In the calculations of Sec.~\ref{sec:fs-fe-chern} we
had identified the connected BZ subvolumes where band~$n$ is occupied.
Since in bcc~Fe each subvolume is bounded by a single Fermi sheet
$S_{na}$ (this is not true in general), that information can be used
to decompose the Berry flux as $\phi_{nj}(i)=\sum_a \phi_{naj}(i)$ by
assigning each plaquette to the appropriate subvolume.  Having done
this for every slice we then set $\varphi_{naj}(i)=\phi_{naj}(i)+2\pi
N_{naj}(i)$, choosing nonzero integers $N_{naj}(i)$ at the critical
slices so as to cancel the jumps in $\phi_{naj}(i)$. The procedure is
the same for the hole-like sheets in bands five to eight, except that
we switch from occupied to empty subvolumes and flip the signs of the
Berry fluxes.  As in Sec.~\ref{sec:bbb}, calculations only need to be
carried out for $j=3$.

\begin{table}
\begin{center}
\begin{ruledtabular}
\begin{tabular}{ccccr@{\hspace{-0.5cm}}c}
Band & Sheet      & Group & Distance to a PN&\multicolumn{2}{c}{AHC}\\ 
\dent{n}  & \dent{a}        & label      & \dent{(2\pi/a)} &  
\multicolumn{2}{c}{(S/cm)}\\
\hline
5    & 1          &   IV      & 0.30    &    9 & \phantom{$\times 1$}   \\
6    & 1          &   III     & 0.02    &  $-$274 &\phantom{$\times 1$}  \\
7    & 1          &   V       & 0.06    &   459 & \phantom{$\times 1$}  \\
7    & 2,3,4,5    &   VIII(a) & 0.01    & $-$203& $\times 4$\\
7    & 6,7        &   VIII(b) & 0.09    & 100   & $\times 2$\\
8    & 1          &   II      & 0.03    & 242 & \phantom{$\times 1$}\\
9    & 1          &   I       & 0.02    & 714 & \phantom{$\times 1$}\\
10   & 1          &   VI      & 0.10    & 58 & \phantom{$\times 1$}\\
10   & 2,3,4,5    &   VII(a)  & 0.31    & $-$1 & $\times 4$\\
10   & 6,7        &   VII(b)  & 0.01    & 167 & \phantom{$\times 1$}\\
Total          &&&& 759 & \phantom{$\times 1$}\\
\end{tabular}
\end{ruledtabular}
\caption{\label{table5}Decomposition of the AHC of bcc Fe into
  nonquantized Fermi-sheet contributions. Symmetry-related sheets are
  grouped in the same row, and they contribute equal amounts. The two
  pockets $(10_6,10_7)$ [group VII(b)] with opposite Chern
  numbers are treated as a single
  ``composite sheet'' and assigned a joint AHC contribution. The
  shortest distance from each Fermi sheet to a chiral point node (PN)
  on the same band is also indicated.}
\end{center}
\end{table}

The Fermi-sheet contributions to the AHC are listed in
Table~\ref{table5}.  Because these are nonquantized contributions,
they are only defined modulo $e^2/ha=1350$~S/cm.  This ambiguity
arises from the freedom to choose the branch cut of the Berry phase
$\varphi_{na3}$ on some reference slice (Sec.~\ref{sec:fermi-loops}),
and below we describe the choices we have made in order to arrive at
the values in the table.

\begin{figure}
\centering\includegraphics[width=1.0\columnwidth]{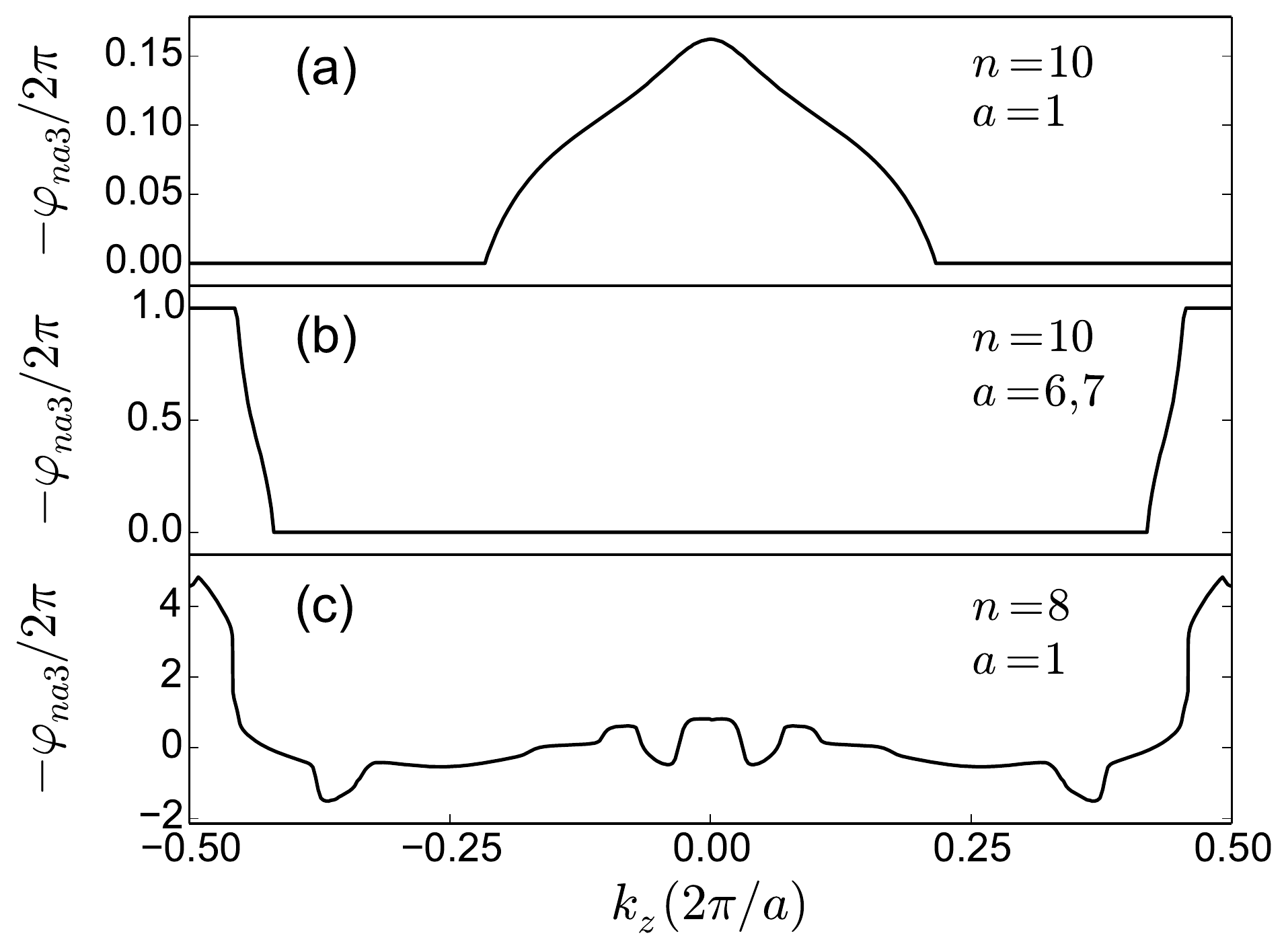}
\caption{Evolution with $k_z$ of the Berry phase along the
  intersection loops between a Fermi sheet and a Brillouin-zone slice,
  enforcing continuity from one slice to the next.  (a)~Fermi
  pocket~$10_1$. (b)~Topological pockets $(10_6,10_7)$, treated as a
  composite Fermi sheet. (c)~Fermi sheet~$8_1$.}
\label{fig15}
\end{figure}

For the sheets in groups~I, III, IV, V, VI, VII(a), VIII(a), and
VIII(b), all of which have zero Chern numbers, we arbitrarily set
$\varphi_{na3}=0$ on an ``empty'' slice below the Fermi sheet. As an
example, Fig.~\ref{fig15}(a) shows the evolution of the phase
$-\varphi_{na3}$ for $n=10$ and $a=1$ [the large pocket centered
at~$\Gamma$ in Fig.~\ref{fig10}(f)].  Upon hitting the pocket it first
rises continuously from zero, reaches a maximum, and then starts to
decrease, returning to zero at the top of the pocket.

The two pockets $(10_6,10_7)$ have nonzero Chern numbers. It is
therefore not possible to set the Berry phase to zero both at the
bottom and at the top of each of them, while at the same time
insisting on a continuous evolution across each
pocket.\cite{vanderbilt-prb14} Following Sec.~\ref{sec:fermi-loops} we
treat the two pockets as a composite Fermi sheet with an overall zero
Chern number, and assign them a joint AHC contribution. There are two
possibilities for setting $\varphi_{na3}=0$ on empty slices between
the two pockets: either in the wide region around $k_z=0$, or in the
narrow region around $k_z=\pi/a$. We have arbitrarily chosen the
former, and the resulting Berry-phase curve is shown in
Fig.~\ref{fig15}(b) (with the latter choice the entire curve would be
shifted downwards by one). The phase $-\varphi_{na3}(k_z)$ rises
quickly from zero to $2\pi$ while traversing the small pocket with
$C=+1$ on the right side of the panel, stays constant at $2\pi$ in the
narrow region between the two pockets, and finally drops rapidly back
to zero while traversing the periodic image of the pocket with $C=-1$
on the left side.

The third case we have encountered is that of the tubular Fermi sheet
in band eight. As in the first case considered above the Chern number
vanishes, but here there are no empty slices that can be used to set
the Berry phase to zero [see Fig.~\ref{fig10}(d)].  Instead, we have
made in Fig.~\ref{fig15}(c) the unique branch choice leading to the
correct total intrinsic AHC when summing the contributions from all
the Fermi sheets.  (We have verified that the curve $-\varphi_3(k_z)$
obtained by summing $-\varphi_{na3}(k_z)$ over all Fermi sheets agrees
with Fig.~5 of Ref.~\onlinecite{wang-prb07}.)  Note the very steep
variation in the Berry phase around $k_z=0.465\times 2\pi/a$: it is
caused by the tiny avoided crossing between sheets~$(8,9)$ in
Fig.~\ref{fig12} that was mentioned at the end of
Sec.~\ref{sec:fs-fe-overview}.

Regarding the magnitude of the nonquantized AHC contributions,
inspection of Table~\ref{table5} and Fig.~\ref{fig10} reveals that the
 most significant contributions tend to come
from large Fermi sheets with chiral PNs closeby. This makes
  intuitive sense, as can be seen by considering a typical Fermi
  pocket with zero Chern number and that can be completely enclosed
  inside a BZ cell.  Using Haldane's formulation with such a
  choice of BZ cell, its AHC contribution equals the dipole moment of
the surface-normal Berry-curvature [the term $\Kaomega$ in
\eq{Ksummary}]. This quantity scales with the pocket size, and is
enhanced by the presence of nearby sources and sinks of Berry
curvature. As a specific example, compare the pockets
  $(7_2,7_3,7_4,7_5)$ in group VIII(a) with the pockets $(7_6,7_7)$ in
  group VIII(b). Although they are almost identical in size and
shape, the former contribute twice as much AHC per pocket, by virtue
of being much closer to chiral PNs.

The joint AHC contribution from the topological pockets $(10_6,10_7)$
is considerable (about 20\% of the total), despite the fact that the
pockets themselves are rather small. In this case the AHC contribution
scales not with the size of the pockets but with the distance between
them, as can be seen from Fig.~\ref{fig15}(b).  An estimate is
provided by the $k$-space dipole moment between the mirror-symmetric
WPs enclosed by each of the two pockets,
$$(+1)\times0.428 + (-1)\times(-0.428)-1=-0.144$$ 
in units of $2\pi/a$ (note that the WP at $k_z=0.428\times 2\pi/a$ in
Table~\ref{table2} acts as a sink of Berry curvature in band nine, and
therefore it acts as a source in band ten).  This corresponds to a
nonquantized AHC contribution of $-0.144\times(-e^2/ha)=194$~S/cm,
close to the {\it ab initio} value of 167~S/cm in Table~\ref{table5};
the small overestimation has to do with the finite size of the
pockets.

We conclude by analyzing the evolution of the Fermi-sheet Chern
numbers and AHC upon varying the Fermi level. Inspection of
Fig.~\ref{fig13}(a) and Table~\ref{table2} shows that at the critical
value $\Delta\ef=0.07$~eV the Fermi level coincides with the energy of
a double-Weyl node between bands nine and ten, located along the
four-fold axis $\Delta$ at $k_z=0.48\times 2\pi/a$.  The evolution
with $\Delta\ef$ of sheets~9 and $10_7$ is depicted in the
Supplemental Material.\cite{Note9} At the critical value the two
sheets touch at the node, and there is a simultaneous touching between
sheets~9 and~$10_6$ at the mirror symmetric node in the lower half of
the BZ. Since the chiral charges at the touching points are $\pm 2$,
sheet~9 donates to~$10_7$ a Chern number of $+2$, and it
simultaneously receives a compensating amount from $10_6$. The net
result is that the Chern numbers on pockets $10_6$ and $10_7$ flip
sign, while that of sheet~9 remains at zero.  As for the AHC, the
joint contribution from the pair $(10_6,10_7)$ changes abruptly by a
nonquantized amount, but the total AHC remains continuous because of a
compensating discontinuity in the contribution from
sheet~9.\footnote{We do not observe any discernible effect on the AHC
  as the Fermi level passes through a Weyl node. Rapid variations in
  $\sigma_{xy}(\ef)$ are not generally associated with Weyl nodes
  at~$\ef$ as stated in Ref.~\onlinecite{chen-prb13}, but with avoided
  crossings.}
This example suggests that a measurement of the AHC cannot by
itself resolve the Berry topology of the Fermi sheets.

\section{Summary}
\label{sec:summary}

In summary, we have used first-principles methods to survey the
degeneracies in the spinor bandstructure of bcc~Fe, and to calculate
the chiral charges of those degeneracies. From the census of chiral
band touchings we then determined the Chern numbers of the individual
Fermi sheets, paying attention to the subtleties that arise when Fermi
sheets are not isolated, but glued together along nodal rings. We
found that when the magnetization points along the easy axis [001]
most Fermi sheets in bcc~Fe are topologically trivial, except for two
low-symmetry electron pockets along~$\Delta$ with Chern numbers~$\pm
1$, and four others along~$\Delta'$ with ill-defined Chern numbers due
to Fermi-gluing.

The systematic relations we derived between the Fermi-sheet Chern
numbers and the enclosed chiral charges are generally applicable to
any metal with broken $PT$ symmetry. In particular, we have considered
in our formal discussion the case of complex Fermi surfaces with
nested sheet structures, as well as the case where sheets with a
Luttinger anomaly are present. Combined with the efficient
steepest-descent strategy that was used to locate band degeneracies,
our algorithm for determining the FS Chern numbers could be useful in
high-throughput {\it ab initio} searches for topological metals.

The role played by chiral degeneracies in the intrinsic AHC was
carefully examined, confirming that they do not pose an impediment to
a bulk Fermi-surface formulation for the nonquantized part.  We
identified two different ways of decompositing the AHC (band by band
and in terms of Fermi sheets), and showed how the two decompositions
are related by dipole moments of the distribution inside the BZ of
chiral band touchings below the Fermi level. We carried out both
decompositions numerically for bcc~Fe, and found the Fermi-surface
decomposition to be particularly informative and physically
transparent: chiral degeneracies act as sources and sinks of Berry
curvature in $k$ space, and the distribution of Berry curvature across
the Fermi sheets in turn governs the nonquantized AHC response. So,
for example, Fermi sheets with chiral PNs very nearby tend to
contribute more to the AHC than otherwise similar Fermi sheets that
are farther from chiral PNs.

By showing that two out of eighteen Fermi sheets in bcc~Fe are
topologically nontrivial we have established that nonzero Chern
numbers are not only possible in principle in $T$-broken Fermi
surfaces, but that they actually occur in realistic ferromagnetic
bandstructures.  Further studies on other ferromagnets will be needed
before it becomes clear how common they are. Since symmetry was seen
to play a an important role, it would be interesting to explore
materials with other symmetries, e.g., a six-fold axis.  The fact that
parity imposes zero Chern numbers on most Fermi sheets in bcc~Fe
suggests that topologically nontrivial sheets may be more common in
$P$-broken metals. Known examples include MnSi and related
ferromagnetic compounds with the B20 structure, and also metals such
as LiOsO$_3$ that undergo ferroelectric-like polar
distortions.\cite{shi-natmat13}

By varying either the magnetization direction or the Fermi level, we
were able to change the Chern numbers on the Fermi surface of bcc
Fe. The process consisted of concerted chiral touching events
involving three electron-like Fermi sheets, with a large,
topologically trivial sheet in band nine mediating the transfer of
Chern number between two small enclosed pockets in band ten. The
experimental consequences of such topological transitions remain
largely unexplored. The manner in which the topological properties can
change as a function of other external parameters such as pressure,
e.g., via discrete Fermi-sheet reconnection or PN pair annihilation or
creation events, is also an attractive subject for future
investigation.

\begin{acknowledgments}

  This work was supported by grants No.~MAT2012-33720 from the
  Ministerio de Econom\'ia y Competitividad (Spain), No.~CIG-303602
  from the European Commission, by NSF Grant DMR-14-08838, and by a
  grant from the Simons Foundation (\#305025 to D.~Vanderbilt).
  
\end{acknowledgments}

\appendix

\section{Symmetry constraints}
\label{app:symm}

In this appendix we list the constraints imposed on several $k$-space
quantities by the spatial inversion symmetry $P$ of bcc~Fe, and also
by the additional mirror symmetry $M_z$ that is present when the
magnetization points along the easy axis $[001]$. The quantities of
interest are the energy eigenvalues $E_n(\k)$, the Berry curvature
$\bomega_n(\k)$, the chiral charge $\chi_{n\alpha}$, and the slice
Chern numbers $C_n(k_z)$ and $\wt{C}_n(k_z)$.

Inversion symmetry implies
\bea
\label{eq:eig-P}
E_n(-\k)&=&E_n(\k) \,, \\
\label{eq:curv-P}
\bomega_n(-\k)&=&\bomega_n(\k) \,, \\
\label{eq:chi-P}
\chi_{n\alpha'}&=&-\chi_{n\alpha}\,,
\eea
where \eq{chi-P} follows from \eqs{div}{curv-P}, and the node
$n\alpha'$ in \eq{chi-P} is the parity-reflected partner of $n\alpha$.
Evaluating \eq{C-n-kz} at $-k_z$ we find
\beq
\label{eq:C-about-0}
C_n(-k_z)=\frac{1}{2\pi}\int d^2k\,\Omega_{n,z}(-\k)=C_n(k_z)\,,
\eeq
where we first made the change of variables $k_x\rightarrow -k_x$,
$k_y\rightarrow -k_y$, and then used \eq{curv-P}.  Because of the
periodicity condition (\ref{eq:per}) we also have
\beq
\label{eq:C-about-pi}
C_n(\pi/a-k_z)=C_n(\pi/a+k_z)\,,
\eeq
so that $C_n(k_z)$ is even with respect to both $k_z=0$ and
$k_z=\pi/a$, and the same is of course true for $\wt{C}_n(k_z)$.

The presence of mirror symmetry $M_z$ implies
\bea
E_n(M_z\k)&=&E_n(\k) \,, \\
\Omega_{n,x}(M_z\k)&=&-\Omega_{n,x}(\k) \,, \\
\Omega_{n,y}(M_z\k)&=&-\Omega_{n,y}(\k) \,, \\
\Omega_{n,z}(M_z\k)&=&\Omega_{n,z}(\k) \,, \\
\label{eq:chi-mz}
\chi_{n\alpha'}&=&-\chi_{n\alpha}\,,
\eea
where $M_z(k_x,k_y,k_z)=(k_x,k_y,-k_z)$ and $n\alpha'$
is the mirror-reflected partner of $n\alpha$.

%

\end{document}